\documentclass[a4paper]{article}%
\usepackage{amssymb}
\usepackage{graphicx}
\usepackage{amsbsy}
\usepackage{amsmath}
\usepackage{cite}
\usepackage{slashed}
\usepackage{amsfonts}
\usepackage{hyperref}
\hypersetup{
    colorlinks   = true,
    linkcolor    = blue,
    citecolor    = blue,
}

\begin{document}
\thispagestyle{empty} \setcounter{page}{0} \begin{flushright}%

October 2018\\
\end{flushright}

\vskip          4.1 true cm

\begin{center}
{\huge The effective action for gauge bosons}\\[1.9cm]

\textsc{Jeremie Quevillon}$^{1}$
\textsc{, Christopher Smith}$^{2}$ 
\textsc{and Selim Touati}$^{3}$\vspace{0.5cm}\\[9pt]\smallskip
{\small \textsl{\textit{Laboratoire de Physique Subatomique et de Cosmologie,
}}}\linebreak{\small \textsl{\textit{Universit\'{e} Grenoble-Alpes,
CNRS/IN2P3, Grenoble INP, 38000 Grenoble, France}.}}
\\[1.9cm]\textbf{Abstract}\smallskip
\end{center}

\begin{quote}
\noindent 

By treating the vacuum as a medium, H. Euler and W. Heisenberg estimated the non-linear interactions between photons well before the advent of Quantum Electrodynamics. In a modern language, their result is often presented as the archetype of an Effective Field Theory (EFT). In this work, we develop a similar EFT for the gauge bosons of some generic gauge symmetry, valid for example for $SU(2)$, $SU(3)$, various grand unified groups, or mixed $U(1)\otimes SU(N)$ and $SU(M)\otimes SU(N)$ gauge groups. Using the diagrammatic approach, we perform a detailed matching procedure which remains manifestly gauge invariant at all steps, but does not rely on the equations of motion hence is valid off-shell. We provide explicit
analytic expressions for the Wilson coefficients of the dimension four, six, and eight operators as induced by massive scalar, fermion, and vector fields in generic representations of the gauge group. These expressions rely on a careful analysis of the quartic Casimir invariants, for which we provide a review using conventions adapted to Feynman diagram calculations. Finally, our computations show that at one loop, some operators are redundant whatever the representation or spin of the particle being integrated out, reducing the apparent complexity of the operator basis that can be constructed solely based on symmetry arguments.

\let\thefootnote\relax\footnotetext{$^{1}\;$jeremie.quevillon@lpsc.in2p3.fr}%
\ \footnotetext{$^{2}\;$chsmith@lpsc.in2p3.fr} \footnotetext{$^{3}%
\;$touati@lpsc.in2p3.fr}
\end{quote}

\newpage

\setcounter{tocdepth}{2}
\tableofcontents

\section{Introduction}

In 1936, H. Euler and W. Heisenberg calculated the non-linear interactions among photons for a constant Maxwell field, as induced by a spinor loop~\cite{EulerH36}. This has been an important step in the development of QED, and their result remains as the canonical example of an Effective Field Theory (EFT). That is, the idea that at energies below some cut-off scale $\Lambda$, all the effects of the degrees of freedom more massive than $\Lambda$ can be encoded as new interactions among the fields remaining active below $\Lambda$. This concept is central to modern phenomenology. The Fermi theory of the weak interactions~\cite{Fermi:1934sk} and the chiral Lagrangian of pions~\cite{chiralpion} have played an important role in the development of the Standard Model \cite{Weinberg:2009bg,Georgi:1994qn}. The methodology has since been used to define many other frameworks to either simplify the problem at hand, or to parametrize possible New Physics effects, for example for neutrino~\cite{Weinberg:1979sa}, nuclear~\cite{Kaplan:2000qu}, flavour~\cite{EFTflavour}, electroweak~\cite{EFTelectroweak}, and Higgs physics~\cite{EFThiggs}, or more globally for the Standard Model (SMEFT)~\cite{SMEFT}. Few developments have also been done regarding EFTs for dark matter~\cite{EFTdarkmatter}, inflation~\cite{EFTinflation} and cosmology~\cite{EFTcosmology}.

The purpose of the present paper is to generalize the Euler-Heisenberg (EH) result for photons to the gauge bosons of an arbitrary gauge group, with their effective interactions induced by loops of heavy fields in generic representations and of spin $0$, $1/2$, or $1$. Let us thus first recall a few facts about the EH Lagrangian, and some of its applications.

In QED, for energies below the electron mass $m_{e}$, the photons can interact between themselves only indirectly via virtual loops of electron-positron extracted from the vacuum. These interactions are suppressed by inverse powers of the electron mass compared with the Maxwell term and are thus very small. Integrating out the electron field in the QED Lagrangian lead to a tower of new photon interactions which should be Lorentz and gauge invariant, and respect parity invariance. The first non-trivial photon interaction corresponds to dimension-eight operators, the Euler-Heisenberg Lagrangian, and reads%
\begin{equation}
\mathcal{L}_{\text{EH}}=-\mathcal{F}+\frac{8}{45}\left( \frac{\alpha ^{2}}{%
m_{e}^{4}}\right) \mathcal{F}^{2}+\frac{14}{45}\left( \frac{\alpha ^{2}}{%
m_{e}^{4}}\right) \mathcal{G}^{2},  \label{EH}
\end{equation}%
with 
\begin{equation}
\mathcal{F}=\frac{1}{4}F_{\mu \nu }F^{\mu \nu }=\frac{1}{2}({\mathbf{B}}^{2}-%
{\mathbf{E}}^{2}),\quad \mathcal{G}=\frac{1}{8}\epsilon ^{\mu \nu \lambda
\rho }F_{\mu \nu }F^{\lambda \rho }={\mathbf{E}}\cdot {\mathbf{B}}\ ,
\end{equation}%
where ${\mathbf{B}}$ and ${\mathbf{E}}$ are the magnetic and electric fields, $\alpha =e^{2}/4\pi $ the fine structure constant, $e$ the electron electric charge, and $\epsilon _{\mu \nu \lambda\rho }$ the totally antisymmetric tensor. The first term in Eq.~(\ref{EH}), quadratic in the fields, is the classical Lagrangian corresponding to Maxwell's equations in vacuum. From it, one concludes that electromagnetic waves propagating in the vacuum cannot interact with each other, the superposition principle holds, and colliding light-by-light will not give rise to any scattering. However, this does not remain true once the corrections induced by the two last terms in Eq.~(\ref{EH}) are included. At the loop level, electrodynamics is nonlinear even in vacuum. In that sense, observing e.g. the scattering of light by light would be a tremendous confirmation of the quantum nature of QED.

Another consequence of the non-linearities in Eq.~(\ref{EH}) is the so-called vacuum magnetic birefringence. Two photons interact with an external field and this leads, in vacuum, to magnetic birefringence, namely to different indices of refraction for light polarized parallel and perpendicular to an external magnetic field. This property of the vacuum has never been observed, despite many dedicated searches. For example, attempts were made to measure the change of the polarization of a laser beam passing through an external strong magnetic field~\cite{Cadene:2013bva,DellaValle:2015xxa,Yamazaki:2016zfx}. The PVLAS \cite{Mignani:2016fwz} experiment is another approach to detect the vacuum birefringence, by measuring the degree of polarization of visible light from a Magnetar, i.e., a neutron star whose magnetic field is presumably very large ($B\sim 10^{13}G$). In that case, there is also an interesting interplay with well-motivated axion-like scenarios that could enhance the QED predictions (see for example~\cite{Capparelli:2017mlv}).

When discussing the Euler-Heisenberg result, one should also mention that the Born-Infeld (BI) electrodynamics~\cite{BI} contains similar nonlinear corrections to the Maxwell theory, at least from a classical point of view. It was motivated by the idea that there should be an upper limit on the strength of the electromagnetic field. Nevertheless, BI electrodynamics is peculiar, since BI-type effective actions arise in many different contexts in superstring theory~\cite{Tseytlin:1999dj}. In heavy-ion collisions, the ATLAS data on light-by-light scattering can exclude the QED BI scale $\sim $100 GeV~\cite{Ellis:2017edi}. It has been subsequently shown in Ref.~\cite{Ellis:2018cos} that the ATLAS data on $gg\rightarrow \gamma \gamma $ scattering enhances the sensitivity to $\lesssim 1$ TeV for the analogous dimension-8 operator scales (containing other combinations of gluon and electromagnetic fields). Searches for $\gamma \gamma $ production at possible future proton-proton colliders are an example of how one should complement the searches via dimension-6 SMEFT operators.

Returning to the purpose of this paper, generalizing EH to non-abelian gauge bosons present several challenges. As a first step, all the effective interactions up to dimension-eight can be constructed solely relying on gauge invariance. The non-linear nature of the field strength permits to construct many more operators than for QED. Operators involving three field strengths arise already at the dimension-six level, and were constructed some time ago in Ref.~\cite{SMEFT}. The most general basis of operators for QCD, up to dimension eight and without imposing the gluon Equation of Motion (EOM), was described in Ref.~\cite{Gracey17}.

Remains the task of actually computing the coupling constants of these operators, as induced by loops of heavy particles. To our knowledge, this has never been done before. To tackle this problem, there are two different approaches. First, the heavy particle field can be genuinely integrated out of the path integral. Several techniques are available to perform this integration and obtain the effective action at the one-loop level~\cite{Gaillard:1985uh,Cheyette:1987qz,EffAction,Drozd:2015rsp,Ellis:2016enq,Ellis:2017jns,Zhang:2016pja}. Though most powerful, the calculation has only been pushed up to the coefficients of dimension-six operators~\cite{EffAction,Drozd:2015rsp}. Another approach, which we will adopt in the present paper, is to actually compute the loop amplitudes, expand them in inverse power of the heavy particle mass, and match the result with that computed using effective interactions. Though most straightforward, several issues have to be addressed. Since the EOM should not be imposed to reproduce the generic effective action, loop amplitudes have to be computed off-shell. But then, gauge invariance is not automatic since the amplitudes are not physical, and special care is needed to ensure a proper matching onto gauge invariant operators.

To illustrate our procedure and explain in details how to deal with these aspects, we start in the next section by re-building the well-known effective interactions of photons, as induced by loops of massive fermions, scalars, or vector bosons. In particular, we point out that using a non-linear gauge is compulsory for the matching to succeed for massive vector fields, in agreement with Ref.~\cite{Boudjema86}. Then, we generalize this computation to gluonic effective interactions in Section~3, as induced by loops of massive fermions, scalars, or vector bosons in the fundamental representation of QCD. For the latter case, we use as prototypes the leptoquarks of the $SU(5)$ GUT, quantized using a non-linear gauge condition. As this is not fully standard, that construction is detailed in Appendix~\ref{AppGUT}.

Once the QCD case with heavy fields in the fundamental representation is fully under control, it is a simple matter to first generalize to arbitrary representations, and then to generic gauge groups. This is done in Section~4, where we discuss first the $SU(N)$ case, then show how to recover the previous results for $U(1)$ and $SU(3)$, and finally derive the mixed operators and their coefficients for non-simple gauge group like $U(1)\otimes SU(N)$ or $SU(N)\otimes SU(M)$. The most striking result of that section is that some operator combinations are never induced at one-loop, no matter the spin or representation of the heavy particle. For QCD, this means four instead of six operators are required to describe the four-gluon interaction, while only two instead of four operators are needed for the two gluon-two photon interaction. Throughout this section, the only technical difficulty is related to quartic Casimir invariants, which arise in the reduction of traces of four generators. From a group theory perspective, these invariants have been described in details before~\cite{Okubo81,vanRitbergenSV99}, but a more user-oriented review seems to be lacking. Therefore, we collect in Appendix~\ref{AppCasimir} all the relevant information, as well as the explicit values of the quartic invariant for simple Lie algebras of interest for particle physics.

\section{Photon effective interactions}

In the path integral formalism, the effective action is obtained by
integrating out some heavy fields~\cite{Dobado}. In general, this generates an infinite
number of effective couplings among the remaining light fields.
Renormalizability ensures that the effective couplings of dimension less than
four can be absorbed into the light-field Lagrangian free parameters, while
the other couplings are all finite and can be organized as a series in powers
of the inverse of the heavy mass~\cite{ApplequistC75}.

To set the stage, consider the QED generating functional%
\begin{equation}
Z_{QED}\left[  J^{\mu},\eta,\overline{\eta}\right]  =\int DA^{\mu}D\psi
D\overline{\psi}\;\exp i\int dx(\mathcal{L}_{QED}+\overline{\eta}%
\psi+\overline{\psi}\eta+J^{\mu}A_{\mu})\ ,
\end{equation}
with%
\begin{equation}
\mathcal{L}_{QED}=-\frac{1}{4}F_{\mu\nu}F^{\mu\nu}+\overline{\psi}(i \slashed{D}-m)\psi\ ,
\end{equation}
and $D^{\mu}$ the usual covariant derivative. We omit the gauge fixing term
and its associated ghosts. At very low energy, below $m$, only the photons are
active. To construct the effective theory valid in that limit, the fermion
field is integrated out. This can easily be carried out since the fermionic
path integral is gaussian when the sources $\eta,\overline{\eta}$ are set to
zero:%
\begin{align}
Z_{QED}\left[  J^{\mu},0,0\right]   & =\int DA^{\mu}\;\exp i\int dx\left\{
-\frac{1}{4}F_{\mu\nu}F^{\mu\nu}+J^{\mu}A_{\mu}\right\}  \times\det(i \slashed{D}-m)\\
& \equiv\int DA^{\mu}\;\exp i\int dx(\mathcal{L}_{eff}+J^{\mu}A_{\mu})\ .
\end{align}
Exponentiating the determinant, the QED effective Lagrangian is then
\begin{equation}
\mathcal{L}_{eff}=-\frac{1}{4}F_{\mu\nu}F^{\mu\nu}-iTr\ \ln(i \slashed{D}-m)\ .
\end{equation}
At this stage, several techniques are available to actually compute $\det(i \slashed{D}-m)$ perturbatively, as an inverse mass expansion. 

Probably the most universal and powerful way is using functional methods. For this approach, Gaillard \cite{Gaillard:1985uh} and Cheyette \cite{Cheyette:1987qz} introduced a manifestly gauge-covariant method of performing the calculation, using a Covariant Derivative Expansion (CDE). This elegant method simplifies evaluating the quadratic term of the heavy fields in the path integral to obtain the low-energy EFT, and was revived recently in Ref.~\cite{EffAction}. In particular, this work pointed out that under the assumption of degenerate particle masses one could evaluate the momentum dependence of the coefficients that factored out of the trace over the operator matrix structure, without specifying the specific UV model. In Ref.~\cite{Drozd:2015rsp}, it has been shown that this universality property can be extended without any assumptions on the mass spectrum, to obtain a universal result for the one-loop effective action for up to dimension-six operators. There the loop integrals have been computed for a general mass spectrum once and for all. This Universal One-Loop Effective Action~\cite{Drozd:2015kva,Drozd:2015rsp,Ellis:2016enq,Zhang:2016pja,Ellis:2017jns} is a general expression that may then be applied in any context where a one-loop path integral needs to be computed, as for example in matching new physics models to the Standard Model (SM) EFT.

\begin{figure}[t]
\begin{center}
\includegraphics[height=1.2315in,width=3.0346in]{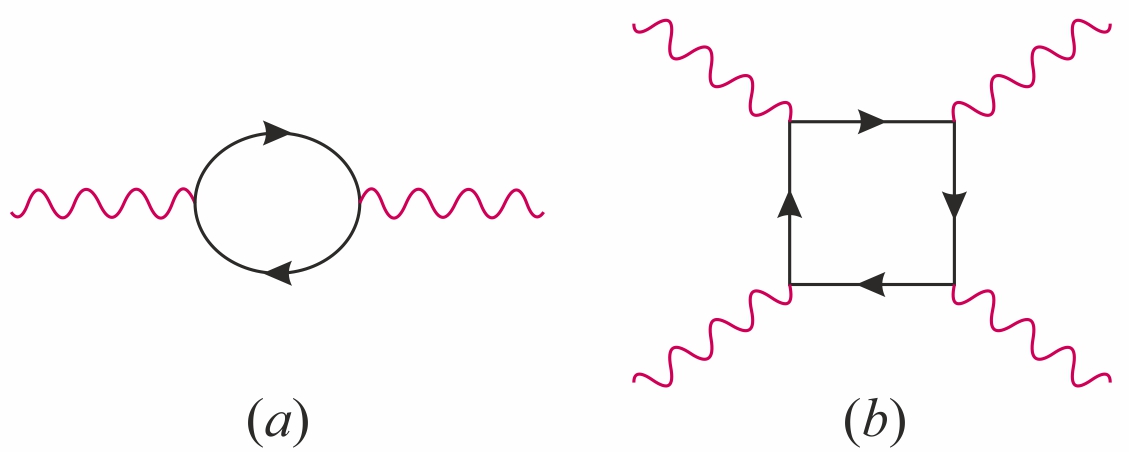}
\caption{Fermionic one-loop 1PI amplitudes generating the QED effective action up to dimension-eight operators. The six permutations of the photons are understood for diagram ($b$).}%
\label{EHphot}
\end{center}
\end{figure}

However, in the present work, we wish to
stick to the more pedestrian diagrammatic approach with external gauge fields, in which case one expands
$\det(i \slashed{D}-m)$ as%
\begin{equation}
\mathcal{L}_{eff}=-\frac{1}{4}F_{\mu\nu}F^{\mu\nu}+i\sum_{n=1}^{\infty}%
\frac{e^{n}}{n}Tr\left(  \frac{1}{i \slashed{\partial}-m}\slashed{A}\right)  ^{n}\ .
\end{equation}
Graphically, this series is represented by the tower of one-loop 1PI diagrams
shown in Fig.~\ref{EHphot}. The main advantage of expressing the effective
action in terms of 1PI diagrams is that well-tested automatic tools are
available to actually compute these loop amplitudes. In the present work, we
will rely on the Mathematica packages \textit{FeynArts}~\cite{FeynArts},
\textit{FeynCalc}~\cite{FeynCalc}, and \textit{Package X}~\cite{PackageX} (as
implemented through \textit{FeynHelpers}~\cite{FeynHelper}).%

For QED, all the diagrams with an odd number of photons vanish because they
are odd under charge conjugation (Furry's theorem~\cite{Furry37}). Let us
construct the effective couplings up to order $m^{-4}$. First, the
inverse-mass expansion of a charge-one fermion (of mass $m$ and quadri-momentum $p^{\mu}$) contribution to the photon
vacuum polarization is%
\begin{equation}
\Pi^{\mu\nu}(p^{2})=i\frac{8e^{2}}{(4\pi)^{2}}\left(  g^{\mu\nu}p^{2}-p^{\mu
}p^{\nu}\right)  \left\{  \frac{1}{6}D_{\varepsilon}+\frac{p^{2}}{30m^{2}%
}+\frac{p^{4}}{280m^{4}}+\mathcal{O}\left(  p^{6}/m^{6}\right)  \right\}  \;,
\end{equation}
with $D_{\varepsilon}=2/\varepsilon-\gamma+\log4\pi\mu^{2}/m^{2}$. The
corresponding effective interactions with two photons are%
\begin{equation}
\mathfrak{L}_{eff}^{(0+2)}=-\frac{1}{4}\left\{  1+\frac{\alpha}{3\pi
}D_{\varepsilon}\right\}  F_{\mu\nu}F^{\mu\nu}+\frac{\alpha}{60\pi m^{2}%
}F_{\mu\nu}\Box F^{\mu\nu}-\frac{\alpha}{560\pi m^{4}}F_{\mu\nu}\Box^{2}%
F^{\mu\nu}+\mathcal{O}(m^{-6})\;.
\end{equation}
With four photons, the amplitude matches onto the two couplings%
\begin{equation}
\mathfrak{L}_{eff}^{(4)}=\frac{\alpha^{2}}{90m^{4}}(F_{\mu\nu}F^{\mu\nu}%
)^{2}+\frac{7\alpha^{2}}{360m^{4}}(F_{\mu\nu}\tilde{F}^{\mu\nu})^{2}%
+\mathcal{O}(m^{-6})\;,
\end{equation}
where the dual field strength is defined as $\tilde{F}^{\mu\nu}= \frac{1}{2} \varepsilon
^{\mu\nu\rho\sigma}F_{\rho\sigma}$, so that $(F_{\mu\nu}\tilde{F}^{\mu\nu
})^{2}=2(F_{\mu\nu}F^{\mu\nu})^{2}-4F_{\mu\nu}F^{\nu\rho}F_{\rho\sigma
}F^{\sigma\mu}$. The divergent term is the usual photon wavefunction
renormalization, the first derivative term yields the Uehling
interaction~\cite{Uehling35}, and the Euler-Heisenberg effective
couplings~\cite{EulerH36} are the non-derivative $\mathcal{O}(m^{-4})$ terms.

A word is in order concerning the derivative coupling. In most operator
bases~\cite{SMEFT}, it is eliminated using the equation of motion
(EOM) as%
\begin{equation}
F_{\mu\nu}\Box F^{\mu\nu}=F_{\mu\nu}\partial^{\rho}\partial_{\rho}F^{\mu\nu
}=F_{\mu\nu}\partial^{\rho}\partial^{\mu}F_{\rho}^{\,\,\,\nu}+F_{\mu\nu
}\partial^{\rho}\partial^{\nu}F_{\,\,\,\rho}^{\mu}=-2\partial^{\mu}F_{\mu\rho
}\partial_{\nu}F^{\nu\rho}=-2j_{\nu}j^{\nu}\;,
\end{equation}
where the Jacobi identity $\partial_{\mu}F_{\rho\nu}-\partial_{\rho}F_{\mu\nu
}+\partial_{\nu}F_{\mu\rho}=0$ has been used in the first equality, followed by
integration by part, and finally the equation of motion $\partial_{\mu}%
F^{\mu\nu}=j^{\nu}$. This makes sense physically, since the only impact of the
Uehling potential is on the interaction between currents, at non-zero momentum
transfer. Yet, in the QED effective theory considered here, all the fermions
have been integrated out and $\partial_{\mu}F^{\mu\nu}=0$. This illustrate a
generic feature of the effective action formalism, where all the effect of the
heavy fields are encoded into effective couplings among light fields at the
path integral (i.e., quantum) level. At no stage are the light fields assumed
on-shell. So, some effective interactions may actually never contribute to
physical processes, even though they are required to fully encode the
underlying dynamics of the heavy field.

\begin{figure}[t]
\begin{center}
\includegraphics[height=1.2315in]{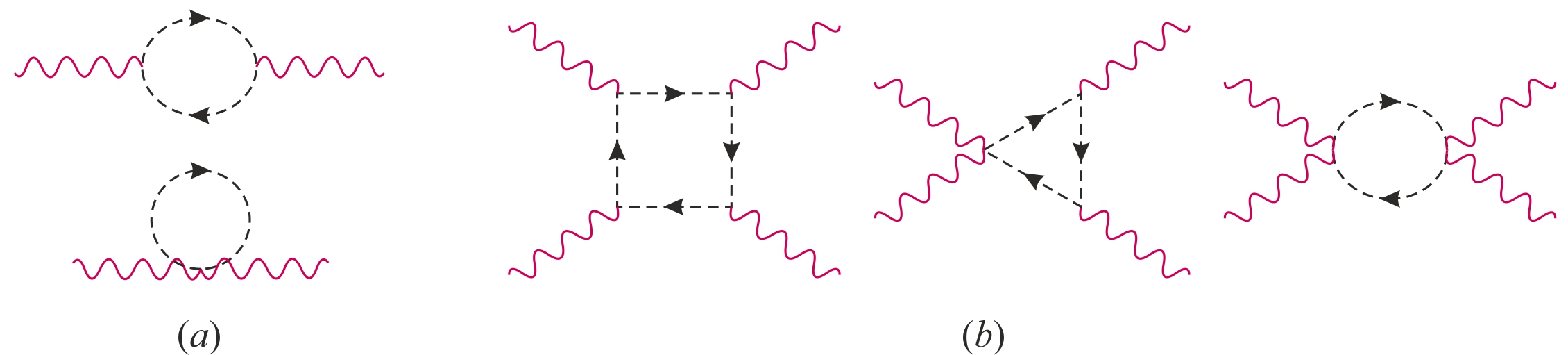}
\caption{Scalar one-loop 1PI amplitudes generating the QED effective action up to dimension-eight operators. Permutations of the photons are understood for diagrams ($b$). For massive vector bosons, the topologies are the same but one should also include the appropriate would-be Goldstone and ghost diagrams.}%
\label{EHphotSV}
\end{center}
\end{figure}

At this stage, it should also be clear that the effective couplings can be
constructed a priori. Using only the requirement of QED gauge invariance, the
most general basis is%
\begin{align}
\mathfrak{L}_{eff}  & =-\frac{1}{4}\left\{  1+\alpha_{0}\frac{e^{2}}{4!\pi
^{2}}\right\}  F_{\mu\nu}F^{\mu\nu}+\alpha_{2}\frac{e^{2}}{5!\pi^{2}m^{2}%
}\partial^{\mu}F_{\mu\nu}\partial_{\rho}F^{\rho\nu}+\alpha_{4}\frac{e^{2}%
}{6!\pi^{2}m^{4}}\partial^{\mu}F_{\mu\nu}\Box\partial_{\rho}F^{\rho\nu
}\nonumber\\
& +\gamma_{4,1}\frac{e^{4}}{6!\pi^{2}m^{4}}(F_{\mu\nu}F^{\mu\nu})^{2}%
+\gamma_{4,2}\frac{e^{4}}{6!\pi^{2}m^{4}}(F_{\mu\nu}\tilde{F}^{\mu\nu}%
)^{2}+\mathcal{O(}m^{-6})\;.\label{EffPhotons}%
\end{align}
The derivative operators are rewritten in a form that makes the EOM manifest.
This will prove useful when comparing with the non-abelian results in the next
section, for which this choice of operator basis is far more convenient. The
nomenclature adopted throughout the paper is to denote by $\alpha_{i}$,
$\beta_{i}$, $\gamma_{i}$ the two, three, four-field strength couplings of
inverse mass dimension $i$. Only the specific values of these coefficients
encode the information about the heavy field, and we give in
Table~\ref{TableU1} the results for a scalar, fermion, and vector boson.
Note that the sole purpose of the rather unconventional normalization of
the coefficients in Eq.~(\ref{EffPhotons}) is to increase the readability of Table~\ref{TableU1}. It is designed to make the coefficients appear as simple $\mathcal{O}(1)$ fractions for the fermion case.%

\begin{table}[t] \centering
$
\begin{tabular}[c]{cccccc}\hline
& $\alpha_{0}$ & $\alpha_{2}$ & $\alpha_{4}$ & $\gamma_{4,1}$ & $\gamma_{4,2}%
$\smallskip\\\hline
\multicolumn{1}{r}{Scalar} & $\dfrac{1}{2}D_{\varepsilon}Q^{2}$ & $-\dfrac
{1}{8}Q^{2}$ & $\dfrac{3}{56}Q^{2}$ & $\dfrac{7}{32}Q^{4}$ & $\dfrac{1}%
{32}Q^{4}\rule[-0.18in]{0in}{0.4in}$\\
\multicolumn{1}{r}{Fermion} & $2D_{\varepsilon}Q^{2}$ & $-Q^{2}$ & $\dfrac
{9}{14}Q^{2}$ & $\dfrac{1}{2}Q^{4}$ & $\dfrac{7}{8}Q^{4}\rule[-0.18in]%
{0in}{0.4in}$\\
\multicolumn{1}{r}{Vector} & $-\dfrac{21D_{\varepsilon}+2}{2}Q^{2}$ &
$\dfrac{37}{8}Q^{2}$ & $-\dfrac{159}{56}Q^{2}$ & $\dfrac{261}{32}Q^{4}$ &
$\dfrac{243}{32}Q^{4}\rule[-0.18in]{0in}{0.4in}$\\\hline
\end{tabular}
$
\caption{Wilson coefficients of the effective photon operators for a scalar, fermion, and vector boson of electric charge $Q$. For the latter, the matching of 1PI amplitudes onto the $U(1)$-gauge-invariant operators of Eq.~(\ref{EffPhotons}) is possible only when using a non-linear gauge for the massive vector bosons, and the quoted values for $\alpha_{1,2,3}$ are specific to that gauge ($\kappa=1$ in Eq.~(\ref{LNLG}) and~(\ref{GaugeDep})).}
\label{TableU1}
\end{table}

The calculation in the scalar case is very similar to that for fermions and
present no particular difficulty (see Fig.~\ref{EHphotSV}). On the other hand, that for vectors
circulating in the loop is far less straightforward. Let us take the SM, where
the electroweak gauge bosons acquire their masses through the Higgs mechanism.
Working in the 't Hooft-Feynman gauge, the amplitude does not satisfy the QED
ward identities when the photons are off-shell. Consequently, the four-photon
amplitude matches onto the local $\mathcal{O}(m^{-4})$ effective operators
only when the four photons are on-shell~\cite{FanchiottiGS72}, and the usual
procedure to construct the effective action breaks down. The problem
originates in the gauge-fixing procedure. In the usual $R_{\xi}$ gauge, one
adds the term%
\begin{equation}
\mathfrak{L}_{gauge-fixing}^{R_{\xi},linear}=-\frac{1}{\xi}|\partial^{\mu}W_{\mu}^{+}+\xi
M_{W}\phi^{+}|^{2}\ ,\label{Rksi}%
\end{equation}
with $\phi^{\pm}$ the would-be Goldstone (WBG) scalars associated to $W^{\pm}%
$, and this explicitly breaks $U(1)_{QED}$. Though the photon vacuum
polarization remains transverse and matches onto the effective operators in
Eq.~(\ref{EffPhotons}), the off-shell four-photon amplitude is not gauge
invariant and requires more operators already at the $\mathcal{O}(m^{-2}%
)$~\cite{DongJZ93}. Of course, physical processes have to be gauge invariant,
so this should have no consequence. But in practice, adding non-gauge
invariant operators in the effective Lagrangian is not very appealing. One
could attempt to solve this problem by working in the unitary gauge, for which
the $W$ couplings to the photon derive from%
\begin{equation}
\mathcal{L}_{U-gauge}=-\frac{1}{2}(D_{\mu}W_{\nu}^{+}-D_{\nu}W_{\nu}%
^{+})(D^{\mu}W^{-\nu}-D^{\nu}W^{-\mu})+ieF^{\mu\nu}W_{\mu}^{+}W_{\nu}%
^{-}+M_{W}^{2}W_{\mu}^{+}W^{-\mu}\ ,
\end{equation}
where $D_{\mu}W_{\nu}^{\pm}=\partial_{\mu}W_{\nu}^{\pm}\mp ieA_{\mu}W_{\nu
}^{\pm}$. The magnetic moment term $F^{\mu\nu}W_{\mu}^{+}W_{\nu}^{-}$,
gauge-invariant by itself, is fixed by the underlying $SU(2)_{L}$ gauge
symmetry. As shown in Ref.~\cite{g2}, its presence ensures a proper
high-energy behavior for scattering amplitudes. However, this is not
sufficient to ensure a correct behavior off-shell, and the matching fails
again~\cite{PreucilH17}.

A better way to proceed is to enforce a non-linear gauge condition where
$\partial^{\mu}W_{\mu}^{\pm}\rightarrow D^{\mu}W_{\mu}^{\pm}=\partial^{\mu
}W_{\mu}^{\pm}\pm ieA^{\mu}W_{\mu}^{\pm}$ in Eq.~(\ref{Rksi}).\ This closely
parallels the constraint one needs to impose to construct the
CDE~\cite{EffAction}. In the diagrammatic approach, as shown in
Ref.~\cite{Boudjema86}, the four-photon amplitude is then gauge invariant,
even off-shell. We checked this explicitly using the dedicated FeynArts model
file~\cite{FeynArtsNLG} for the SM in the non-linear gauge, and indeed found a
consistent off-shell matching on the Euler-Heisenberg operators. The result in
that gauge for all the coefficients is shown in Table~\ref{TableU1}. It should
be clear though that the first three coefficients are gauge-dependent, and
only $\gamma_{4,1}$ and $\gamma_{4,2}$ are physical. To investigate this
feature, let us set the gauge fixing term as~\cite{BaceH75,GavelaGMS81}%
\begin{equation}
\mathfrak{L}_{gauge-fixing}^{non-linear}=-\frac{1}{\xi}|\partial^{\mu}W_{\mu}^{+}+i\kappa
eA^{\mu}W_{\mu}^{+}+\xi M_{W}\phi^{+}|^{2}\ ,
\label{LNLG}
\end{equation}
which permits to interpolate between the linear ($\kappa=0$) and the
$U(1)$-gauge-invariant non-linear ($\kappa=1$) gauge. The inverse-mass expansion 
of the photon vacuum polarization in the 't Hooft-Feynman gauge 
($\xi=1$) then gives $\kappa$-dependent coefficients:%
\begin{equation}
\alpha_{0}=-\dfrac{12\kappa+9}{2}D_{\varepsilon}-1\ ,\ \ \alpha_{2}%
=\dfrac{20\kappa+17}{8}\ ,\ \ \alpha_{4}=-\dfrac{84\kappa+75}{56}%
\ .\label{GaugeDep}%
\end{equation}
Of course, these gauge dependences are unphysical. At very low energy, when the photon remains as the only active degree of freedom, the first coefficient is absorbed into the photon field as the wavefunction renormalization constant while the other two do not contribute since $\partial_{\mu}F^{\mu\nu}=0$. If some fields remain active such that $\partial_{\mu}F^{\mu\nu}=j^{\nu}\neq0$, then other types of processes are also present. In that case, the $\alpha_{2}$ operator should be eliminated in favor of the dimension-six $j_{\mu}j^{\mu}/m^{2}$ operator, for which other diagrams occur. In the SM,
even if the fields in the current $j^{\mu}$ are not coupled directly to the $W^{\pm}$, they are necessarily coupled to the $Z$ boson. The $\kappa$ dependence of the $W^{\pm}$ contributions to the $Z\gamma$ and $ZZ$ vacuum polarization~\cite{GRACE} must cancel that of $\alpha_{2}$, so that the coefficient of the $j_{\mu}j^{\mu}/m^{2}$ operator ends up gauge-invariant and physical. The conclusion is thus that in the SM, it is not consistent to define the Uehling potential in terms of the $F_{\mu\nu}\Box F^{\mu\nu}$ operator, and one must use the effective four-fermion operators instead. After
all, this is rather natural since the Uehling potential is only relevant when some fermion fields remain active.

\section{Gluon effective interactions}

\begin{figure}[t]
\begin{center}
\includegraphics[height=1.2012in,width=4.4399in]{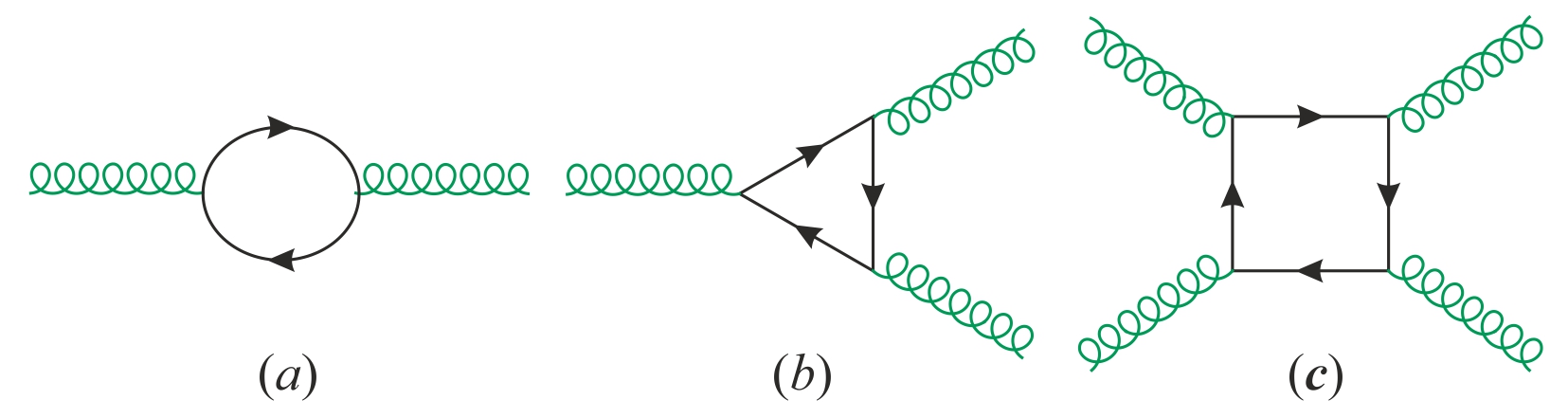}
\caption{Fermionic one-loop 1PI amplitudes generating the gluonic effective action. Permutations of the gluons are understood for diagram $(b)$ and $(c)$. As for QED in Fig.~\ref{EHphotSV}, additional diagrams are understood for the scalar and vector case.}
\label{EHglue}
\end{center}
\end{figure}

The effective action for gluon fields is constructed in the same way as for
photons, using the diagrammatic approach. For example, integrating out a heavy
fermion generates%
\begin{align}
\mathcal{L}_{eff} &  =-\frac{1}{4}G_{\mu\nu}^{a}G^{a,\mu\nu}-iTr\ \ln(i \slashed{D}-m)\nonumber\\
&  =-\frac{1}{4}G_{\mu\nu}^{a}G^{a,\mu\nu}+i\sum_{n=1}^{\infty}\frac{e^{n}}%
{n}Tr\left(  \frac{1}{i \slashed{\partial}-m} \slashed{G}^{a}T^{a}\right)  ^{n}\ ,
\end{align}
where $T^{a}$ are the $SU(3)$ generators, and the trace carries over both
Dirac and color space. This generates the series of 1PI diagrams shown in
Fig.~\ref{EHglue} where, contrary to QED, the odd-number of gluon amplitudes
do not vanish. Another difference with respect to QED is the non-linear
nature of the field strength, which blurs the relationship between the leading
inverse-mass power of a given diagram and the number of external gluons. The
most striking consequence is that the three and four-gluon diagrams are not
finite. Actually, since these infinities both correspond to the
renormalization of the same operator $G_{\mu\nu}^{a}G^{a,\mu\nu}$, they must
be coherent with that obtained from the two-gluon vacuum polarization. Let us
see how this happens in more details.

As a first step in the calculation of the effective action, let us construct
the most general basis of operators up to $\mathcal{O}(m^{-4})$. With two
field strengths, the operators are simple generalizations of those for QED:%
\begin{align}
\mathfrak{L}_{eff}^{(0+2)}  & =-\frac{1}{4}\left\{  1+\alpha_{0}\frac
{g_{S}^{2}}{4!\pi^{2}}\right\}  G_{\mu\nu}^{a}G^{a,\mu\nu}\nonumber\\
& +\alpha_{2}\frac{g_{S}^{2}}{5!\pi^{2}m^{2}}D^{\nu}G_{\nu\mu}^{a}D_{\rho
}G^{a,\rho\mu}+\alpha_{4}\frac{g_{S}^{2}}{6!\pi^{2}m^{4}}D^{\nu}G_{\nu\mu}%
^{a}D^{2}D_{\rho}G^{a,\rho\mu}\;,
\end{align}
where $G_{\mu\nu}^{a}=\partial_{\mu}A_{\nu}^{a}-\partial_{\nu}A_{\mu}%
^{a}+gf^{abc}A_{\mu}^{b}A_{\nu}^{c}$ and $D_{\rho}G_{\mu\nu}^{a}%
=(\partial_{\rho}\delta^{ac}+gf^{abc}G_{\rho}^{b})G_{\mu\nu}^{c}$. To see that
there can be only one derivative operator per inverse-mass
order~\cite{SMEFT}, first remark that all the derivatives can be move
to act on one of the field strength by partial integration. Then, only one
ordering of the covariant derivatives is relevant since commuting them
generates an additional field strength, $[D^{\rho},D^{\sigma}]G_{\mu\nu}%
^{a}=gf^{abc}G_{\rho\sigma}^{b}G_{\mu\nu}^{c}$. Finally, combining this with
the Bianchi identity%
\begin{equation}
D_{[\mu}G_{\rho\sigma]}^{a}=D_{\mu}G_{\rho\sigma}^{a}+D_{\rho}G_{\sigma\mu
}^{a}+D_{\sigma}G_{\mu\rho}^{a}=0\;,
\end{equation}
these operators can be written as manifestly vanishing under the EOM for the
field strength, $D^{\mu}G_{\mu\nu}^{a}=0$. Let us stress though that the EOM
are not used at any stage, since using them would render the matching impossible.

With three-gluon field strengths, there is only one operator at $\mathcal{O}%
(m^{-2})$ but many at $\mathcal{O}(m^{-4})$. However, upon partial
integration, use of the Bianchi identity, and discarding terms involving four
or more field strengths, only two inequivalent contractions
remain~\cite{Simmons89}. Here again, we choose them to be manifestly vanishing
under the field strength EOM:%
\begin{align}
\mathfrak{L}_{eff}^{(3)}  & =\beta_{2}\frac{g_{S}^{3}}{5!\pi^{2}m^{2}}%
f^{abc}G_{\mu}^{a\;\nu}G_{\nu}^{b\;\rho}G_{\rho}^{c\;\mu}\\
& +\beta_{4,1}\frac{g_{S}^{3}}{6!\pi^{2}m^{4}}f^{abc}G^{a,\mu\nu}D^{\alpha
}G_{\mu\nu}^{b}D^{\beta}G_{\alpha\beta}^{c}+\beta_{4,2}\frac{g_{S}^{3}}%
{6!\pi^{2}m^{4}}f^{abc}G^{a,\mu\nu}D^{\alpha}G_{\alpha\mu}^{b}D^{\beta
}G_{\beta\nu}^{c}\;.
\end{align}

At the four-field strength level, the operators up to $\mathcal{O}(m^{-4})$
contain no covariant derivatives. To reach a minimal number of operators, we
use the generalization of the QED identity:%
\begin{equation}
G_{\mu\nu}^{a}\tilde{G}^{b,\mu\nu}G_{\rho\sigma}^{c}\tilde{G}^{d,\rho\sigma
}=G_{\mu\nu}^{a}G^{c,\mu\nu}G_{\rho\sigma}^{b}G^{d,\rho\sigma}+G_{\mu\nu}%
^{a}G^{d,\mu\nu}G_{\rho\sigma}^{b}G^{c,\rho\sigma}-4G_{\mu\nu}^{a}G^{c,\nu
\rho}G_{\rho\sigma}^{b}G^{d,\sigma\mu}\;,\label{IdGGGG}%
\end{equation}
and note that no contractions with the totally symmetric tensor $d^{abc}$
occurs because those are reduced using (see Appendix~\ref{AppCasimir})%
\begin{equation}
3d^{abe}d^{cde}=\delta^{ac}\delta^{bd}-\delta^{ab}\delta^{cd}+\delta
^{ad}\delta^{bc}+f^{ace}f^{bde}+f^{ade}f^{bce}\;.\label{IddSU3}%
\end{equation}
Contractions with both $f$ and $d$ tensors vanish identically owing to their
mixed symmetry properties. This leaves six $\mathcal{O}(m^{-4})$ operators for
$\mathfrak{L}_{eff}^{(4)}$:%
\begin{align}
\mathfrak{L}_{eff}^{(4)}  & =\gamma_{4,1}\frac{g_{S}^{4}}{6!\pi^{2}m^{4}%
}G_{\mu\nu}^{a}G^{a,\mu\nu}G_{\rho\sigma}^{b}G^{b,\rho\sigma}+\gamma
_{4,2}\frac{g_{S}^{4}}{6!\pi^{2}m^{4}}G_{\mu\nu}^{a}\tilde{G}^{a,\mu\nu
}G_{\rho\sigma}^{b}\tilde{G}^{b,\rho\sigma}\nonumber\\
& +\gamma_{4,3}\frac{g_{S}^{4}}{6!\pi^{2}m^{4}}G_{\mu\nu}^{a}G^{b,\mu\nu
}G_{\rho\sigma}^{a}G^{b,\rho\sigma}+\gamma_{4,4}\frac{g_{S}^{4}}{6!\pi
^{2}m^{4}}G_{\mu\nu}^{a}\tilde{G}^{b,\mu\nu}G_{\rho\sigma}^{a}\tilde
{G}^{b,\rho\sigma}\nonumber\\
& +\gamma_{4,5}\frac{g_{S}^{4}}{6!\pi^{2}m^{4}}f^{abe}f^{cde}G_{\mu\nu}%
^{a}G^{c,\mu\nu}G_{\rho\sigma}^{b}G^{d,\rho\sigma}+\gamma_{4,6}\frac{g_{S}%
^{4}}{6!\pi^{2}m^{4}}f^{abe}f^{cde}G_{\mu\nu}^{a}\tilde{G}^{c,\mu\nu}%
G_{\rho\sigma}^{b}\tilde{G}^{d,\rho\sigma}\;.\label{EffSUNA}%
\end{align}
This basis corresponds to that in Ref.~\cite{Gracey17}, but for a slightly
different numbering and replacement of dual tensors via Eq.~(\ref{IdGGGG}).%

\begin{figure}[t]
\begin{center}
\includegraphics[height=0.9608in,width=5.4578in]{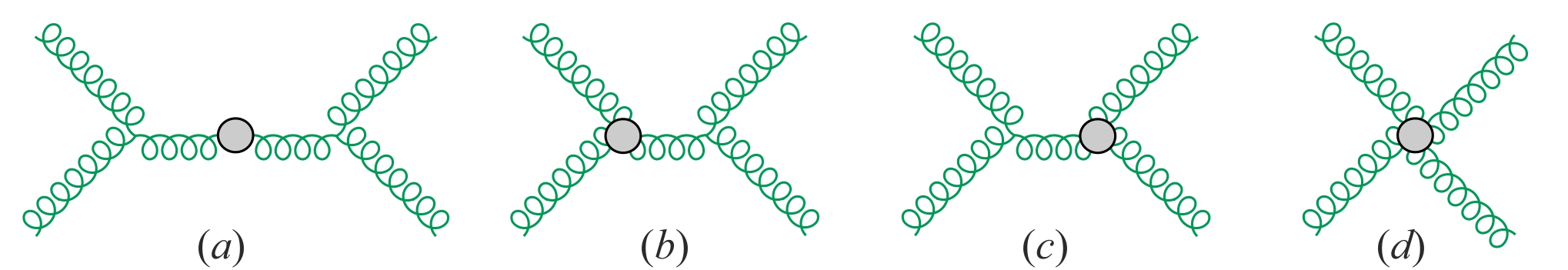}
\caption{The four basic $s$-channel topologies for the gluon-gluon scattering
amplitude. That for the $t$- and $u $-channel are understood. The grey disks
represent the insertion of the effective action vertices.}
\label{EHEffGlue}
\end{center}
\end{figure}

The non-abelian nature of QCD makes the effective action expansion quite
different from the QED case. The operators vanishing under the EOM have to be
kept because they contribute to several off-shell 1PI diagrams. For example,
the $D^{\nu}G_{\nu\mu}^{a}D_{\rho}G^{a,\rho\mu}$ operator occurs in the two,
three, and four-gluon off-shell 1PI diagrams of Fig.~\ref{EHglue} simply
because of the non-abelian terms present in the gluon field strengths. On the
other hand, for a physical process involving external on-shell gluons, these
operators should not contribute, and the basis could be simplified. Let us
check this in the simplest case, which is the gluon-gluon scattering
amplitude
\begin{equation}
\mathcal{A}(g(p_{1},\varepsilon_{p_{1}}^{\mu_{1}})g(p_{2},\varepsilon_{p_{2}%
}^{\mu_{2}})\rightarrow g(p_{3},\varepsilon_{p_{3}}^{\mu_{3}})g(p_{4}%
,\varepsilon_{p_{4}}^{\mu_{4}}))=\varepsilon_{p_{1}}^{\mu_{1}}\varepsilon
_{p_{2}}^{\mu_{2}}\varepsilon_{p_{3}}^{\mu_{3}\ast}\varepsilon_{p_{4}}%
^{\mu_{4}\ast}\mathcal{M}_{\mu_{1}\mu_{2}\mu_{3}\mu_{4}}\;.
\end{equation}
Computing this amplitude using the effective Lagrangian up to $\mathcal{O}%
(m^{-4})$, the basic topologies to consider are shown in Fig.~\ref{EHEffGlue}.
Besides the four point local terms, we must add the non-local contributions
from the three-gluon and two-gluon operators, as well as the wavefunction
corrected tree-level term. We observe:

\begin{itemize}
\item The wavefunction correction is automatically accounted for through a
rescaling of the field and coupling constant $g_{S}$.

\item The $\mathfrak{L}_{eff}^{(2)}$ operators contribute to all topologies,
$\mathfrak{L}_{eff}^{(3)}$ operators to $(b-d)$ topologies, and $\mathfrak{L}%
_{eff}^{(4)}$ to the $(d)$ topology only.

\item For the EOM operators, these topologically distinct contributions
precisely cancel each other. These operators thus play no role for physical processes.

\item Independently for each non-EOM operator $Q_{i}$, the sum of the
contributions $\mathcal{M}_{\mu_{1}\mu_{2}\mu_{3}\mu_{4}}(Q_{i})$ satisfy the
four Ward identities $p_{k}^{\mu_{k}}\mathcal{M}_{\mu_{1}\mu_{2}\mu_{3}\mu
_{4}}(Q_{i})=0$, $k=1,2,3,4$.
\end{itemize}

The fact that EOM operators drop out of the full physical amplitude can be
easily understood qualitatively. For example, taking the dimension-six
$\mathfrak{L}_{eff}^{(2)}$ operator $D^{\nu}G_{\nu\mu}^{a}D_{\rho}G^{a,\rho
\mu}$ and expanding the covariant derivatives, we get%
\begin{align}
D^{\nu}G_{\nu\mu}^{a}D_{\rho}G^{a,\rho\mu}  & =\partial^{\nu}G_{\nu\mu}%
^{a}\partial_{\rho}G^{a,\rho\mu}+gf^{abc}G^{b,\nu}G_{\nu\mu}^{c}\partial
_{\rho}G^{a,\rho\mu}\nonumber\\
& +gf^{abc}\partial^{\nu}G_{\nu\mu}^{a}G_{\rho}^{b}G^{c,\rho\mu}+g^{2}%
f^{abc}f^{ade}G^{b,\nu}G_{\nu\mu}^{c}G_{\rho}^{d}G^{e,\rho\mu}%
\;.\label{DGDGcancel}%
\end{align}
Replacing the field strength as $G_{\mu\nu}^{a}\rightarrow\partial_{\mu}%
G_{\nu}^{a}-\partial_{\nu}G_{\mu}^{a}$, these four terms are precisely those
entering the four topologies in Fig.~\ref{EHEffGlue}. We can see that the
cancellation occurs because the gluon propagator poles are precisely
compensated by the momentum factors arising from the LO three-gluon vertex
and from the derivatives in the first three terms of Eq.~(\ref{DGDGcancel}). A
similar reasoning can be applied to the non-abelian terms in the field
strengths, which cancel out similarly.%

\begin{table}[t] \centering
$
\begin{tabular}
[c]{ccccccc}\hline
& $\alpha_{0}$ & $\alpha_{2}$ & $\alpha_{4}$ & $\beta_{2}$ & $\beta_{4,1}$ &
$\beta_{4,2}$\smallskip\\\hline
\multicolumn{1}{r}{Scalar} & $\dfrac{1}{4}D_{\varepsilon}$ & $-\dfrac{1}{16} $
& $\dfrac{3}{112}$ & $\dfrac{1}{48}$ & $-\dfrac{1}{28}$ & $0\rule[-0.18in]%
{0in}{0.4in}$\\
\multicolumn{1}{r}{Fermion} & $D_{\varepsilon}$ & $-\dfrac{1}{2}$ & $\dfrac
{9}{28}$ & $-\dfrac{1}{24}$ & $\dfrac{1}{14}$ & $-\dfrac{3}{4}\rule[-0.18in]%
{0in}{0.4in}$\\
\multicolumn{1}{r}{Vector} & $-\dfrac{21D_{\varepsilon}+2}{4}$ & $\dfrac
{37}{16}$ & $-\dfrac{159}{112}$ & $\dfrac{1}{16}$ & $-\dfrac{3}{28}$ &
$3\rule[-0.18in]{0in}{0.4in}$\\\hline
\multicolumn{1}{r}{} & $\gamma_{4,1}$ & $\gamma_{4,2}$ & $\gamma_{4,3}$ &
$\gamma_{4,4}$ & $\gamma_{4,5}$ & $\gamma_{4,6}$\smallskip\\\hline
\multicolumn{1}{r}{Scalar} & $\dfrac{7}{768}$ & $\dfrac{1}{768}$ & $\dfrac
{7}{384}$ & $\dfrac{1}{384}$ & $\dfrac{1}{96}$ & $\dfrac{1}{672}%
\rule[-0.18in]{0in}{0.4in}$\\
\multicolumn{1}{r}{Fermion} & $\dfrac{1}{48}$ & $\dfrac{7}{192}$ & $\dfrac
{1}{24}$ & $\dfrac{7}{96}$ & $\dfrac{1}{96}$ & $\dfrac{19}{672}\rule[-0.18in]%
{0in}{0.4in}$\\
\multicolumn{1}{r}{Vector} & $\dfrac{87}{256}$ & $\dfrac{81}{256}$ &
$\dfrac{87}{128}$ & $\dfrac{81}{128}$ & $-\dfrac{3}{32}$ & $-\dfrac{27}%
{224}\rule[-0.18in]{0in}{0.4in}$\\\hline
\end{tabular}
$
\caption
{Wilson coefficients of the effective gluonic operators for a scalar, fermion, and vector boson in the
fundamental representation. This corresponds for example to the contributions of squarks in the MSSM,
or heavy quarks in the SM. For the coefficients in the vector case, we use the leptoquark gauge fields of
the minimal $SU(5)$ GUT model, quantized using a non-linear gauge fixing procedure (see Appendix~\ref
{AppGUT}).}
\label{TableSU3}
\end{table}

Let us now compute explicitly the coefficients of the effective operators for
a fermion, scalar, or vector in the fundamental representation. Generically,
the procedure is as follow. Starting with the vacuum polarization graph
(Fig.~\ref{EHglue}$a$), we fix the $\alpha_{0,2,4}$ coefficients. Then, the
three-point 1PI loop amplitudes (Fig.~\ref{EHglue}$b$) generate again the
$\mathfrak{L}_{eff}^{(2)}$ operators together with that of $\mathfrak{L}%
_{eff}^{(3)}$ and thus fix $\beta_{2}$, $\beta_{4,1}$, and $\beta_{4,2}$. As a
side effect, the basis chosen for $\mathfrak{L}_{eff}^{(0+2)}$ thus affects
the three Wilson coefficients of $\mathfrak{L}_{eff}^{(3)}$. Finally, the
four-point 1PI amplitudes (Fig.~\ref{EHglue}$c$) match over the local
four-gluon terms extracted from $\mathfrak{L}_{eff}^{(0+2+3+4)}$, and given
the coefficients obtained in the first two steps, fix the six $\gamma_{4,i}$
coefficients. The final results for the coefficients are given in
Table~\ref{TableSU3}. They agree with Ref.~\cite{EffAction} for dimension-six operators.

This procedure is rather straightforward for scalars and fermions circulating
in the loop, and only marginally more complicated than in the QED case of
Section 2. We checked our computation using the SM and MSSM FeynArts models,
using quarks or squarks as representative particles in the fundamental
representation. For vector particles, the calculation is far more challenging.
First, we must construct a consistent model involving a massive vector field
in the fundamental representation of QCD. Second, we know from the QED case
that working in the unitary gauge does not work, and even introducing an
appropriate Higgs mechanism to make these vectors massive is not sufficient.
Some generalization of the non-linear gauge has to be designed to preserve the
QCD symmetry throughout the quantization, otherwise the 1PI off-shell
amplitudes cannot be matched onto gauge invariant operators. This is
particularly annoying here since the three gluon 1PI amplitudes kinematically
vanish on-shell.

To proceed, our strategy is to use the minimal $SU(5)$ GUT model,
spontaneously broken by an adjoint Higgs scalar down to the (unbroken) SM
gauge group. Twelve of the $SU(5)$ gauge bosons become massive in the process,
and those fields have precisely the quantum numbers we need. The weak doublet
of leptoquarks $(X,Y)$ transforms as color antitriplets, so integrating them
out generate the effective gluonic operators. Note that we do not need the
second breaking stage down to $SU(3)_{C}\otimes U(1)_{em}$. In
Appendix~\ref{AppGUT}, we describe in some details the minimal $SU(5)$ GUT
model, along with its quantization using non-linear gauge fixing terms for the
$X$ and $Y$ gauge bosons. Denoting by $H_{X}^{k}$ and $H_{Y}^{k}$ the WBG
scalars associated to $X_{\mu}^{k}$ and $Y_{\mu}^{k}$, the main point is to
modify the usual $R_{\xi}$ gauge fixing terms%
\begin{equation}
\mathcal{L}_{\text{gf}}=-\frac{1}{\xi}|\partial^{\mu}X_{\mu}^{k+}-i\xi
M_{XY}H_{X}^{k+}|^{2}-\frac{1}{\xi}|\partial^{\mu}Y_{\mu}^{k+}-i\xi
M_{XY}H_{Y}^{k+}|^{2}+....
\end{equation}
by replacing the derivative by%
\begin{align}
\partial^{\mu}X_{\mu}^{i+}  & \rightarrow\partial^{\mu}X_{\mu}^{i+}%
-ig_{5}\left(  -\alpha_{G}X_{\nu}^{j+}T_{ji}^{a}G_{\mu}^{a}+\frac{\alpha_{W}%
}{2}W_{\mu}^{3}X_{\nu}^{i+}+\frac{\alpha_{W}}{\sqrt{2}}W_{\mu}^{+}Y_{\nu}%
^{i+}+\alpha_{B}\sqrt{\frac{\ 5}{12}}B_{\mu}X_{\nu}^{i+}\right)  \ ,\\
\partial^{\mu}Y_{\nu}^{i+}  & \rightarrow\partial^{\mu}Y_{\mu}^{i\pm}%
-ig_{5}\left(  -\alpha_{G}Y_{\nu}^{j+}T_{ji}^{a}G_{\mu}^{a}-\frac{\alpha_{W}%
}{2}W_{\mu}^{3}Y_{\nu}^{i+}+\frac{\alpha_{W}}{\sqrt{2}}W_{\mu}^{-}X_{\nu}%
^{i+}+\alpha_{B}\sqrt{\frac{\ 5}{12}}B_{\mu}Y_{\nu}^{i+}\right)  \ ,
\end{align}
where $T^{\alpha}$ are the $SU(3)$ generators for the fundamental
representation, and $i,j,k$ the corresponding indices. The gauge parameters
$\alpha_{G}$, $\alpha_{W}$, $\alpha_{B}$ interpolate between the 't
Hooft-Feynman gauge $\alpha_{G}=\alpha_{W}=\alpha_{B}=0$ and the non-linear
gauge $\alpha_{G}=\alpha_{W}=\alpha_{B}=1$, when the above terms coincide with
$D^{\mu}X_{\mu}^{i+}$ and $D^{\mu}Y_{\mu}^{i+}$. In that limit, the SM gauge
symmetries are preserved, exactly like the $U(1)_{em}$ in the SM in the
non-linear gauge. Technically, it should be remarked also that this gauge has
the nice feature of drastically reducing the number of diagrams for a given
process~\cite{GavelaGMS81}. Indeed, remember that the purpose of the usual
$R_{\xi}$ gauge is to get rid of the mixing terms like $X_{\mu}^{k}%
\partial^{\mu}H_{X}^{k}$. But when the vector is charged under some remaining
unbroken symmetries, this term is necessarily of the form $X_{\mu}^{k}D^{\mu
}H_{X}^{k}$ since it arises from the Higgs scalar kinetic term which is
invariant under the unbroken symmetries. With the non-linear gauge, all these
terms cancel out, leaving no $X-V_{SM}-H_{X}$ couplings. As a result, all the
mixed loops where the massive vector occurs alongside its WBG boson disappear,
and given the large number of diagrams, this is very welcome.

To actually perform the computation, we again use FeynArts~\cite{FeynArts} but
with a custom $SU(5)$ model file. The calculation then proceeds without
particular difficulty, and gives the coefficients quoted in
Table~\ref{TableSU3}. Several comments are in order:

\begin{itemize}
\item The matching works only for $\alpha_{G}=\alpha_{W}=\alpha_{B}=1$.
Without this condition, non-gauge-invariant operators are required. Note that
out of a total of 207 irreducible four-gluon diagrams, the gauge conditions
$\alpha_{G}=\alpha_{W}=\alpha_{B}=1$ leaves only 21 gauge-boson loops, 21 WBG
loops, and 42 ghost loops. The disappearance of mixed loops therefore reduces
the number of diagrams by more than a factor of two.

\item Many of the properties discovered in Ref.~\cite{Boudjema86} for photons
survive to the non-abelian generalization: the ghost and WBG contributions are
separately gauge invariant when $\alpha_{G}=\alpha_{W}=\alpha_{B}=1$.
Actually, matching separately the $H_{X}^{k}$ contributions on the effective
operators reproduce the coefficients for the scalar case in
Table~\ref{TableSU3}, while matching the $c_{X}$ and $c_{X}^{\dagger}$ ghost
contributions gives $-2$ times the scalar coefficients of Table~\ref{TableSU3}%
. With the non-linear gauge, the ghosts behave exactly like scalar particles,
but for the fermi statistics.

\item As a check, we computed the full physical gluon-gluon scattering
amplitude keeping the gauge parameter $\alpha_{G}$ arbitrary. On-shell and
when both 1PI and non-1PI topologies are included, the only remaining
$\alpha_{G}$ dependence can be absorbed into a wavefunction correction. In
other words, the inverse-mass expansion of the full amplitude matches onto the
non-EOM operators, and except for $\alpha_{0}$, their coefficients are
gauge-independent and physical, as they should.

\item To further check our results, we computed the 1PI diagrams with two,
three, and four external $SU(2)_{L}$ bosons. Since $SU(2)_{L}$ is kept
unbroken, and since $(X,Y)$ form an $SU(2)_{L}$ doublet, we can use the same
operator basis as for gluons, up to obvious substitutions, and found again the
coefficients in Table~\ref{TableSU3}.

\item Finally, we also computed the effective operators involving two and four
$U(1)_{Y}$ gauge bosons, and recover the same results as in
Table~\ref{TableU1} for the $W^{\pm}$ contribution in the non-linear gauge to
photon effective operators.
\end{itemize}

To close this section, the same cautionary remark as for the Uehling operator
should be repeated here for EOM gluonic operators. Those play no role for
on-shell gluon processes, but do contribute when other fields like light
quarks remain active. However, in that case, it is compulsory to include also
all the effective operators involving quark fields. Though the EOM operators
are gauge invariant by construction, their coefficients are not gauge
invariant by themselves. For instance, the gauge chosen for the $X_{\mu}^{k}$
and $Y_{\mu}^{k}$ fields does affect their values (Eq.~(\ref{GaugeDep})
remains valid for the gluonic vacuum polarization). In a phenomenological
study, it would thus make no sense to consider for example the $D^{\nu}%
G_{\nu\mu}^{a}D_{\rho}G^{a,\rho\mu}$ operator without including all the
four-quark operators. Taking again $SU(5)$, it is clear that $X_{\mu}^{k}$ and
$Y_{\mu}^{k}$ loops would contribute to both $D^{\nu}G_{\nu\mu}^{a}D_{\rho
}G^{a,\rho\mu}$ and four-quark operators, and only their combination would
result in a gauge-invariant physical result at the dimension-six level. As an
aside, it should be mentioned also that the gauge-dependent coefficient of
the $D^{\nu}G_{\nu\mu}^{a}D_{\rho}G^{a,\rho\mu}$ operator quoted in
Table~\ref{TableSU3} agrees with that in Ref.~\cite{EffAction}; the CDE
computation being done in the same non-linear gauge.

\section{$SU(N)$ effective interactions}

The computation done in the case of QCD can be extended to arbitrary
representations of other Lie groups. For that, it suffices to replace the
traces over the fundamental generators of $SU(3)$ occurring for each of the
1PI diagrams of the previous section by traces over generators in some generic
representation $\mathbf{R}$. Our notations along with various group-theoretic
results are collected in Appendix~\ref{AppCasimir}. In this section, for
definiteness, we refer to $SU(N)$ gauge group, but the results are trivially
extended to other Lie algebras.

Specifically, the vacuum polarization is tuned by $\operatorname*{Tr}%
(T_{\mathbf{R}}^{a}T_{\mathbf{R}}^{b})=I_{2}(\mathbf{R})\delta^{ab}$ with
$I_{2}(\mathbf{R})$ the quadratic invariant, so the $\alpha_{i}$ coefficients
are simply $I_{2}(\mathbf{R})/I_{2}(\mathbf{F})= 2I_{2}(\mathbf{R})$ times those in Table~\ref{TableSU3}.
Similarly, the three-boson diagrams are proportional to
\begin{equation}
\operatorname*{Tr}(T_{\mathbf{R}}^{a}[T_{\mathbf{R}}^{b},T_{\mathbf{R}}%
^{c}])=iI_{2}(\mathbf{R})f^{abc}\;.\label{TrTTT}%
\end{equation}
The fact that both the two and three-boson amplitudes are proportional to the
same $I_{2}(\mathbf{R})$ coefficient ensures a proper matching. In particular,
the divergence of the three-boson diagrams is correctly accounted for by the
$\mathfrak{L}_{eff}^{(2)}$ couplings.

The situation is more involved for the four-boson amplitude. The 1PI loops in
either the fermion, scalar, or vector case are equivalent two-by-two under the
reversing of the loop momentum, so the total amplitudes can always be brought
to the form%
\begin{equation}
\mathcal{M}^{abcd}=C_{1}^{abcd}\mathcal{M}_{1}+C_{2}^{abcd}\mathcal{M}%
_{2}+C_{3}^{abcd}\mathcal{M}_{3}\;,\ \ \left\{
\begin{array}
[c]{c}%
C_{1}^{abcd}=\operatorname*{Tr}(T_{\mathbf{R}}^{a}T_{\mathbf{R}}%
^{b}T_{\mathbf{R}}^{d}T_{\mathbf{R}}^{c})+\operatorname*{Tr}(T_{\mathbf{R}%
}^{a}T_{\mathbf{R}}^{c}T_{\mathbf{R}}^{d}T_{\mathbf{R}}^{b})\;,\\
C_{2}^{abcd}=\operatorname*{Tr}(T_{\mathbf{R}}^{a}T_{\mathbf{R}}%
^{b}T_{\mathbf{R}}^{c}T_{\mathbf{R}}^{d})+\operatorname*{Tr}(T_{\mathbf{R}%
}^{a}T_{\mathbf{R}}^{d}T_{\mathbf{R}}^{c}T_{\mathbf{R}}^{b})\;,\\
C_{3}^{abcd}=\operatorname*{Tr}(T_{\mathbf{R}}^{a}T_{\mathbf{R}}%
^{c}T_{\mathbf{R}}^{b}T_{\mathbf{R}}^{d})+\operatorname*{Tr}(T_{\mathbf{R}%
}^{a}T_{\mathbf{R}}^{d}T_{\mathbf{R}}^{b}T_{\mathbf{R}}^{c})\;.
\end{array}
\right. \label{Decomp4p}%
\end{equation}
Expanding $\mathcal{M}^{abcd}$ in the mass of the heavy particle circulating within the loop, only two independent combinations of traces
occur at $\mathcal{O}(m^{0})$ and $\mathcal{O}(m^{-2})$, which can be
expressed entirely in terms of the quadratic invariants as
\begin{subequations}
\begin{align}
D_{1}^{abcd}  & =2C_{1}^{abcd}-C_{2}^{abcd}-C_{3}^{abcd}=I_{2}(\mathbf{R}%
)(2f^{ace}f^{bde}-f^{ade}f^{bce})\;,\\
D_{2}^{abcd}  & =2C_{2}^{abcd}-C_{1}^{abcd}-C_{3}^{abcd}=I_{2}(\mathbf{R}%
)(2f^{ade}f^{bce}-f^{ace}f^{bde})\;,\\
D_{3}^{abcd}  & =2C_{3}^{abcd}-C_{1}^{abcd}-C_{2}^{abcd}=I_{2}(\mathbf{R}%
)(-f^{ade}f^{bce}-f^{ace}f^{bde})=-D_{1}^{abcd}-D_{2}^{abcd}\;,
\end{align}
where we used $[T_{\mathbf{R}}^{a},T_{\mathbf{R}}^{b}]=if^{abc}T_{\mathbf{R}%
}^{c}$ together with Eq.~(\ref{TrTTT}) and imposed the Jacobi identity
$f^{abe}f^{cde}=f^{ace}f^{bde}-f^{ade}f^{bce}$. Thanks to this reduction,
$\mathcal{M}^{abcd}$ matches the four-boson amplitude obtained from the
$\mathfrak{L}_{eff}^{(2)}$ and $\mathfrak{L}_{eff}^{(3)}$ couplings at the
$\mathcal{O}(m^{0})$ and $\mathcal{O}(m^{-2})$.

At $\mathcal{O}(m^{-4})$, these same combinations $D_{1,2,3}^{abcd}$ induce
the operators tuned by $\gamma_{4,5}$ and $\gamma_{4,6}$, which involve the
structure constants. The rest is proportional to the fully symmetrized trace
\end{subequations}
\begin{equation}
D_{0}^{abcd}=C_{1}^{abcd}+C_{2}^{abcd}+C_{3}^{abcd}=\frac{1}{4}%
S\operatorname*{Tr}(T_{\mathbf{R}}^{a}T_{\mathbf{R}}^{b}T_{\mathbf{R}}%
^{c}T_{\mathbf{R}}^{d})\ .\label{D01}%
\end{equation}
As detailed in Appendix~\ref{AppCasimir}, for a general $SU(N)$ algebra, the
fully symmetrized trace decomposes into quadratic and quartic invariants.
Plugging Eq.~(\ref{GenQuartic}) in Eq.~(\ref{D01}),%
\begin{equation}
D_{0}^{abcd}=6I_{4}(\mathbf{R})d^{abcd}+6\Lambda(\mathbf{R})(\delta^{ab}%
\delta^{cd}+\delta^{ac}\delta^{bd}+\delta^{ad}\delta^{bc})\;,\label{D02}%
\end{equation}
where $d^{abcd}$ is the fully symmetric fourth-order symbol normalized such
that $I_{4}(\mathbf{F})=1$ for the defining representation, and%
\begin{equation}
\Lambda(\mathbf{R})=\left(  \frac{N(\mathbf{A})I_{2}(\mathbf{R})}%
{N(\mathbf{R})}-\frac{I_{2}(\mathbf{A})}{6}\right)  \frac{I_{2}(\mathbf{R}%
)}{2+N(\mathbf{A})}\ ,\label{ConvLambda}%
\end{equation}
where $\mathbf{A}$ denotes the adjoint representation and $N(\mathbf{R})$ the
dimension of the representation $\mathbf{R}$. The term proportional to
$\Lambda(\mathbf{R})$ matches onto the operators tuned by $\gamma_{4,1}$ to
$\gamma_{4,4}$, while that proportional to $d^{abcd}$ requires to extend
$\mathfrak{L}_{eff}^{(4)}$ of Eq.~(\ref{EffSUNA}) with two extra operators.
The total effective Lagrangian is then:%
\begin{align}
\mathfrak{L}_{eff}^{(4)}  & =\gamma_{4,1}\frac{g_{S}^{4}}{6!\pi^{2}m^{4}%
}G_{\mu\nu}^{a}G^{a,\mu\nu}G_{\rho\sigma}^{b}G^{b,\rho\sigma}+\gamma
_{4,2}\frac{g_{S}^{4}}{6!\pi^{2}m^{4}}G_{\mu\nu}^{a}\tilde{G}^{a,\mu\nu
}G_{\rho\sigma}^{b}\tilde{G}^{b,\rho\sigma}\nonumber\\
& +\gamma_{4,3}\frac{g_{S}^{4}}{6!\pi^{2}m^{4}}G_{\mu\nu}^{a}G^{b,\mu\nu
}G_{\rho\sigma}^{a}G^{b,\rho\sigma}+\gamma_{4,4}\frac{g_{S}^{4}}{6!\pi
^{2}m^{4}}G_{\mu\nu}^{a}\tilde{G}^{b,\mu\nu}G_{\rho\sigma}^{a}\tilde
{G}^{b,\rho\sigma}\nonumber\\
& +\gamma_{4,5}\frac{g_{S}^{4}}{6!\pi^{2}m^{4}}f^{abe}f^{cde}G_{\mu\nu}%
^{a}G^{c,\mu\nu}G_{\rho\sigma}^{b}G^{d,\rho\sigma}+\gamma_{4,6}\frac{g_{S}%
^{4}}{6!\pi^{2}m^{4}}f^{abe}f^{cde}G_{\mu\nu}^{a}\tilde{G}^{c,\mu\nu}%
G_{\rho\sigma}^{b}\tilde{G}^{d,\rho\sigma}\nonumber\\
& +\gamma_{4,7}\frac{g_{S}^{4}}{6!\pi^{2}m^{4}}d^{abcd}G_{\mu\nu}^{a}%
G^{b,\mu\nu}G_{\rho\sigma}^{c}G^{d,\rho\sigma}+\gamma_{4,8}\frac{g_{S}^{4}%
}{6!\pi^{2}m^{4}}d^{abcd}G_{\mu\nu}^{a}\tilde{G}^{b,\mu\nu}G_{\rho\sigma}%
^{c}\tilde{G}^{d,\rho\sigma}\;.\label{EffSUNB}%
\end{align}
The need of a total of eight operators for $SU(N)$ and their connection with
the quartic tensor structure is in agreement with Ref.~\cite{Gracey17}. Note,
however, that the definition of $\Lambda(\mathbf{R})$ is a matter of
convention, and it indirectly affects the definition of all the operators but
those tuned by $\gamma_{4,3}$ and $\gamma_{4,6}$. Yet, adopting the convention
in Eq.~(\ref{ConvLambda}) for $\Lambda(\mathbf{R})$ looks optimal since it
ensures $I_{4}(\mathbf{R})=0$ for all $SU(2)$ and $SU(3)$ representations, as
it should since these algebras have no irreducible invariant tensor of rank four.
As said before, all these results stay valid for $SO(N)$ algebras, but for a single exception. 
As explained in Appendix~\ref{AppCasimir}, $SO(8)$ has the unique feature of having two quartic symbols, and an additional term occurs in Eq.~(\ref{D02}). In that case, two extra operators are
required, tuned by the second quartic symbol of Eq.~(\ref{SymTensor}).%

\begin{figure}[t]
\begin{center}
\includegraphics[height=1.2272in,width=3.5189in]{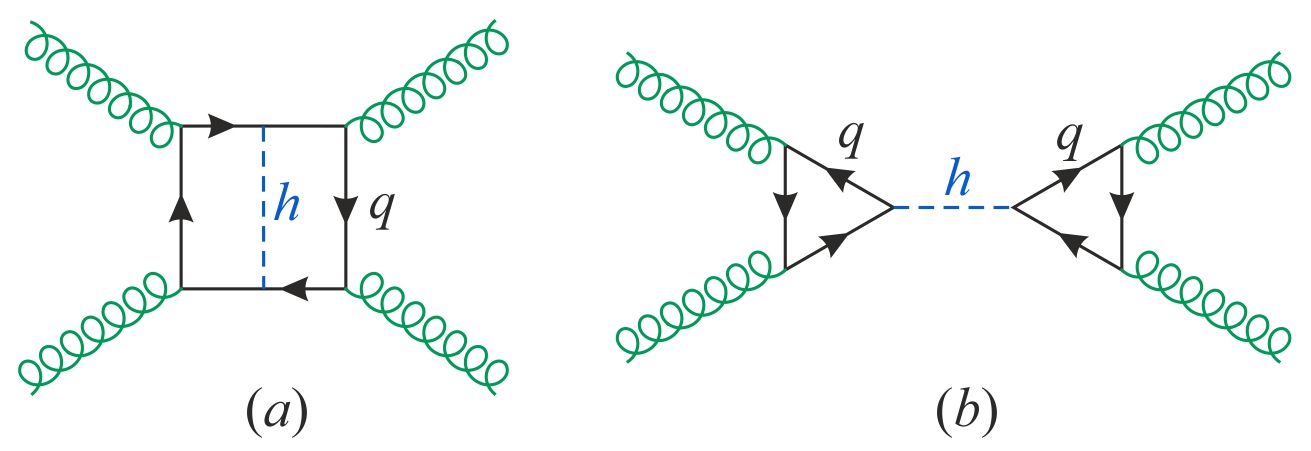}
\caption{Examples of two-loop diagrams in the SM that preserves ($a$) or
violate ($b$) the one-loop predictions Eq.~(\ref{CoeffCoh}) among the gluonic
operators. The particle circulating in the loops are heavy quarks, and the
dashed lines denote the Higgs boson.}%
\label{Fig2LH}%
\end{center}
\end{figure}

Now, even if a total of eight (or ten for $SO(8)$) independent operators can be constructed in
general, our specific computations show that at one loop, most of these
operators derive from a single symmetrized trace and are thus always
correlated. In particular, no matter the representation or spin of the
particle in the loop:
\begin{subequations}
\label{CoeffCoh}%
\begin{align}
\gamma_{4,1}  & =\frac{1}{2}\gamma_{4,3}\ ,\\
\gamma_{4,2}  & =\frac{1}{2}\gamma_{4,4}\ .
\end{align}
There are thus two operator combinations that never occur in the one-loop
effective action. From an effective theory point of view, this should remain
true in most cases since it derives from the symmetry of the amplitude. A
necessary condition beyond one-loop is the absence of diagrams where the color
flow is disconnected, that is, where a product of traces over the generators
occurs instead of a single trace. This never happens if only one heavy state
is integrated out, but could arise in more general settings. For example, in
the SM, integrating out heavy quarks together with the Higgs boson, the
diagrams in Fig.~\ref{Fig2LH} arise at two loops. Since the CP-conserving
effective Higgs coupling to two gluons is of the form $h^{0}G_{\mu\nu}%
^{a}G^{a,\mu\nu}$, it is clear that the Higgs boson exchange in
Fig.~\ref{Fig2LH}$b$ contribute to $\gamma_{4,1}$ but not to $\gamma_{4,3}$.%

\begin{table}[t] \centering
$
\begin{tabular}
[c]{ccccccc}\hline
& $\alpha_{0}$ & $\alpha_{2}$ & $\alpha_{4}$ & $\beta_{2}$ & $\beta_{4,1}$ &
$\beta_{4,2}$\smallskip\\\hline
\multicolumn{1}{r}{Scalar} & $\dfrac{1}{2}I_{2}(\mathbf{R})D_{\varepsilon}$ &
$-\dfrac{1}{8}I_{2}(\mathbf{R})$ & $\dfrac{3}{56}I_{2}(\mathbf{R})$ &
$\dfrac{1}{24}I_{2}(\mathbf{R})$ & $-\dfrac{1}{14}I_{2}(\mathbf{R})$ &
$0\rule[-0.18in]{0in}{0.4in}$\\
\multicolumn{1}{r}{Fermion} & $2I_{2}(\mathbf{R})D_{\varepsilon}$ &
$-I_{2}(\mathbf{R})$ & $\dfrac{9}{14}I_{2}(\mathbf{R})$ & $-\dfrac{1}{12}%
I_{2}(\mathbf{R})$ & $\dfrac{1}{7}I_{2}(\mathbf{R})$ & $-\dfrac{3}{2}%
I_{2}(\mathbf{R})\rule[-0.18in]{0in}{0.4in}$\\
\multicolumn{1}{r}{Vector} & $-\dfrac{21D_{\varepsilon}+2}{2}I_{2}%
(\mathbf{R})$ & $\dfrac{37}{8}I_{2}(\mathbf{R})$ & $-\dfrac{159}{56}%
I_{2}(\mathbf{R})$ & $\dfrac{1}{8}I_{2}(\mathbf{R})$ & $-\dfrac{3}{14}%
I_{2}(\mathbf{R})$ & $6I_{2}(\mathbf{R})\rule[-0.18in]{0in}{0.4in}$\\\hline
\multicolumn{1}{r}{} & $\gamma_{4,1}=\gamma_{4,3}/2$ & $\gamma_{4,2}%
=\gamma_{4,4}/2$ & $\gamma_{4,5}$ & $\gamma_{4,6}$ & $\gamma_{4,7}$ &
$\gamma_{4,8}$\smallskip\\\hline
\multicolumn{1}{r}{Scalar} & $\dfrac{7}{32}\Lambda(\mathbf{R})$ & $\dfrac
{1}{32}\Lambda(\mathbf{R})$ & $\dfrac{1}{48}I_{2}(\mathbf{R})$ & $\dfrac
{1}{336}I_{2}(\mathbf{R})$ & $\dfrac{7}{32}I_{4}(\mathbf{R})$ & $\dfrac{1}%
{32}I_{4}(\mathbf{R})\rule[-0.18in]{0in}{0.4in}$\\
\multicolumn{1}{r}{Fermion} & $\dfrac{1}{2}\Lambda(\mathbf{R})$ & $\dfrac
{7}{8}\Lambda(\mathbf{R})$ & $\dfrac{1}{48}I_{2}(\mathbf{R})$ & $\dfrac
{19}{336}I_{2}(\mathbf{R})$ & $\dfrac{1}{2}I_{4}(\mathbf{R})$ & $\dfrac{7}%
{8}I_{4}(\mathbf{R})\rule[-0.18in]{0in}{0.4in}$\\
\multicolumn{1}{r}{Vector} & $\dfrac{261}{32}\Lambda(\mathbf{R})$ &
$\dfrac{243}{32}\Lambda(\mathbf{R})$ & $-\dfrac{3}{16}I_{2}(\mathbf{R})$ &
$-\dfrac{27}{112}I_{2}(\mathbf{R})$ & $\dfrac{261}{32}I_{4}(\mathbf{R})$ &
$\dfrac{243}{32}I_{4}(\mathbf{R})\rule[-0.18in]{0in}{0.4in}$\\\hline
\end{tabular}
$
\caption{Wilson coefficients of the effective operators for $SU(N)$ or $SO(N\neq8)$ gauge bosons, as induced by a set of complex fields of spin 0, 1/2, and 1 transforming under the representation \textbf{R}. For real fields, all the coefficients should be halved.}
\label{TableSUN}
\end{table}

The coefficients for a complex field (fermion, scalar, vector particle) circulating
in the loops are given in Table~\ref{TableSUN}. Those for a self-conjugate particle are half of those quoted there. Indeed, when the propagator is not oriented, some Feynman diagrams get an extra
symmetry factor 1/2, while for others, the loop momentum cannot be reversed
and runs in only one direction. This latter situation also brings a factor 1/2
because $(T_{\mathbf{R}}^{a})^{T}=-T_{\mathbf{R}}^{a}$ for a real
representation. For example, instead of Eq.~(\ref{TrTTT}), the triangle
diagrams are now tuned by
\end{subequations}
\begin{equation}
\left.  \operatorname*{Tr}(T_{\mathbf{R}}^{a}T_{\mathbf{R}}^{b}T_{\mathbf{R}%
}^{c})\right\vert _{self-conjugate}=\frac{1}{2}\operatorname*{Tr}%
(T_{\mathbf{R}}^{a}[T_{\mathbf{R}}^{b},T_{\mathbf{R}}^{c}])=\frac{1}{2}%
iI_{2}(\mathbf{R})f^{abc}\ .
\end{equation}
Similarly, the coefficients for the four-point amplitude satisfy%
\begin{equation}
\left.  C_{1}^{abcd}\right\vert _{self-conjugate}=\operatorname*{Tr}%
(T_{\mathbf{R}}^{a}T_{\mathbf{R}}^{b}T_{\mathbf{R}}^{d}T_{\mathbf{R}}%
^{c})=\frac{1}{2}(\operatorname*{Tr}(T_{\mathbf{R}}^{a}T_{\mathbf{R}}%
^{c}T_{\mathbf{R}}^{d}T_{\mathbf{R}}^{b})+\operatorname*{Tr}(T_{\mathbf{R}%
}^{a}T_{\mathbf{R}}^{c}T_{\mathbf{R}}^{d}T_{\mathbf{R}}^{b}))=\frac{1}{2}%
C_{1}^{abcd}\ .
\end{equation}
We checked this property of the coefficients for two physically relevant cases: the
contribution to the gluon coefficients of the $SU(5)$ Higgs bosons $H_{G}^{a}
$ and of the MSSM gluinos, both self-conjugate fields transforming in the
adjoint representation of $SU(3)_{C}$.

\subsection{Reduction to SU(3) and SU(2)}

\begin{figure}[t]
\begin{center}
\includegraphics[height=1.9026in,width=6.3261in]{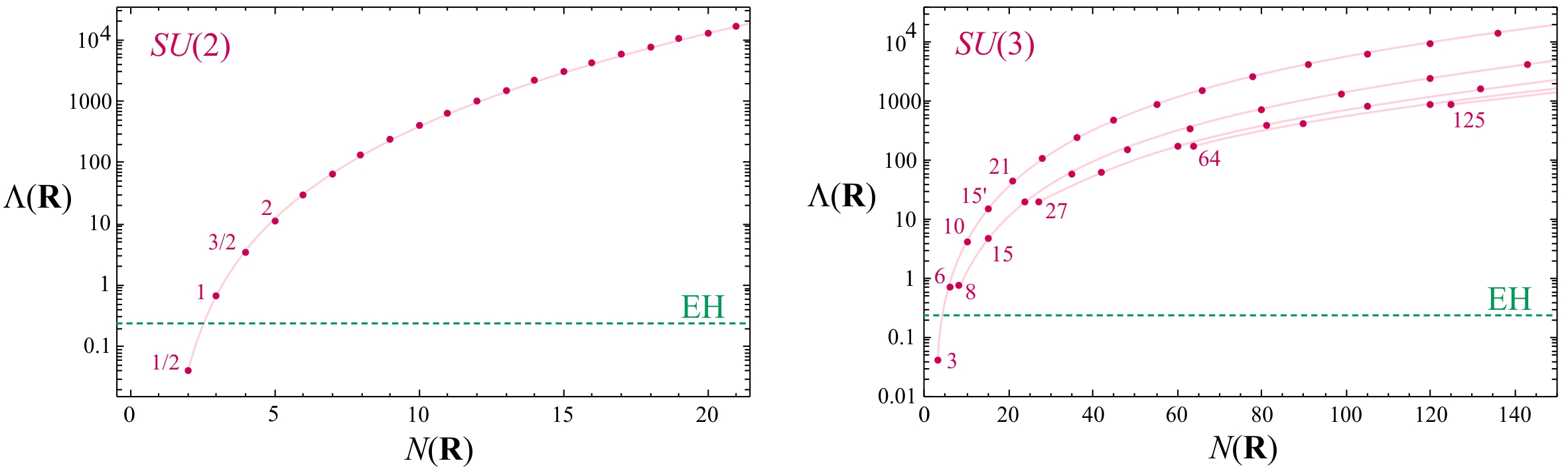}
\caption{Evolution of $\Lambda(\mathbf{R}) $ as a function of the dimension $N(\mathbf{R})$ for $SU(2)$ and $SU(3)$. In the former case, we denote the first few representations by the corresponding isospin. In the $SU(3)$ case, several branches are apparent, each starting with a real representation. The horizontal dashed lines depict the Euler-Heisenberg value, identified as $\Lambda(\mathbf{1})=1/3$ for a charge-one loop particle from Eq.~(\ref{ReducU1}).}%
\label{LambaR}
\end{center}
\end{figure}

The general basis of effective operators reduces immediately to $SU(3)$ by
removing the quartic invariant operators, i.e., by setting $\gamma_{4,7}$ and
$\gamma_{4,8}$ to zero. For the fundamental representation, $I_{2}%
^{SU(3)}(\mathbf{F})=1/2$ and $\Lambda^{SU(3)}(\mathbf{F})=1/24$, and we
recover the results of Table~\ref{TableSU3}. But, an interesting feature
appears for more general representations. A priori, as the representation get
larger, one would expect the strength of the effective interactions to
increase mechanically due to the increased number of particles circulating in
the loop. However, we show in Fig.~\ref{LambaR} that $\Lambda(\mathbf{R}) $
grows much faster than $N(\mathbf{R})$. The fastest growth happens for
representations which are the symmetric tensor products of the fundamental
representations, for which $\Lambda(\mathbf{R})\sim N(\mathbf{R})^{3}$. For
instance, $\Lambda(\mathbf{3})=1/24$ but $\Lambda(\mathbf{6}=\mathbf{3}%
\otimes_{S}\mathbf{3})=17/24$, $\Lambda(\mathbf{10}=\mathbf{3}\otimes
_{S}\mathbf{3}\otimes_{S}\mathbf{3})=99/24$, and $\Lambda(\mathbf{15}%
=\mathbf{3}\otimes_{S}\mathbf{3}\otimes_{S}\mathbf{3}\otimes_{S}%
\mathbf{3})=371/24$. The adjoint representation is not on this series, but the
effective interactions are nevertheless stronger than naively expected from
the dimension since $\Lambda^{SU(3)}(\mathbf{8})=3/4=18\times\Lambda
^{SU(3)}(\mathbf{3})$. Interestingly, this corresponds to physically sensible
scenarios, for example that of the gluinos in the MSSM for which (including
the 1/2 factor for self-conjugate particles):%
\begin{equation}
\frac{1}{2}\times\frac{g_{S}^{4}}{6!\pi^{2}m_{\tilde{g}}^{4}}\gamma
_{4,1}=-\frac{1}{2}\times\dfrac{1}{2}18\frac{g_{S}^{4}}{6!\pi^{2}m_{\tilde{g}%
}^{4}}=\frac{\alpha_{S}}{10m_{\tilde{g}}^{4}}\ ,
\end{equation}
which is an order of magnitude larger than the coefficient of the effective
photon interactions of the Euler-Heisenberg Lagrangian.%

For $SU(2)$, the effective Lagrangian gets simpler thanks to the identity%
\begin{equation}
f^{abe}f^{cde}\rightarrow\varepsilon^{abe}\varepsilon^{cde}=\delta^{ac}%
\delta^{bd}-\delta^{ad}\delta^{bc}\ ,
\end{equation}
which permits to get rid of two operators. Expressing the remaining four
operators explicitly in terms of the $SU(2)$ triplet states denoted as
$\{W_{\mu}^{-},W_{\mu}^{3},W_{\mu}^{+}\}$:%
\begin{align}
\mathfrak{L}_{eff,SU(2)_{L}}^{(4)}  & =\frac{(\gamma_{4,1}+\gamma_{4,3})g^{4}%
}{6!\pi^{2}m^{4}}(W_{\mu\nu}^{3}W^{3,\mu\nu})^{2}+\frac{(\gamma_{4,2}%
+\gamma_{4,4})g^{4}}{6!\pi^{2}m^{4}}(W_{\mu\nu}^{3}\tilde{W}^{3,\mu\nu}%
)^{2}\nonumber\\
& +\frac{4(\gamma_{4,1}+\gamma_{4,5})g^{4}}{6!\pi^{2}m^{4}}W_{\mu\nu}%
^{3}W^{3,\mu\nu}W_{\rho\sigma}^{+}W^{-,\rho\sigma}+\frac{4(\gamma_{4,2}%
+\gamma_{4,6})g^{4}}{6!\pi^{2}m^{4}}W_{\mu\nu}^{3}\tilde{W}^{3,\mu\nu}%
W_{\rho\sigma}^{+}\tilde{W}^{-,\rho\sigma}\nonumber\\
& +\frac{4(\gamma_{4,3}-\gamma_{4,5})g^{4}}{6!\pi^{2}m^{4}}|W_{\mu\nu}%
^{3}W^{+,\mu\nu}|^{2}+\frac{4(\gamma_{4,4}-\gamma_{4,6})g^{4}}{6!\pi^{2}m^{4}%
}|W_{\mu\nu}^{3}\tilde{W}^{+,\mu\nu}|^{2}\nonumber\\
& +\frac{2(2\gamma_{4,1}+\gamma_{4,3}+\gamma_{4,5})g^{4}}{6!\pi^{2}m^{4}%
}(W_{\mu\nu}^{+}W^{-,\mu\nu})^{2}+\frac{2(\gamma_{4,4}-\gamma_{4,6})g^{4}%
}{6!\pi^{2}m^{4}}|W_{\mu\nu}^{+}\tilde{W}^{+,\mu\nu}|^{2}\nonumber\\
& +\frac{2(\gamma_{4,3}-\gamma_{4,5})g^{4}}{6!\pi^{2}m^{4}}|W_{\mu\nu}%
^{+}W^{+,\mu\nu}|^{2}+\frac{2(2\gamma_{4,2}+\gamma_{4,4}+\gamma_{4,6})g^{4}%
}{6!\pi^{2}m^{4}}(W_{\mu\nu}^{+}\tilde{W}^{-,\mu\nu})^{2}\;.\label{LargEffSU2}%
\end{align}
These operators and coefficients are obtained from the effective action, and
are independent of the invariant mass of the external states. Thus, they
remain valid for massive external weak bosons, at least as long as $m$ is
sufficiently large compared to $M_{Z,W}$. An important caveat though, of
relevance for the SM, is the presence of chiral fermions. Those cannot be
massive without breaking the gauge symmetry, so the inverse mass expansion is
defined only in the broken phase. Non-gauge invariant operators can then
arise, at both the $\mathcal{O}(m^{0})$ and $\mathcal{O}(m^{-2})$ level.

Concerning the strength of the effective interactions, here also
$\Lambda(\mathbf{R})$ grows much faster than $N(\mathbf{R})$. Actually, as the
$SU(2)$ representations are smaller than those of $SU(3)$, the increase is
much more pronounced, with $\Lambda(\mathbf{R})\sim N(\mathbf{R})^{5}$, see
Fig.~\ref{LambaR}. So, while $\Lambda(\mathbf{F})=1/24$, it is already an
order of magnitude stronger for the adjoint representation, $\Lambda
(\mathbf{3})=2/3=16\times\Lambda(\mathbf{2})$.

To close this section, it is instructive to look at the application of the
$SU(N)$ result from a group-theoretic perspective. Up to now, the $SU(2)$ and
$SU(3)$ effective Lagrangians are obtained simply by setting $N=2$ or $N=3$ in
the general result. But, if $SU(N)$ is large enough to contain an $SU(2)$ or
$SU(3)$ subalgebra, we could also ask where these pieces are in the general
$SU(N)$ Lagrangian. More generally, consider the effective Lagrangian for a
representation $\mathbf{R}_{M}$ of $SU(M)$. These $N(\mathbf{R}_{M})$ states
organize themselves into representations of $SU(N)\subset SU(M)$, that is,
$\mathbf{R}_{M}$ branches into a direct sum of $SU(N)$ representations
$\mathbf{R}_{N}$. So, from the $SU(N)$ perspective, the $SU(M)$ coefficients
encode the circulation of a collection of states in the loop. Since these
contributions simply add up, the $SU(M)$ coefficients must be the sum over the
$SU(N)$ coefficients for all the $\mathbf{R}_{N}$ representations present in
the representation $\mathbf{R}_{M}$. Going back to Eq.~(\ref{D02}), we must
thus have%
\begin{align}
\frac{1}{6}D_{0}^{abcd}  & =I_{4}(\mathbf{R}_{M})d_{M}^{abcd}+\Lambda
_{N}(\mathbf{R}_{M})(\delta^{ab}\delta^{cd}+\delta^{ac}\delta^{bd}+\delta
^{ad}\delta^{bc})\nonumber\\
& =\sum_{\mathbf{R}_{N}\subset\mathbf{R}_{M}}I_{4}(\mathbf{R}_{N})d_{N}%
^{abcd}+\sum_{\mathbf{R}_{N}\subset\mathbf{R}_{M}}\Lambda_{N}(\mathbf{R}%
_{N})(\delta^{ab}\delta^{cd}+\delta^{ac}\delta^{bd}+\delta^{ad}\delta^{bc})\ ,
\end{align}
where the indices $a,b,c,d$ are understood to denote those $SU(M)$ generators
that correspond to the $SU(N)$ subalgebra. The main difficulty though is that
even restricted to those particular generators, $d_{M}^{abcd}\neq d_{N}%
^{abcd}$ because the definition of the quartic invariant involves different
functions $\Lambda_{N}$ and $\Lambda_{M}$. To proceed, let us assume that the
fundamental representation has the branching rule $\mathbf{F}_{M}%
\rightarrow\mathbf{F}_{N}$. Knowing that by definition, $I_{4}(\mathbf{F}%
_{M})=I_{4}(\mathbf{F}_{N})=1$, we find
\begin{subequations}
\label{IdI4I2}%
\begin{align}
I_{4}(\mathbf{R}_{M})  & =\sum_{\mathbf{R}_{N}\subset\mathbf{R}_{M}}%
I_{4}(\mathbf{R}_{N})\ ,\\
I_{4}(\mathbf{R}_{M})(\Lambda_{N}(\mathbf{F}_{N})-\Lambda_{M}(\mathbf{F}%
_{M}))+\Lambda_{M}(\mathbf{R}_{M})  & =\sum_{\mathbf{R}_{N}\subset
\mathbf{R}_{M}}\Lambda_{N}(\mathbf{R}_{N})\ .
\end{align}
Using the numbers quoted in Appendix~\ref{AppCasimir} and the branching rules
in Ref.~\cite{Slansky}, one can check that the two formulas are valid for
$SU(3)\subset SU(4)$ and $SU(4)\subset SU(5)$. The second one also applies to
$SU(2)\subset SU(3)$ in which case it becomes a sum rule for the $\Lambda$
functions since $I_{4}(\mathbf{R})=0$ in $SU(3)$. From a calculation point of
view, once the branching rules of the $SU(M)$ representations are known, these
equations are particularly powerful, with the second one even allowing to
compute $I_{4}(\mathbf{R}_{M})$ in terms of $\Lambda_{N}$ and $\Lambda_{M}$,
that is, entirely in terms of the quadratic invariants $I_{2}(\mathbf{R}_{N})$
and $I_{2}(\mathbf{R}_{M})$.

Thanks to the convention Eq.~(\ref{ConvLambda}), the branching rule for the
$I_{4}$ invariant is very simple~\cite{Okubo81}, but there is a price to pay.
Some part of the $\gamma_{4,7}$ and $\gamma_{4,8}$ operators of $SU(M)$ are
moved into the $\gamma_{4,1}$ to $\gamma_{4,4}$ operators of $SU(N<M)$. This
is due to the very definition of the operators in terms of different quartic
symbols, and not related to the loop structure of the amplitude or the
specific branching rules. For example, if for some unification group a
specific mechanism is found that generates only $\gamma_{4,7}$ and
$\gamma_{4,8}$, the four operators tuned by $\gamma_{4,1}$ to $\gamma_{4,4}$
are in general present once the symmetry is spontaneously broken simply
because the $d^{abcd}$ symbol is defined differently within the surviving subalgebra.

\subsection{Reduction to U(1)}

Comparing the $SU(N)$ coefficients $\gamma_{4,i}$ in Table~\ref{TableSUN} with
the Euler-Heisenberg results in Table~\ref{TableU1}, the two clearly appear
related. Heuristically, it is simple to understand this relationship by
adapting the decomposition Eq.~(\ref{Decomp4p}) to the $U(1)$ case. When only
a single generator occurs, $C_{1}=C_{2}=C_{3}=2Q^{4}$. This ensures the
cancellation of the UV divergence, and more generally the absence of all the
operators tuned by the structure constants. The whole amplitude is then
proportional to
\end{subequations}
\begin{equation}
D_{0}\equiv C_{1}+C_{2}+C_{3}=6Q^{4}\ .\label{D03}%
\end{equation}
Since the same factor of $6$ occurs in the $SU(N)$ result in Eq.~(\ref{D02}),
it is clear that $\gamma_{4,1}^{EH}$ and $\gamma_{4,2}^{EH}$ can be obtained
equivalently from $\gamma_{4,1}^{SU(N)}$, $\gamma_{4,2}^{SU(N)}$ with
$\Lambda(\mathbf{R})\rightarrow Q^{4}$ or from $\gamma_{4,7}^{SU(N)}$,
$\gamma_{4,8}^{SU(N)}$ with $I_{4}(\mathbf{R})\rightarrow Q^{4}$, in agreement
with Table~\ref{TableSUN} and Table~\ref{TableU1}. Obviously, this line of
reasoning is a naive identification of the coefficients of the loop
functions, not a group-theoretic reduction of $SU(N)$ down to one of its
$U(1)$ subgroup.

To perform a true reduction, let us denote $T^{a}$ one of the diagonal
generators of the Cartan algebra of $SU(N)$. This generator induces a
$U(1)_{\alpha}\subset SU(N)$ for which the $SU(N)$ effective Lagrangian
reduces to
\begin{align}
\mathfrak{L}_{eff}^{(4)}(U(1)_{\alpha}\overset{}{\subset}SU(N))  &
=(\gamma_{4,1}+\gamma_{4,3}+d^{\alpha\alpha\alpha\alpha}\gamma_{4,7}%
)\frac{g_{S}^{4}}{6!\pi^{2}m^{4}}G_{\mu\nu}^{\alpha}G^{\alpha,\mu\nu}%
G_{\rho\sigma}^{\alpha}G^{\alpha,\rho\sigma}\nonumber\\
& +(\gamma_{4,2}+\gamma_{4,4}+d^{\alpha\alpha\alpha\alpha}\gamma_{4,8}%
)\frac{g_{S}^{4}}{6!\pi^{2}m^{4}}G_{\mu\nu}^{\alpha}\tilde{G}^{\alpha,\mu\nu
}G_{\rho\sigma}^{\alpha}\tilde{G}^{\alpha,\rho\sigma}\ .
\end{align}
The Euler-Heisenberg result must arise from a combination of six of the eight
$SU(N)$ operators, including those involving the quartic invariant. Looking
back at their values in Table~\ref{TableSUN} for a given representation
$\mathbf{R}$, this reduction matches the results in Table~\ref{TableU1} for
scalar, fermion, and vector provided a single condition is satisfied:%
\begin{equation}
3\Lambda(\mathbf{R})+d^{\alpha\alpha\alpha\alpha}I_{4}(\mathbf{R}%
)=\sum_{q_{\alpha}\in\mathbf{R}}q_{\alpha}^{4}\ .\label{ReducU1}%
\end{equation}
The sum on the right-hand side is carried over all the states in the
representation $\mathbf{R}$. To see that this condition holds in general, it
suffices to go back to the very definition of the quartic invariant,
Eq.~(\ref{GenQuartic}), which becomes for a single generator:%
\begin{equation}
\frac{1}{4!}S\operatorname*{Tr}(T_{\mathbf{R}}^{\alpha}T_{\mathbf{R}}^{\alpha
}T_{\mathbf{R}}^{\alpha}T_{\mathbf{R}}^{\alpha})=\operatorname*{Tr}%
((T_{\mathbf{R}}^{\alpha})^{4})=I_{4}(\mathbf{R})d^{\alpha\alpha\alpha\alpha
}+3\Lambda(\mathbf{R})\ .
\end{equation}
Since $T_{\mathbf{R}}^{\alpha}$ is diagonal, the trace collapses to a sum over
the quartic power of its eigenvalues, i.e., over the quartic power of the
$U(1)_{\alpha}$ charges of the states of the representation $\mathbf{R}$. The
final step to match Table~\ref{TableU1} is to rescale the generator
$T_{\mathbf{R}}^{\alpha}$ to properly normalize the $U(1)_{\alpha}$ charge in
units of $Q$. Note that this relation can be trivially generalized to other
Casimir invariants. In particular, for the dimension-four and six operators,
$I_{2}(\mathbf{R})=\operatorname*{Tr}((T_{\mathbf{R}}^{\alpha})^{2}%
)=\sum_{q_{\alpha}\in\mathbf{R}}q_{\alpha}^{2}$, showing that the $\alpha_{i}$
coefficients for $SU(N)$ reduce to those for QED under the naive substitution
$I_{2}(\mathbf{R})\rightarrow Q^{2}$ in Table~\ref{TableSUN}.

Numerical applications to illustrate this formula are in
Appendix~\ref{AppCasimir}. Note that for both $SU(2)$ and $SU(3)$, there is no
quartic invariant and the Euler-Heisenberg coefficients for a single unit
charge state are formally obtained setting $\Lambda(\mathbf{1})=1/3$ in
Eq.~(\ref{ReducU1}). This value is plotted in Fig.~\ref{LambaR} for comparison.

\subsection{Reduction to factor groups}%

\begin{figure}[t]
\begin{center}
\includegraphics[height=1.1485in,width=3.0346in]{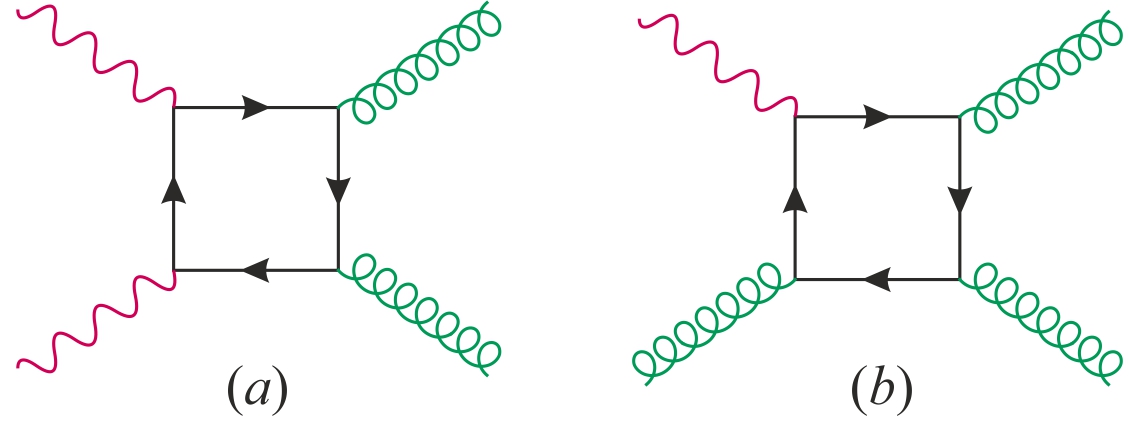}
\caption{Quark loops generating the effective dimension-eight photon-gluon
interactions.}
\label{EHphotglue}
\end{center}
\end{figure}

The general result also reduces to mixed interactions, involving the gauge
bosons of two different algebras. Before investigating this reduction, let us
directly compute them using FeynArts models. For that, we consider the
photon-gluon interactions induced by quark, squark, or $SU(5)$ leptoquark
loops in the non-linear gauge (see Fig.~\ref{EHphotglue}). It is then a simple
matter to generalize the results obtained for the fundamental $SU(3)_{C}$
representation to that for generic $SU(N)$ representations. The loops are
finite and the effective interactions start at the dimension-eight level,%
\begin{align}
\mathfrak{L}_{eff}^{(4)}(U(1)\otimes SU(N))  & =\alpha_{1}\frac{g_{1}^{2}%
g_{n}^{2}}{6!\pi^{2}m^{4}}F_{\mu\nu}F^{\mu\nu}G_{\rho\sigma}^{a}%
G^{a,\rho\sigma}+\alpha_{2}\frac{g_{1}^{2}g_{n}^{2}}{6!\pi^{2}m^{4}}F_{\mu\nu
}\tilde{F}^{\mu\nu}G_{\rho\sigma}^{a}\tilde{G}^{a,\rho\sigma}\nonumber\\
& +\alpha_{3}\frac{g_{1}^{2}g_{n}^{2}}{6!\pi^{2}m^{4}}F_{\mu\nu}G^{a,\mu\nu
}F_{\rho\sigma}G^{a,\rho\sigma}+\alpha_{4}\frac{g_{1}^{2}g_{n}^{2}}{6!\pi
^{2}m^{4}}F_{\mu\nu}\tilde{G}^{a,\mu\nu}F_{\rho\sigma}\tilde{G}^{a,\rho\sigma
}\nonumber\\
& +\beta_{1}\frac{g_{1}g_{n}^{3}}{6!\pi^{2}m^{4}}d^{abc}F_{\mu\nu}G^{a,\mu\nu
}G_{\rho\sigma}^{b}G^{c,\rho\sigma}+\beta_{2}\frac{g_{1}g_{n}^{3}}{6!\pi
^{2}m^{4}}d^{abc}F_{\mu\nu}\tilde{G}^{a,\mu\nu}G_{\rho\sigma}^{b}\tilde
{G}^{c,\rho\sigma}\;,\label{EffU1SUN}%
\end{align}
where $g_{1}$ and $g_{n}$ denote the $U(1)$ and $SU(N)$ coupling constants,
respectively. The numerical values of the Wilson coefficients are in
Table~\ref{TableMixed}. They are invariant under charge conjugation since
$Q(\mathbf{R}^{\ast})=-Q(\mathbf{R})$, $I_{2}(\mathbf{R}^{\ast})=+I_{2}%
(\mathbf{R})$, and $I_{3}(\mathbf{R}^{\ast})=-I_{3}(\mathbf{R})$, and they
obviously vanish for a real representation. Note in particular that the
$SU(5)$ leptoquarks give $\beta_{i}<0$ since the electric charge of the
antitriplet is positive, $Q(\mathbf{\bar{3}})=+\sqrt{5/12}$.

The first four interactions are immediately extended to the case of two
$SU(N)$ and two $SU(M)$ gauge bosons. Specifically, the operators are then%
\begin{align}
\mathfrak{L}_{eff}^{(4)}(SU(M)\otimes SU(N))  & =\alpha_{1}\frac{g_{m}%
^{2}g_{n}^{2}}{6!\pi^{2}m^{4}}W_{\mu\nu}^{i}W^{i,\mu\nu}G_{\rho\sigma}%
^{a}G^{a,\rho\sigma}+\alpha_{2}\frac{g_{m}^{2}g_{n}^{2}}{6!\pi^{2}m^{4}}%
W_{\mu\nu}^{i}\tilde{W}^{i,\mu\nu}G_{\rho\sigma}^{a}\tilde{G}^{a,\rho\sigma
}\nonumber\\
& +\alpha_{3}\frac{g_{m}^{2}g_{n}^{2}}{6!\pi^{2}m^{4}}W_{\mu\nu}^{i}%
G^{a,\mu\nu}W_{\rho\sigma}^{i}G^{a,\rho\sigma}+\alpha_{4}\frac{g_{m}^{2}%
g_{n}^{2}}{6!\pi^{2}m^{4}}W_{\mu\nu}^{i}\tilde{G}^{a,\mu\nu}W_{\rho\sigma}%
^{i}\tilde{G}^{a,\rho\sigma}\ ,
\end{align}
where $g_{m}$ and $g_{n}$ denote the $SU(M)$ and $SU(N)$ coupling constants,
respectively. Looking at Fig.~\ref{EHphotglue}$a$, it is easy to realize that
the coefficients are obtained from those for $U(1)$ in Table~\ref{TableMixed}
by replacing $Q(\mathbf{R})^{2}I_{2}(\mathbf{R})\rightarrow I_{2}%
^{M}(\mathbf{R}_{M})I_{2}^{N}(\mathbf{R}_{N})$ when the particles in the loop
are in the $(\mathbf{R}_{M},\mathbf{R}_{N})$ representation of $SU(M)\otimes
SU(N)$.

For the SM, the case $SU(2)_{L}\otimes SU(3)_{C}$ is immediately obtained in
the $\{W_{\mu}^{-},W_{\mu}^{3},W_{\mu}^{+}\}$ basis by replacing $W_{\mu\nu
}^{i}W^{i,\mu\nu}=W_{\mu\nu}^{3}W^{3,\mu\nu}+2W_{\mu\nu}^{+}W^{-,\mu\nu}$ and
$g_{n}\rightarrow g$, $g_{m}\rightarrow g_{S}$. Note however that the same
caveat as for the effective interactions in Eq.~(\ref{LargEffSU2}) applies. In
the presence of chiral fermions, these interactions are not leading and
dimension-six operators of $\mathcal{O}(m^{-2})$ appear, like for example
$\tilde{G}_{\mu\nu}^{a}G^{a,\nu\rho}Z^{\mu\rho}$ or $Z_{\mu}Z_{\rho}G_{\mu\nu
}^{a}G^{a,\rho\nu}$ inducing $Z\rightarrow ggg$~\cite{LaursenMS85} and
$gg\rightarrow ZZ$~\cite{ZZgg}. The only exceptions are the $Z\rightarrow
gg\gamma$~\cite{LaursenSTS83} and $Z\rightarrow\gamma\gamma\gamma
$~\cite{Z3phot} interactions for on-shell gluons and photons, which still
start at $\mathcal{O}(m^{-4})$ for chiral fermions because the $\gamma_{5}$
term of the $Z$ boson coupling to fermions cancels out. On-shell, these
effective interactions are simply obtained from the $\gamma\gamma\rightarrow
gg$ and $\gamma\gamma\rightarrow\gamma\gamma$ results by rescaling of one
photon couplings to match that of the $Z$ boson.

Because $U(1)\otimes SU(N)\subset SU(M\geqslant N+1)$, the $\alpha_{i}$,
$\beta_{i}$ coefficients in Table~\ref{TableMixed} are directly related to the
$\gamma_{4,i}$ in Table~\ref{TableSUN}, which is not very surprising comparing
their values. As for the reduction down to $U(1)$ in the previous section,
this can be understood looking at the coefficients of the loop functions.\ For
the $\alpha_{i}$ coefficients, the decomposition Eq.~(\ref{Decomp4p}) becomes
$C_{1}^{ab}=C_{2}^{ab}=C_{3}^{ab}=2I_{2}(\mathbf{R})Q^{2}\delta^{ab}$, hence
$D_{0}^{ab}=6I_{2}(\mathbf{R})Q^{2}\delta^{ab}$. Comparing with Eq.~(\ref{D02}%
), we see that $\alpha_{i}=2\gamma_{4,i}$ with the replacement $\Lambda
(\mathbf{R}_{M})\rightarrow Q(\mathbf{R}_{N})^{2}I_{2}(\mathbf{R}_{N})$ in
Table~\ref{TableSUN}. The factor of two comes from the two ways of identifying
the $U(1)$ and $SU(N)$ gauge bosons, e.g. $(G_{\mu\nu}^{a}G^{a,\mu\nu}%
)_{M}^{2}\rightarrow2(F_{\rho\sigma}F^{\rho\sigma})(G_{\mu\nu}^{a}G^{a,\mu\nu
})_{N}$. A similar reasoning can be done for the $\beta_{i}$ coefficients.

To go beyond a naive identification of the loop functions, let us denote
$T^{a}$ the Cartan generator of $SU(M)$ generating $U(1)$ and $T^{i},$
$i=2,...,N^{2}-1$ those generating $SU(N)$. Because $[T^{\alpha},T^{i}]=0$
implies $f^{\alpha ia}=0$, the UV divergent contributions disappear and the
$\gamma_{4,5}^{SU(M)}$ and $\gamma_{4,6}^{SU(M)}$ operators do not contribute
to the $U(1)\otimes SU(N)$ effective operators. For the other coefficients,
consider a specific representation of $SU(M)$ with branching rule
$\mathbf{R}_{M}\rightarrow\sum\mathbf{R}_{N}$, and denote $q_{\alpha
}(\mathbf{R}_{N})$ the $U(1)_{\alpha}$ charge of the states of the
representation $\mathbf{R}_{N}$. Mathematically, this branching rule means
$N^{2}$ of the $T_{\mathbf{R}_{M}}$ generators of $SU(M)$ can be brought to a
block diagonal form. Those corresponding to $SU(N)$ have blocks containing the
$SU(N)$ generators in the representation $\mathbf{R}_{N}$, while the
$T^{\alpha}$ generator is a diagonal matrix containing all the $q_{\alpha
}(\mathbf{R}_{N})$ charges, which are constant over each block since
$[T^{\alpha},T^{i}]=0$. The fully symmetrized trace with two or three $SU(N)$
generators then necessarily take the form%
\begin{align}
\frac{1}{4!}S\operatorname*{Tr}(T_{\mathbf{R}}^{\alpha}T_{\mathbf{R}}^{\alpha
}T_{\mathbf{R}}^{i}T_{\mathbf{R}}^{j})  & =\Lambda(\mathbf{R}_{M})\delta
^{ij}+d^{\alpha\alpha ij}I_{4}(\mathbf{R}_{M})=\sum_{\mathbf{R}_{N}%
\subset\mathbf{R}_{M}}q_{\alpha}(\mathbf{R}_{N})^{2}I_{2}(\mathbf{R}%
_{N})\delta^{ij}\ ,\\
\frac{1}{4!}S\operatorname*{Tr}(T_{\mathbf{R}}^{\alpha}T_{\mathbf{R}}%
^{i}T_{\mathbf{R}}^{j}T_{\mathbf{R}}^{k})  & =d^{\alpha ijk}I_{4}%
(\mathbf{R}_{M})=\frac{1}{4}\sum_{\mathbf{R}_{N}\subset\mathbf{R}_{M}%
}q_{\alpha}(\mathbf{R}_{N})I_{3}(\mathbf{R}_{N})d^{ijk}\ .\label{BranchingI4}%
\end{align}
This shows how the $\alpha_{i}$ and $\beta_{i}$ coefficients of $U(1)\otimes
SU(N)$ arise from the $\gamma_{4,i}$ coefficients of the general
$SU(M\geqslant N+1)$ effective Lagrangian. Computationally, to check these
identities requires first to work out the relationship between the symmetric
symbols. In general, all we can say from the block-diagonal structure of the
generators is that $d_{M}^{\alpha\alpha ij}=\eta_{1}\delta^{ij}$ and
$d_{M}^{\alpha ijk}=\eta_{2}d_{N}^{ijk}$ (see Eq.~(\ref{D4D3D3})), but the
proportionality constants $\eta_{1}$ and $\eta_{2}$ depend on how $U(1)\otimes
SU(N)$ is embedded into $SU(M)$. This is illustrated in
Appendix~\ref{AppCasimir}, where Eq.~(\ref{BranchingI4}) is used to derive the
quartic Casimir invariant $I_4$ of $SU(5)$ out of the anomaly coefficients $I_3$ of $SU(3)$.

As an interesting corrolary of this exact reduction, the identities in
Eq.~(\ref{CoeffCoh}) remain valid and imply $\alpha_{1,2}=\alpha_{3,4}/2$. So,
there are only two indepedent operators at the one loop level, no matter the
spin and representation of the particle in the loop. As before, this is not
true in general if more than a single field is integrated out. For example,
the analogue of the Higgs boson exchange shown in Fig.~\ref{Fig2LH}$b$
contributes to $\alpha_{1}$ only since the effective Higgs boson couplings to
photons and gluons are $h^{0}F_{\mu\nu}F^{\mu\nu}$ and $h^{0}G_{\mu\nu}%
^{a}G^{a,\mu\nu}$.%

\begin{table}[t] \centering
$
\begin{tabular}
[c]{ccccc}\hline
& $\alpha_{1}=\alpha_{3}/2$ & $\alpha_{2}=\alpha_{4}/2$ & $\beta_{1}$ &
$\beta_{2}$\smallskip\\\hline
Scalar & $\dfrac{7}{16}Q(\mathbf{R})^{2}I_{2}(\mathbf{R})$ & $\dfrac{1}%
{16}Q(\mathbf{R})^{2}I_{2}(\mathbf{R})$ & $\dfrac{7}{32}Q(\mathbf{R}%
)I_{3}(\mathbf{R})$ & $\dfrac{1}{32}Q(\mathbf{R})I_{3}(\mathbf{R}%
)\rule[-0.18in]{0in}{0.4in}$\\
Fermion & $Q(\mathbf{R})^{2}I_{2}(\mathbf{R})$ & $\dfrac{7}{4}Q(\mathbf{R}%
)^{2}I_{2}(\mathbf{R})$ & $\dfrac{1}{2}Q(\mathbf{R})I_{3}(\mathbf{R})$ &
$\dfrac{7}{8}Q(\mathbf{R})I_{3}(\mathbf{R})\rule[-0.18in]{0in}{0.4in}$\\
Vector & $\dfrac{261}{16}Q(\mathbf{R})^{2}I_{2}(\mathbf{R})$ & $\dfrac
{243}{16}Q(\mathbf{R})^{2}I_{2}(\mathbf{R})$ & $\dfrac{261}{32}Q(\mathbf{R}%
)I_{3}(\mathbf{R})$ & $\dfrac{243}{32}Q(\mathbf{R})I_{3}(\mathbf{R}%
)\rule[-0.18in]{0in}{0.4in}$\\\hline
\end{tabular}
$
\caption{Wilson coefficients of the effective operators for the mixed operators, as induced by a complex field (scalar, fermion, vector boson) in the representation \textbf{R} of $SU(N)$ with $U(1)$ charge $Q(\mathbf{R})$. The $\alpha_i$ coefficients for two $SU(N)$ and two $SU(M)$ gauge bosons are obtained by replacing $Q(\mathbf{R})^{2}I_{2}(\mathbf{R}) \rightarrow I_{2}^{M}(\mathbf{R}_M) I_{2}^{N}(\mathbf{R}_N)$.}
\label{TableMixed}
\end{table}

\section{Conclusion}

In this paper, the effective action for gauge theories is revisited.
Integrating out some heavy charged fields, self-interactions among gauge
bosons are encoded into effective operators. Using the diagrammatic approach,
we explicitly constructed these interactions up to the dimension-eight level,
and computed their coefficients as induced by loops of heavy
particles of spin 0, 1/2, or 1. More specifically,

\begin{itemize}
\item To set the stage and identify possible issues, we first reviewed in
details the construction of the off-shell effective couplings for photons. In
the diagrammatic approach, integrating out fermions or scalars is
straightforward and we recover the usual Euler-Heisenberg result. For heavy
vector fields, the matching does not proceeds as trivially. Indeed, in the 't
Hooft-Feynman gauge, the gauge-fixing term required for the massive vector
fields breaks the $U(1)$ gauge invariance. Consequently, the off-shell
four-photon amplitude fails to satisfy the QED Ward identities, and the usual
procedure to construct the effective action breaks down. To solve this
problem, we adopted the strategy of Ref.~\cite{Boudjema86} and quantized the SM
in the non-linear gauge. Matching is then consistent off-shell, and the
diagrammatic approach closely parallels the path integral-based Covariant Derivative Expansion method~\cite{Gaillard:1985uh,Cheyette:1987qz}.
The Wilson coefficients in that gauge
are shown in Table~\ref{TableU1}.

\item The calculation of the photon EFT was then extended to the QCD gluon
EFT. The most general basis of gluonic operators up to dimension-eight is
quite different from the QED case due to the non-abelian nature of
QCD~\cite{Gracey17}. We computed explicitly the coefficients of the effective
operators for a scalar, fermion or vector in the fundamental representation.
The final results for the coefficients are given in Table~\ref{TableSU3}. As
for photons, integrating out heavy vector fields requires dealing with gauge
dependences. Our strategy was to use the minimal $SU(5)$ GUT model,
spontaneously broken by an adjoint Higgs scalar down to the unbroken SM gauge
group, and quantized using a non-linear gauge condition preserving the SM
gauge invariance. Twelve of the $SU(5)$ gauge bosons become massive in the
process, and those fields have precisely the quantum numbers needed to induce
the effective gluon couplings. This construction is detailed in
Appendix~\ref{AppGUT}. Technically, it should be mentioned that this
non-linear gauge has the additional nice feature of drastically reducing the
number of diagrams for a given process.

\item We then extended the computation done in the QCD case to generic Lie
gauge groups, taking $SU(N)$, $U(1)\otimes SU(N)$, and $SU(M)\otimes SU(N)$ as
examples, and allowing the heavy particle to sit in arbitrary representations.
The coefficients for a complex field of spin 0, 1/2, or 1 circulating in the
loops are given in Table~\ref{TableSUN} for $SU(N)$ and in
Table~\ref{TableMixed} for non-simple gauge groups. One feature apparent in
these tables is worth stressing. At one loop, some operators are redundant no
matter the representation or spin of the particle circulating in the loops.
From our Eq.~(\ref{CoeffCoh}), we conclude that two operator combinations
never occur in the one-loop effective action for $SU(N)$ gauge bosons. This
implies in particular that only four instead of six operators are required for
QCD, and only two instead of four operators are sufficient to describe the two
gluon-two photon interactions. Finally, it should be mentioned that
generalizing the QCD result to an arbitrary Lie algebra required a careful
analysis of quartic Casimir invariants. While all the needed information can
be dig out of the available literature\cite{Okubo81,vanRitbergenSV99}, it
seems to us a short review detailing all the definitions and conventions, and
with emphasis on practical use in loop calculations, was lacking and so, it is
included in Appendix~\ref{AppCasimir}.

\item On a more technical note, the relationship between effective action and
Feynman diagram matching was carefully analyzed. Specifically, the effective
action can be computed from the one-loop 1PI off-shell amplitudes. In this
way, the coefficients of all the operators, including those vanishing under
the equation of motion, are obtained. However, these coefficients are not
necessarily gauge-invariant. Actually, since the matching is possible only
using a non-linear gauge fixing, they are well-defined in that gauge only.
This is to be compared to the computation of the coefficients using on-shell
processes, where the physical on-shell one-loop amplitudes are matched onto a
subset of operators. Those operators that vanish under the EOM are absent, so
the whole effective action is never reproduced. Further, from a calculation
point of view, matching with on-shell processes requires dealing with both 1PI and non-1PI amplitudes. For example, the
coefficient of the three-gluon-field strength operator $f^{abc}G_{\mu}%
^{a\;\nu}G_{\nu}^{b\;\rho}G_{\rho}^{c\;\mu}$ cannot be obtained from a
three-gluon process since it is kinematically forbidden. Instead, it has to be
extracted alongside all the four-gluon-field strength operators by matching
onto the four-gluon physical amplitudes.
\end{itemize}

Altogether, the construction of the effective gauge-boson Lagrangian up to
dimension-eight is now fully under control in the diagrammatic approach. The
operator bases are confirmed, their group-theoretic properties clarified, and the
coefficients are known for the standard benchmark scenarios of heavy scalars,
fermions, and vector bosons. Phenomenologically, though the four-gluon or four
weak boson effective couplings is unlikely to be ever seen, given the presence
of such a coupling in the tree-level Lagrangian, there may be some room for
$\gamma\gamma\rightarrow gg$. In any case, having laid out a well-defined
strategy to construct fully general effective actions involving gauge bosons
will prove useful in the future.

\newpage
\appendix

\section{SU(5) gauge bosons in the non-linear gauge\label{AppGUT}}

This appendix is not intended as a review of the minimal $SU(5)$ model.
Rather, it is meant as a guide to construct the Lagrangian of $SU(5)$ broken
down to $SU(3)_{C}\otimes SU(2)_{L}\otimes U(1)_{Y}$, quantized using a
non-linear gauge-fixing term, in a form suitable for automatic calculation
tools. The main point is to input all the Lagrangian terms in a consistent and
tractable way. This requires to set a number of conventions and definitions,
so we found it useful to detail them here.

The starting point is to input the $SU(5)$ gauge bosons, and write them in
terms of those of the $SU(3)_{C}\otimes SU(2)_{L}\otimes U(1)_{Y}$ gauge
group. For that, we start from the branching rule of the adjoint
representation $\mathbf{24}$:%
\begin{equation}
\mathbf{24}=(\mathbf{8},\mathbf{1})_{0}+(\mathbf{1},\mathbf{3})_{0}%
+(\mathbf{3},\mathbf{2})_{5}+(\mathbf{\bar{3}},\mathbf{2})_{-5}+(\mathbf{1}%
,\mathbf{1})_{0}\;.
\end{equation}
Denoting by $A,B,...=1,...,24$ the $SU(5)$ adjoint indices, $a,b,...=1,...,8$
the adjoint color indices, and $i,j,...=1,2,3$ the fundamental $SU(3)$
indices, the twenty-four $A_{A}^{\mu}$ gauge bosons are identified as the
octet of gluons $(\mathbf{8}\otimes\mathbf{1})_{0}\sim G_{i}^{\mu}=A_{i}^{\mu
}$, $a=1,...,8$, the triplet of weak bosons $(\mathbf{1}\otimes\mathbf{3}%
)_{0}\sim W^{\pm\mu}=(A_{9}^{\mu}\mp iA_{10}^{\mu})/\sqrt{2}$, $W_{3}^{\mu
}=A_{11}^{\mu}$, and the singlet $(\mathbf{1}\otimes\mathbf{1})_{0}\sim
B^{\mu}=A_{24}^{\mu}$. The remaining fields are the twelve leptoquark gauge
bosons and their conjugate fields in the $(\mathbf{\bar{3}}\otimes
\mathbf{2})_{5/3}$ and $(\mathbf{3}\otimes\mathbf{\bar{2}})_{-5/3}$
representation, respectively. We define these fields as $X_{1}^{\mu\pm
}=(A_{12}^{\mu}\pm iA_{13}^{\mu})/\sqrt{2}$, $Y_{1}^{\mu\pm}=(A_{18}^{\mu}\pm
iA_{19}^{\mu})/\sqrt{2}$ and so on. Note that leptoquarks are charged under
all the SM gauge groups, and those with positive hypercharge transform like
antiquarks under $SU(3)_{C}$.

Since the adjoint is contained in $\mathbf{5}\otimes\mathbf{\bar{5}%
}=\mathbf{24}\oplus\mathbf{1}$, all these identifications of the gauge fields
can be put together to construct a traceless $5\times5$ matrix for the $SU(5)$
gauge fields:%
\begin{equation}
\mathbf{A}^{\mu}=A_{A}^{\mu}T^{A}=\left(
\begin{array}
[c]{ccc}%
T_{ij}^{a}G_{\mu}^{a}-\frac{1}{\sqrt{15}}B_{\mu}\delta_{ij} & \frac{1}%
{\sqrt{2}}X_{\mu}^{i-} & \frac{1}{\sqrt{2}}Y_{\mu}^{i-}\\
\frac{1}{\sqrt{2}}X_{\mu}^{j+} & \frac{1}{2}W_{\mu}^{3}+\frac{3}{2\sqrt{15}%
}B_{\mu} & \frac{1}{\sqrt{2}}W_{\mu}^{+}\\
\frac{1}{\sqrt{2}}Y_{\mu}^{j+} & \frac{1}{\sqrt{2}}W_{\mu}^{-} & -\frac{1}%
{2}W_{\mu}^{3}+\frac{3}{2\sqrt{15}}B_{\mu}%
\end{array}
\right)  \,,
\label{SU5Boson}
\end{equation}
where $T^{A}$ are the conventional $SU(5)$ generators in the fundamental
representation, normalized as $\operatorname*{Tr}(T^{A}T^{B})=\delta^{AB}/2$.
This identification is compatible with the eigenstates of the electric charge
operator,%
\begin{equation}
Q=T^{11}+\sqrt{5/3}T^{24}\ ,\ \ \left[  Q,\mathbf{A}_{\mu}\right]  =\frac
{1}{\sqrt{2}}\left(
\begin{array}
[c]{ccc}%
0 & -4/3X_{\mu}^{i-} & -1/3Y_{\mu}^{i-}\\
4/3X_{\mu}^{j+} & 0 & +W_{\mu}^{+}\\
1/3Y_{\mu}^{j+} & -W_{\mu}^{-} & 0
\end{array}
\right)  \ ,
\end{equation}
with the normalization of the hypercharge operator $Y=2\sqrt{5/3}T^{24}$. In
practice, we have used the Mathematica package \textit{FeynArts}~\cite{FeynArts} and
\textit{FeynCalc}~\cite{FeynCalc}. Both allow to keep the summations over the
$SU(3)$ indices as implicit, so $\mathbf{A}^{\mu}$ is truly input as the $3\times3$ matrix of Eq.~(\ref{SU5Boson}). 
Once all the relevant pieces of the Lagrangian are encoded, it is then a simple
matter to extract the Feynman rules and export them to \textit{FeynArts}. Let us now review the Lagrangian terms of relevance to us.

\subsection*{Gauge interactions}

The gauge self-couplings derive from the Yang-Mills kinetic term%
\begin{equation}
\mathcal{L}_{\text{gauge}}=-\frac{1}{2}\langle\mathbf{A}_{\mu\nu}%
\mathbf{A}^{\mu\nu}\rangle=-\frac{1}{4}A_{\mu\nu}^{A}A^{A,\mu\nu}\;,
\end{equation}
with the field strength%
\begin{equation}
\mathbf{A}_{\mu\nu}=\partial_{\mu}\mathbf{A}_{\nu}-\partial_{\nu}%
\mathbf{A}_{\mu}-ig_{5}[\mathbf{A}_{\mu},\mathbf{A}_{\nu}]=(\partial_{\mu
}A_{\nu}^{A}-\partial_{\nu}A_{\mu}^{A}+gf^{ABC}A_{\mu}^{B}A_{\nu}^{C})T^{A}\ .
\end{equation}
The $SU(5)$ structure constants are defined as $[T^{A},T^{B}]=if^{ABC}T^{C}$.
An explicit calculation shows that there are 68 non-zero $f^{ABC}$, plus
antisymmetric permutations of the indices. Among them there are the nine
$f^{abc}$ of $SU(3)$ and the single $\varepsilon^{ijk}$ of $SU(2)$, which
reproduce the QCD and electroweak self-interactions. All the other non-zero
structure constants are $f^{ABC}$ with $A,B=12,...,23$ and $C=1,...,11,24$. In
other words, they involve twice the leptoquark fields, as can be expected
since these particles are charged under the three SM gauge groups. The same
$g_{5}$ occurs for all the interactions between gauge bosons. In explicit
form,%
\begin{align}
\mathcal{L}_{\text{gauge}}  & =-\frac{1}{2}\langle(\partial_{\mu}%
\mathbf{A}_{\nu}-\partial_{\nu}\mathbf{A}_{\mu})(\partial^{\mu}\mathbf{A}%
^{\nu}-\partial^{\nu}\mathbf{A}^{\mu})+4ig_{5}\mathbf{A}_{\mu}\mathbf{A}_{\nu
}(\partial^{\mu}\mathbf{A}^{\nu}-\partial^{\nu}\mathbf{A}^{\mu})-2g_{5}%
^{2}\mathbf{A}_{\mu}\mathbf{A}_{\nu}[\mathbf{A}^{\mu},\mathbf{A}^{\nu}%
]\rangle\nonumber\\
& =-\frac{1}{4}G_{\mu\nu}^{a}G^{a,\mu\nu}-\frac{1}{2}W_{\mu\nu}^{+}W^{-,\mu
\nu}-\frac{1}{4}W_{\mu\nu}^{3}W^{3,\mu\nu}-\frac{1}{4}B_{\mu\nu}B^{\mu\nu
}\nonumber\\
& -\frac{1}{2}(D_{\mu}X_{\nu}^{+}-D_{\nu}X_{\mu}^{+})^{i}(D^{\mu}X^{-\nu
}-D^{\nu}X^{-\mu})^{i}-\frac{1}{2}(D_{\mu}Y_{\nu}^{+}-D_{\nu}Y_{\mu}^{+}%
)^{i}(D^{\mu}Y^{-\nu}-D^{\nu}Y^{-\mu})^{i}\nonumber\\
& \ \ \ +ig_{5}G_{\mu\nu}^{a}(X_{\mu}^{j+}(-T_{ji}^{a})X_{\nu}^{i-}+Y_{\mu
}^{j+}(-T_{ji}^{a})Y_{\nu}^{i-})+i\frac{g_{5}}{\sqrt{2}}(W^{+,\mu\nu}Y_{\mu
}^{i+}X_{\nu}^{i-}+W^{-,\mu\nu}X_{\mu}^{i+}Y_{\nu}^{i-})\nonumber\\
& \ \ \ +i\frac{g_{5}}{2}W^{3,\mu\nu}(X_{\mu}^{i+}X_{\nu}^{i-}-Y_{\mu}%
^{i+}Y_{\nu}^{i-})+ig_{5}\frac{\sqrt{15}}{6}B^{\mu\nu}(X_{\mu}^{i+}X_{\nu
}^{i-}+Y_{\mu}^{i+}Y_{\nu}^{i-})+\mathcal{O}((X,Y)^{4})\ ,
\end{align}
where the weak and strong field strengths are understood to contain their
respective non-abelian terms, as%
\begin{align*}
G_{\mu\nu}^{a}  & =\partial_{\nu}G_{\mu}^{a}-\partial_{\mu}G_{\nu}^{a}%
+g_{5}f^{abc}G_{\mu}^{b}G_{\nu}^{c}\rightarrow G_{\mu\nu}^{a}T^{a}%
=\partial_{\nu}G_{\mu}^{a}T^{a}-\partial_{\mu}G_{\nu}^{a}T^{a}-ig_{5}[G_{\mu
}^{b}T^{b},G_{\nu}^{c}T^{c}]\ ,\\
W^{i,\mu\nu}  & =\partial_{\nu}W_{\mu}^{i}-\partial_{\mu}W_{\nu}^{i}%
+g_{5}\varepsilon^{ijk}W_{\mu}^{j}W_{\nu}^{k}\ \rightarrow\left\{
\begin{array}
[c]{c}%
W^{3,\mu\nu}=\partial_{\nu}W_{\mu}^{3}-\partial_{\mu}W_{\nu}^{3}+ig_{5}%
(W_{\mu}^{-}W_{\nu}^{+}-W_{\mu}^{+}W_{\nu}^{-})\ ,\\
W^{+,\mu\nu}=\partial_{\nu}W_{\mu}^{+}-\partial_{\mu}W_{\nu}^{+}+ig_{5}%
(W_{\mu}^{+}W_{\nu}^{3}-W_{\mu}^{3}W_{\nu}^{+})\ ,\\
W^{-,\mu\nu}=\partial_{\nu}W_{\mu}^{-}-\partial_{\mu}W_{\nu}^{-}+ig_{5}%
(W_{\mu}^{3}W_{\nu}^{-}-W_{\mu}^{-}W_{\nu}^{3})\ .
\end{array}
\right.
\end{align*}
The covariant derivative $D^{\mu}=\partial^{\mu}\mathbf{1}-ig_{5}T^{A}%
A_{A}^{\mu}$ acting on the twelve leptoquarks living in the $(\mathbf{\bar{3}%
}\otimes\mathbf{2})_{5/3}$ representation is%
\begin{align}
(D_{\mu})_{ij}X_{\nu}^{j+}  & =\partial_{\mu}X_{\nu}^{i+}-ig_{5}\left(
X_{\nu}^{j+}(-T_{ji}^{a})G_{\mu}^{a}+\frac{1}{2}W_{\mu}^{3}X_{\nu}^{i+}%
+\frac{1}{\sqrt{2}}W_{\mu}^{+}Y_{\nu}^{i+}+y\frac{5}{6}B_{\mu}X_{\nu}%
^{i+}\right)  \ ,\\
(D_{\mu})_{ij}Y_{\nu}^{j+}  & =\partial_{\mu}Y_{\nu}^{i\pm}-ig_{5}\left(
Y_{\nu}^{j+}(-T_{ji}^{a})G_{\mu}^{a}-\frac{1}{2}W_{\mu}^{3}Y_{\nu}^{i+}%
+\frac{1}{\sqrt{2}}W_{\mu}^{-}X_{\nu}^{i+}+y\frac{5}{6}B_{\mu}Y_{\nu}%
^{i+}\right)  \ ,
\end{align}
where $y=\sqrt{3/5}$ is the hypercharge normalization. Finally, $\mathcal{O}%
((X,Y)^{4})$ denotes quartic interactions among $X$ and $Y$ gauge bosons which
are of no interest for our purpose. It is interesting to remark that the SM
gauge invariance is satisfied separately for the $X,Y$ kinetic terms (thanks
to the covariant derivatives), the magnetic interactions (the $B_{\mu\nu
}X^{\mu}X^{\nu}$ and similar), and the $\mathcal{O}((X,Y)^{4})$
interactions.\ At the level of the SM, the strength of the magnetic and
$\mathcal{O}((X,Y)^{4}$ interactions are thus unconstrained, and these could
even be absent. On the contrary, here their relative strengths is fixed by the
underlying $SU(5)$ gauge invariance. The situation is similar in the SM, with
the relative strength of the $(D_{\mu}W_{\nu}^{+}-D_{\nu}W_{\mu}^{+})(D^{\mu
}W^{-\nu}-D^{\nu}W^{-\mu})$ and $F_{\mu\nu}W_{\mu}^{+}W^{-\nu}$ interactions
fixed by the underlying $SU(2)_{L}\otimes U(1)_{Y}$ symmetry.

\subsection*{Scalar interactions}

In the present work, we are only interested in the initial breaking stage
\begin{equation}
SU(5)\rightarrow SU(3)_{C}\otimes SU(2)_{L}\otimes U(1)_{Y}\ .
\end{equation}
For that, we need a scalar in the adjoint representation, $\mathbf{\bar{H}%
}_{\mathbf{24}}=\sqrt{2}H_{A}T_{A}$. Note that $\mathbf{\bar{H}}_{\mathbf{24}%
}=\mathbf{\bar{H}}_{\mathbf{24}}^{\dagger}$, since the adjoint is a real
representation, and further assuming a $\mathbf{\bar{H}}_{\mathbf{24}%
}\rightarrow-\mathbf{\bar{H}}_{\mathbf{24}}$ symmetry to get rid of cubic
interactions, the most general Lagrangian is%
\begin{equation}
\mathcal{L}_{\text{scalar}}=\frac{1}{2}\langle D_{\mu}\mathbf{\bar{H}%
}_{\mathbf{24}}D^{\mu}\mathbf{\bar{H}}_{\mathbf{24}}\rangle+\frac{\mu^{2}}%
{2}\langle\mathbf{H}_{\mathbf{24}}^{2}\rangle-\frac{a}{4}\langle
\mathbf{H}_{\mathbf{24}}^{2}\rangle^{2}-\frac{b}{2}\langle\mathbf{H}%
_{\mathbf{24}}^{4}\rangle\ .\label{LagrScalar}%
\end{equation}
The breaking of the $SU(5)$ symmetry arises when $\mathbf{\bar{H}%
}_{\mathbf{24}}$ gets its vacuum expectation value $\langle0|\mathbf{\bar{H}%
}_{\mathbf{24}}|0\rangle\sim v_{5}>0$, which happens for $\mu^{2}>0$. There
are two classes of minima, depending on the sign of $b$. First, it is possible
to find values of $\mu$, $a$, and $b<0$ such that the minimum is of the form
$\langle0|\mathbf{\bar{H}}_{\mathbf{24}}|0\rangle=\operatorname*{diag}%
(v,v,v,v,-4v)$. This corresponds to $SU(5)\rightarrow SU(4)\otimes U(1)$. The
second class occurs for $b>0$ and is such that $\langle0|\mathbf{\bar{H}%
}_{\mathbf{24}}|0\rangle$ commutes with the $SU(3)_{C}$, $SU(2)_{L}$, and
$U(1)_{Y}$ generators:%
\begin{equation}
\mathbf{H}_{\mathbf{24}}^{0}=\langle0|\mathbf{\bar{H}}_{\mathbf{24}}%
|0\rangle=\frac{1}{\sqrt{2}}v_{5}\,\operatorname*{diag}%
(1,1,1,-3/2,-3/2)=-v_{5}\sqrt{15/4}T^{24},\ \;\;\;\;v_{5}^{2}=\dfrac{4\mu^{2}%
}{15a+7b}\;.
\end{equation}
The value of $v_{5}$ is found by requiring that this is a global minimum of
the potential, which asks for $15a+7b>0$.

Plugging this constraint in the scalar potential and writing%
\begin{equation}
\mathbf{H}_{\mathbf{24}}=\mathbf{\bar{H}_{\mathbf{24}}}-\mathbf{H}%
_{\mathbf{24}}^{0}=\sqrt{2}\left(
\begin{array}
[c]{ccc}%
T_{ij}^{a}H_{G}^{a}-\frac{1}{\sqrt{15}}H_{B}^{0}\delta_{ij} & \frac{1}%
{\sqrt{2}}H_{X}^{i-} & \frac{1}{\sqrt{2}}H_{Y}^{i-}\\
\frac{1}{\sqrt{2}}H_{X}^{j+} & \frac{1}{2}H_{W}^{3}+\frac{3}{2\sqrt{15}}%
H_{B}^{0} & \frac{1}{\sqrt{2}}H_{W}^{+}\\
\frac{1}{\sqrt{2}}H_{Y}^{j+} & \frac{1}{\sqrt{2}}H_{W}^{-} & -\frac{1}{2}%
H_{W}^{3}+\frac{3}{2\sqrt{15}}H_{B}^{0}%
\end{array}
\right)  \;,
\end{equation}
the Higgs boson masses are found to be%
\[
M_{H_{W}^{i}}^{2}=4M_{H_{G}^{a}}^{2}=5bv_{5}^{2}\ ,\ \ M_{H_{B}}^{2}=2\mu
^{2}\ ,\ \ M_{H_{X,Y}^{i}}^{2}=0\ .
\]
Note that the $\sqrt{2}$ is conventional; it ensures a correctly normalized
kinetic terms given the Lagrangian in Eq.~(\ref{LagrScalar}). Additional
couplings involving three and four scalars are derived from the potential,
with the former all proportional to $v_{5}$.

To get the scalar couplings to gauge bosons, it then suffices to expand the
covariant derivative, with for the adjoint representation,%
\begin{equation}
D^{\mu}\mathbf{\bar{H}}_{\mathbf{24}}=\partial^{\mu}\mathbf{\bar{H}%
}_{\mathbf{24}}-ig_{5}\left[  \mathbf{A}^{\mu},\mathbf{\bar{H}}_{\mathbf{24}%
}\right]  =\partial^{\mu}\mathbf{H}_{\mathbf{24}}-ig_{5}\left[  \mathbf{A}%
^{\mu},\mathbf{H}_{\mathbf{24}}\right]  -ig_{5}\left[  \mathbf{A}^{\mu
},\mathbf{H}_{\mathbf{24}}^{0}\right]  \;.
\end{equation}
This gives%
\begin{equation}
\frac{1}{2}\langle D_{\mu}\mathbf{\bar{H}}_{\mathbf{24}}D^{\mu}\mathbf{\bar
{H}}_{\mathbf{24}}\rangle\rightarrow\frac{1}{2}\langle\partial_{\mu}%
\mathbf{H}_{\mathbf{24}}\partial^{\mu}\mathbf{H}_{\mathbf{24}}\rangle
+\mathcal{L}_{\text{mass}}+\mathcal{L}_{\text{mix}}+\mathcal{L}%
_{\text{gauge-Higgs}}\ .
\end{equation}
The $\mathcal{L}_{\text{mass}}$ couplings are just the leptoquark mass terms,%
\begin{equation}
\mathcal{L}_{\text{mass}}=-\frac{1}{2}g_{5}^{2}\langle\left[  \mathbf{A}_{\mu
},\mathbf{H}_{\mathbf{24}}^{0}\right]  \left[  \mathbf{A}^{\mu},\mathbf{H}%
_{\mathbf{24}}^{0}\right]  \rangle=\frac{25}{16}g_{5}^{2}v_{5}^{2}\left(
X_{\mu}^{i+}X^{i-\mu}+Y_{\mu}^{i+}Y^{i-\mu}\right)  \ ,
\end{equation}
so $M_{XY}=5g_{5}v_{5}/4$. The $\mathcal{L}_{\text{mix}}$ piece induces
mixings between the $X^{\mu}$ and $Y^{\mu}$ gauge bosons and their associated
WBG bosons,%
\begin{equation}
\mathcal{L}_{\text{mix}}=-ig_{5}\langle\left[  \mathbf{A}_{\mu},\mathbf{H}%
_{\mathbf{24}}^{0}\right]  \partial^{\mu}\mathbf{H}_{\mathbf{24}}%
\rangle=iM_{XY}X_{\mu}^{k-}\partial^{\mu}H_{X}^{k+}+iM_{XY}Y_{\mu}%
^{k-}\partial^{\mu}H_{Y}^{k+}+h.c.\ .
\end{equation}
The other couplings involve gauge and scalar bosons,%
\begin{equation}
\mathcal{L}_{\text{gauge-Higgs}}=-ig_{5}\langle\left[  \mathbf{A}_{\mu
},\mathbf{H}_{\mathbf{24}}\right]  \partial^{\mu}\mathbf{H}_{\mathbf{24}%
}\rangle-g_{5}^{2}\langle\left[  \mathbf{A}_{\mu},\mathbf{H}_{\mathbf{24}}%
^{0}\right]  \left[  \mathbf{A}^{\mu},\mathbf{H}_{\mathbf{24}}\right]
\rangle-\frac{g_{5}^{2}}{2}\langle\left[  \mathbf{A}_{\mu},\mathbf{H}%
_{\mathbf{24}}\right]  \left[  \mathbf{A}^{\mu},\mathbf{H}_{\mathbf{24}%
}\right]  \rangle\ .
\end{equation}
The explicit forms can easily be worked out and will not be given here. Remark
though that because all the SM gauge bosons disappear from $\left[
\mathbf{A}_{\mu},\mathbf{H}_{\mathbf{24}}^{0}\right]  $, $\mathcal{L}%
_{\text{AAH}}$ only couples scalars to the massive gauge bosons, with
couplings proportional to their mass.

\subsection*{Gauge-fixing and ghost interactions}

The next step to quantize this theory is to fix the gauge, and add the
corresponding ghost terms. The general ansatz in linear $R_{\xi}$ gauge is to
define the constraint in terms of the WBG as%
\begin{align}
\mathbf{G}  & =\sqrt{2}\partial_{\mu}\mathbf{A}^{\mu}+\xi M_{XY}\left(
\begin{array}
[c]{ccc}%
0 & iH_{X}^{i-} & iH_{Y}^{i-}\\
-iH_{X}^{j+} & 0 & 0\\
-iH_{Y}^{j+} & 0 & 0
\end{array}
\right) \nonumber\\
& =\left(
\begin{array}
[c]{ccc}%
\sqrt{2}T_{ij}^{a}\partial^{\mu}G_{\mu}^{a}-\sqrt{\frac{2}{15}}\partial^{\mu
}B_{\mu}\delta_{ij} & \partial^{\mu}X_{\mu}^{i-}+i\xi M_{XY}H_{X}^{i-} &
\partial^{\mu}Y_{\mu}^{i-}+i\xi M_{XY}H_{Y}^{i-}\\
\partial^{\mu}X_{\mu}^{j+}-i\xi M_{XY}H_{X}^{j+} & \frac{1}{\sqrt{2}}%
\partial^{\mu}W_{\mu}^{3}+\sqrt{\frac{3}{10}}\partial^{\mu}B_{\mu} &
\partial^{\mu}W_{\mu}^{+}\\
\partial^{\mu}Y_{\mu}^{j+}-i\xi M_{XY}H_{Y}^{j+} & \partial^{\mu}W_{\mu}^{-} &
-\frac{1}{\sqrt{2}}\partial^{\mu}W_{\mu}^{3}+\frac{3}{2\sqrt{15}}\partial
^{\mu}B_{\mu}%
\end{array}
\right)  \ ,\label{Gphys}%
\end{align}
so that%
\begin{align}
\mathcal{L}_{\text{gf}}\overset{}{=}-\frac{1}{2\xi}\langle\mathbf{G}%
^{2}\rangle &  =-\frac{1}{\xi}|\partial^{\mu}X_{\mu}^{k+}-i\xi M_{XY}%
H_{X}^{k+}|^{2}-\frac{1}{\xi}|\partial^{\mu}Y_{\mu}^{k+}-i\xi M_{XY}H_{Y}%
^{k+}|^{2}\\
&  \ \ \ \ -\frac{1}{\xi}|\partial^{\mu}W_{\mu}^{+}|^{2}-\frac{1}{2\xi
}(\partial^{\mu}W_{\mu}^{3})^{2}-\frac{1}{2\xi}\left(  \partial^{\mu}B_{\mu
}\right)  ^{2}-\frac{1}{2\xi}(\partial^{\mu}G_{\mu}^{a})^{2}\ .
\end{align}
Since in practice, all our computations are done in the 't Hooft-Feynman
gauge, a common parameter $\xi$ is introduced for all the gauge bosons.
Obviously, the parameters for $G_{\mu}^{a}$, $B_{\mu}$, $W_{\mu}^{3}$, and
$W_{\mu}^{\pm}$ can be all different since they appear only in the respective
propagator and not in any of the vertices. For $X_{\mu}^{i\pm}$ and $Y_{\mu
}^{i\pm}$, not taking a common parameter would make life more complicated
since those two form an $SU(2)_{L}$ doublet. When the first line is expanded,
the terms linear in $M_{XY}$ precisely cancel those in $\mathcal{L}%
_{\text{mix}}$, while those quardratic imply $M_{H_{XY}}^{2}=\xi M_{XY}^{2}$
as usual. Remember that WBG do not get any mass term from the scalar potential.

The goal of the non-linear gauge fixing of Ref.~\cite{GavelaGMS81} is to
maintain the unbroken gauge symmetries as explicit. This requires general
covariant derivatives in the constraints involving the massive gauge bosons.
To be able to interpolating between the linear and non-linear gauge, we
introduce the parameters $\alpha_{G}$, $\alpha_{W}$, $\alpha_{B}$ and use
\begin{align}
\partial^{\mu}X_{\mu}^{i+}  & \rightarrow\partial^{\mu}X_{\mu}^{i+}%
-ig_{5}\left(  \alpha_{G}X_{\nu}^{j+}(-T_{ji}^{a})G_{\mu}^{a}+\alpha_{W}%
\frac{1}{2}W_{\mu}^{3}X_{\nu}^{i+}+\alpha_{W}\frac{1}{\sqrt{2}}W_{\mu}%
^{+}Y_{\nu}^{i+}+\alpha_{B}y\frac{5}{6}B_{\mu}X_{\nu}^{i+}\right)  \ ,\\
\partial^{\mu}Y_{\nu}^{i+}  & \rightarrow\partial^{\mu}Y_{\mu}^{i\pm}%
-ig_{5}\left(  \alpha_{G}Y_{\nu}^{j+}(-T_{ji}^{a})G_{\mu}^{a}-\alpha_{W}%
\frac{1}{2}W_{\mu}^{3}Y_{\nu}^{i+}+\alpha_{W}\frac{1}{\sqrt{2}}W_{\mu}%
^{-}X_{\nu}^{i+}+\alpha_{B}y\frac{5}{6}B_{\mu}Y_{\nu}^{i+}\right)  \ .
\end{align}
Plugging this in $\mathcal{L}_{\text{gf}}$ generates new contributions to
$\mathcal{L}_{\text{gauge}}$ and $\mathcal{L}_{\text{gauge-Higgs}}$. 
At this stage, one of the interest of this gauge becomes apparent. 
The gauge and gauge-WBG Lagrangian of the previous section must be
invariant under the SM gauge symmetry. This means that among the
WBG-gauge-gauge interactions, there are precisely those needed to promote the
derivatives in $\mathcal{L}_{\text{mix}}$ to covariant ones. But then, having
covariant derivatives in $\mathcal{L}_{\text{gf}}$ cancels them out. As a
result, when $\alpha_{i}=1$, the $A-A-WBG$ couplings get much simpler.

To this constraint corresponds the ghost Lagrangian%
\begin{equation}
\mathcal{L}_{\text{ghost}}=c^{A\dagger}\left(  \left.  (-g_{5})\frac{\delta
G^{A}}{\delta\lambda^{B}}\right\vert _{\lambda=0}\right)  c^{B}\ .
\end{equation}
To get the variation of $G^{A}$ under a gauge transformation, we first need
that of the fields, expressed in the same physical basis as the gauge bosons
and WBG scalars. For the gauge fields, the variation under a gauge
transformation is
\begin{equation}
\delta\mathbf{A}^{\mu}=\frac{1}{g_{5}}D_{\mu}\mathbf{\lambda}=\frac{1}{g_{5}%
}\partial^{\mu}\mathbf{\lambda}-i\left[  \mathbf{A}^{\mu},\mathbf{\lambda
}\right]  \ ,
\end{equation}
where the physical basis parameters are defined from $\mathbf{\lambda}%
=\lambda^{A}T^{A}$ in full analogy to the gauge bosons. In explicit form,
reconstructing the individual field transformation,%
\begin{align}
\delta G_{\mu}^{a}  & =\frac{1}{g_{5}}\partial^{\mu}\lambda_{G}^{a}%
+f^{abc}G_{\mu}^{b}\lambda_{G}^{c}+i(X_{\mu}^{i+}T_{ij}^{a}\lambda_{X}%
^{j-}-\lambda_{X}^{i+}T_{ij}^{a}X_{\mu}^{j-}+Y_{\mu}^{i+}T_{ij}^{a}\lambda
_{Y}^{j-}-\lambda_{Y}^{i+}T_{ij}^{a}Y_{\mu}^{j-})\ ,\\
\delta W_{\mu}^{+}  & =\frac{1}{g_{5}}\partial^{\mu}\lambda_{W}^{+}+iW_{\mu
}^{+}\lambda_{W}^{3}-iW_{\mu}^{3}\lambda_{W}^{+}+\frac{i}{\sqrt{2}}%
(\lambda_{X}^{i+}Y_{\mu}^{i-}-\lambda_{Y}^{i-}X_{\mu}^{i+})\ ,\ \delta W_{\mu
}^{-}=(\delta W_{\mu}^{+})^{\dagger}\ ,\\
\delta W_{\mu}^{3}  & =\frac{1}{g_{5}}\partial^{\mu}\lambda_{W}^{3}+iW_{\mu
}^{-}\lambda_{W}^{+}-iW_{\mu}^{+}\lambda_{W}^{-}+\frac{i}{2}(\lambda_{X}%
^{i+}X_{\mu}^{i-}-\lambda_{X}^{i-}X_{\mu}^{i+}-\lambda_{Y}^{i+}Y_{\mu}%
^{i-}+\lambda_{Y}^{i-}Y_{\mu}^{i+})\ ,\\
\delta B_{\mu}  & =\frac{1}{g_{5}}\partial^{\mu}\lambda_{B}+\frac{i}{2}%
\sqrt{\frac{5}{3}}(\lambda_{X}^{i+}X_{\mu}^{i-}-\lambda_{X}^{i-}X_{\mu}%
^{i+}+\lambda_{Y}^{i+}Y_{\mu}^{i-}-\lambda_{Y}^{i-}Y_{\mu}^{i+})\ ,\\
\delta X_{\mu}^{i+}  & =\frac{1}{g_{5}}\partial^{\mu}\lambda_{X}%
^{i+}-i(\lambda_{X}^{k+}T_{ki}^{a}G_{\mu}^{a}-X_{\mu}^{k+}T_{ki}^{a}%
\lambda_{G}^{a})+\frac{i}{\sqrt{2}}(\lambda_{W}^{+}Y_{\mu}^{i+}-\lambda
_{Y}^{i+}W_{\mu}^{+})\nonumber\\
& +\frac{i}{2}(\lambda_{W}^{3}X_{\mu}^{i+}-\lambda_{X}^{i+}W_{\mu}^{3}%
)+\frac{i}{2}\sqrt{\frac{5}{3}}(\lambda_{B}X_{\mu}^{i+}-\lambda_{X}^{i+}%
B_{\mu})\ ,\ \delta X_{\mu}^{i-}=(\delta X_{\mu}^{i+})^{\dagger}\ ,
\end{align}
\begin{align}
\delta Y_{\mu}^{i+}  & =\frac{1}{g_{5}}\partial^{\mu}\lambda_{Y}%
^{i+}+i(\lambda_{Y}^{k+}T_{ki}^{a}G_{\mu}^{a}-Y_{\mu}^{k+}T_{ki}^{a}%
\lambda_{G}^{a})+\frac{i}{\sqrt{2}}(\lambda_{W}^{-}X_{\mu}^{i+}-\lambda
_{X}^{i+}W_{\mu}^{-})\nonumber\\
& -\frac{i}{2}(\lambda_{W}^{3}Y_{\mu}^{i+}-\lambda_{Y}^{i+}W_{\mu}^{3}%
)+\frac{i}{2}\sqrt{\frac{5}{3}}(\lambda_{B}Y_{\mu}^{i+}-\lambda_{Y}^{i+}%
B_{\mu})\ \ ,\ \delta Y_{\mu}^{i-}=(\delta Y_{\mu}^{i+})^{\dagger}\ .
\end{align}
Similarly, the transformation of the scalar fields in the adjoint
representation $\delta H^{A}=f^{ABC}H^{B}\lambda^{C}$ can be obtained in
matrix form%
\[
g\delta\mathbf{\bar{H}}_{24}=i[\mathbf{\lambda},\mathbf{\bar{H}}%
_{24}]\rightarrow\delta\mathbf{H}_{24}=i[\mathbf{\lambda},\mathbf{H}%
_{24}]+i[\mathbf{\lambda},\mathbf{H}_{24}^{0}]\ .
\]
We only need the transformation rule of the WBG, since the other scalar fields
will not be introduced in the gauge constraints:%
\begin{align}
\delta H_{X}^{i+}  & =-iH_{X}^{k+}T_{ki}^{a}\lambda_{G}^{a}+\frac{i}{\sqrt{2}%
}\lambda_{W}^{+}H_{Y}^{i+}+\frac{i}{2}\lambda_{W}^{3}H_{X}^{i+}+\frac{i}%
{2}\sqrt{\frac{5}{3}}\lambda_{B}H_{X}^{i+}+i\frac{5}{4}v_{5}\lambda_{X}^{i+}\nonumber\\
& +i\lambda_{X}^{k+}T_{ki}^{a}H_{G}^{a}-\frac{i}{\sqrt{2}}H_{W}^{+}\lambda
_{Y}^{i+}-\frac{i}{2}H_{W}^{3}\lambda_{X}^{i+}-\frac{i}{2}\sqrt{\frac{5}{3}%
}H_{B}\lambda_{X}^{i+}\ ,\\
\delta H_{Y}^{i+}  & =-iH_{Y}^{k+}T_{ki}^{a}\lambda_{G}^{a}+\frac{i}{\sqrt{2}%
}\lambda_{W}^{-}H_{X}^{i+}-\frac{i}{2}\lambda_{W}^{3}H_{Y}^{i+}+\frac{i}%
{2}\sqrt{\frac{5}{3}}\lambda_{B}H_{Y}^{i+}+i\frac{5}{4}v_{5}\lambda_{Y}^{i+}\nonumber\\
& +i\lambda_{Y}^{k+}T_{ki}^{a}H_{G}^{a}-\frac{i}{\sqrt{2}}H_{W}^{-}\lambda
_{X}^{i+}+\frac{i}{2}H_{W}^{3}\lambda_{Y}^{i+}-\frac{i}{2}\sqrt{\frac{5}{3}%
}H_{B}\lambda_{Y}^{i+}\ .
\end{align}
Note that these transformation rules imply that only the ghost fields
associated to the massive gauge bosons couple to all the Higgs bosons, as
expected from the absence of direct couplings of the scalar fields to SM gauge bosons.

Once $\mathbf{G}$ is expressed in the physical basis (as in Eq.~(\ref{Gphys})
for the linear gauge), the physical gauge parameters identified from
$\mathbf{\lambda}$, and ghost matrices defined in full analogy as%
\begin{equation}
\mathbf{c}=c_{A}T^{A}=\left(
\begin{array}
[c]{ccc}%
T_{ij}^{a}c_{G}^{a}-\frac{1}{\sqrt{15}}c_{B}\delta_{ij} & \frac{1}{\sqrt{2}%
}c_{X}^{i-} & \frac{1}{\sqrt{2}}c_{Y}^{i-}\\
\frac{1}{\sqrt{2}}c_{X}^{j+} & \frac{1}{2}c_{W}^{3}+\frac{3}{2\sqrt{15}}c_{B}
& \frac{1}{\sqrt{2}}c_{W}^{+}\\
\frac{1}{\sqrt{2}}c_{Y}^{j+} & \frac{1}{\sqrt{2}}c_{W}^{-} & -\frac{1}{2}%
c_{W}^{3}+\frac{3}{2\sqrt{15}}c_{B}%
\end{array}
\right)  \,,
\end{equation}
one can proceed by computing $-\sqrt{2}g_{5}\langle\mathbf{c}^{\dagger
}\mathbf{\ }\delta\mathbf{G}\rangle$ and replacing each $\lambda$ by the
corresponding ghost, i.e., $\lambda_{B}\rightarrow c_{B}$, $\lambda_{G}%
^{a}\rightarrow c_{G}^{a}$, etc. Given the many possible couplings once a
non-linear gauge fixing is imposed, the final expression are very lengthy and
will not be written down here. Let us just remark that only the ghosts
associated to the leptoquarks get massive,%
\begin{align}
\mathcal{L}_{\text{ghost}}  & =c_{G}^{a\dagger}(-\partial^{2})c_{G}%
+c_{B}^{\dagger}(-\partial^{2})c_{B}+c_{W}^{3\dagger}(-\partial^{2})c_{W}%
^{3}+c_{W}^{\dagger+}(-\partial^{2})c_{W}^{-}+c_{W}^{\dagger-}(-\partial
^{2})c_{W}^{+}\nonumber\\
& +c_{X}^{\dagger+}(-\partial^{2}-\xi M_{XY}^{2})c_{X}^{-}+c_{X}^{\dagger
-}(-\partial^{2}-\xi M_{XY}^{2})c_{X}^{+}+c_{Y}^{\dagger+}(-\partial^{2}-\xi
M_{XY}^{2})c_{Y}^{-}+c_{Y}^{\dagger-}(-\partial^{2}-\xi M_{XY}^{2})c_{Y}%
^{+}\nonumber\\
& +\mathcal{L}_{\text{CCV}}+\mathcal{L}_{\text{CCH}}+\mathcal{L}_{\text{CCVV}%
}\ .
\end{align}
Still, the SM ghosts get new interactions with pairs of heavy states (one
ghost, one gauge boson). Note also that $\mathcal{L}_{\text{CCVV}}$ derives
entirely from the non-linear gauge fixing.

\newpage
\section{Casimir invariants of standard Lie algebras\label{AppCasimir}}

The structure constants of a simple Lie algebra are defined as
$[T_{\mathbf{R}}^{a},T_{\mathbf{R}}^{b}]=if^{abc}T_{\mathbf{R}}^{c}$, with
$T_{\mathbf{R}}^{a}$ the generators in the representation $\mathbf{R}$. The
quadratic and cubic Casimir invariants are defined in terms of the fully
symmetrized trace over two and three generators%
\begin{align}
\frac{1}{2!}S\operatorname*{Tr}(T_{\mathbf{R}}^{a}T_{\mathbf{R}}^{b})  &
=\operatorname*{Tr}(T_{\mathbf{R}}^{a}T_{\mathbf{R}}^{b})\equiv I_{2}%
(\mathbf{R})d^{ab}\ ,\\
\frac{1}{3!}S\operatorname*{Tr}(T_{\mathbf{R}}^{a}T_{\mathbf{R}}%
^{b}T_{\mathbf{R}}^{c})  &  =\frac{1}{2}\operatorname*{Tr}(T_{\mathbf{R}}%
^{a}\{T_{\mathbf{R}}^{b},T_{\mathbf{R}}^{c}\})\equiv\frac{1}{4}I_{3}%
(\mathbf{R})d^{abc}\ .
\end{align}
In terms of these two invariants, we can reduce the trace over three
generators as%
\begin{equation}
\operatorname*{Tr}(T_{\mathbf{R}}^{a}T_{\mathbf{R}}^{b}T_{\mathbf{R}}%
^{c})=\frac{1}{2}\operatorname*{Tr}([T_{\mathbf{R}}^{a},T_{\mathbf{R}}%
^{b}]T_{\mathbf{R}}^{c})+\frac{1}{2}\operatorname*{Tr}(\{T_{\mathbf{R}}%
^{a},T_{\mathbf{R}}^{b}\}T_{\mathbf{R}}^{c})=\frac{I_{3}(\mathbf{R})}%
{4}d^{abc}+\frac{iI_{2}(\mathbf{R})}{2}f^{abc}\ . \label{TriTrace}%
\end{equation}

The quadratic invariant defines a metric in the generator space.
$\operatorname*{Tr}(T_{\mathbf{R}}^{a}T_{\mathbf{R}}^{b})$ being positive
definite, it is always possible to choose a basis for the generators so that
$d^{ab}=\delta^{ab}$. By convention, the generators are further normalized so
that $I_{2}(\mathbf{F})\equiv c$, with $\mathbf{F}$ the defining
representation of dimension $N(\mathbf{F})=N$ and the constant $c$ usually set
to $1/2$ or $1$. Note also that once $d^{ab}=\delta^{ab}$, $T_{\mathbf{R}}%
^{a}T_{\mathbf{R}}^{a}$ becomes proportional to the identity, with
$T_{\mathbf{R}}^{a}T_{\mathbf{R}}^{a}=(N(\mathbf{A})I_{2}(\mathbf{R}%
)/N(\mathbf{R}))\mathbf{1}_{N(\mathbf{R})\times N(\mathbf{R})}$ where
$N(\mathbf{R})$ denotes the dimension of the representation $\mathbf{R}$,
while $\mathbf{A}$ stands for the adjoint representation.

The totally symmetric tensor $d^{abc}$ is normalized such that $I_{3}%
(\mathbf{F})\equiv1$ for unitary groups. It is absent for orthogonal groups,
except for $SO(6)$ isomorphic to $SU(4)$. When defined, the coefficient
$I_{3}(\mathbf{R})$ is often called the anomaly coefficient of the representation
$\mathbf{R}$.

\subsection*{Quartic symmetric symbol}

To compute traces over four generators, we need to extend the basis to include
the quartic symmetric symbol and its associated invariant (for more
information, see Ref.~\cite{vanRitbergenSV99}). It is not immediately given by the fully
symmetric trace over four generators because the symmetrized product of two
second-order symmetric symbols is an invariant symmetric tensor with four
indices. Specifically, the most general decomposition is:%
\begin{equation}
\frac{1}{4!}S\operatorname*{Tr}(T_{\mathbf{R}}^{a}T_{\mathbf{R}}%
^{b}T_{\mathbf{R}}^{c}T_{\mathbf{R}}^{d})=I_{4}(\mathbf{R})d^{abcd}%
+\Lambda(\mathbf{R})(\delta^{ab}\delta^{cd}+\delta^{ac}\delta^{bd}+\delta
^{ad}\delta^{bc})\ . \label{GenQuartic}%
\end{equation}
The constant $\Lambda(\mathbf{R})$ is a matter of convention, while $d^{abcd}$
is normalized by fixing $I_{4}(\mathbf{F})=c$ for some chosen constant $c$. To
fix $\Lambda(\mathbf{R})$, we choose to define the tensor $d^{abcd}$ as
orthogonal to the lower rank invariants, i.e., such that $d_{ab}d_{cd}%
d^{abcd}=0$:%
\begin{align}
I_{4}(\mathbf{R})d_{ab}d_{cd}d^{abcd}  &  =\frac{1}{4!}\delta_{ab}\delta
_{cd}S\operatorname*{Tr}(T_{\mathbf{R}}^{a}T_{\mathbf{R}}^{b}T_{\mathbf{R}%
}^{c}T_{\mathbf{R}}^{d})-\delta_{ab}\delta_{cd}\Lambda(\mathbf{R})(\delta
^{ab}\delta^{cd}+\delta^{ac}\delta^{bd}+\delta^{ad}\delta^{bc})\nonumber\\
&  =\operatorname*{Tr}(T_{\mathbf{R}}^{a}T_{\mathbf{R}}^{a}T_{\mathbf{R}}%
^{b}T_{\mathbf{R}}^{b})+\frac{1}{3}\operatorname*{Tr}(T_{\mathbf{R}}%
^{a}[T_{\mathbf{R}}^{b},T_{\mathbf{R}}^{a}]T_{\mathbf{R}}^{b})-\Lambda
(\mathbf{R})(2+N(\mathbf{A}))N(\mathbf{A})\nonumber\\
&  =\left(  \frac{N(\mathbf{A})I_{2}(\mathbf{R})}{N(\mathbf{R})}-\frac
{I_{2}(\mathbf{A})}{6}\right)  I_{2}(\mathbf{R})N(\mathbf{A})-\Lambda
(\mathbf{R})(2+N(\mathbf{A}))N(\mathbf{A})\ ,
\end{align}
where we have used $f^{abc}f^{dbc}=I_{2}(\mathbf{A})\delta^{ad}$,
$f^{abc}f^{abc}=I_{2}(\mathbf{A})N(\mathbf{A})$. Hence, $d_{ab}d_{cd}d^{abcd}$
vanishes provided
\begin{equation}
\Lambda(\mathbf{R})=\left(  \frac{N(\mathbf{A})I_{2}(\mathbf{R})}%
{N(\mathbf{R})}-\frac{I_{2}(\mathbf{A})}{6}\right)  \frac{I_{2}(\mathbf{R}%
)}{2+N(\mathbf{A})}\ . \label{LambdaDef}%
\end{equation}
This convention ensures $d^{abcd}$ has no left-over part proportional to the
quadratic symbol. This is particularly convenient because $I_{4}(\mathbf{R})$
then vanishes for all $\mathbf{R}$ of $SU(2)$ and $SU(3)$. Remember that a
tensor $d^{abcd}$ such that $d_{ab}d_{cd}d^{abcd}=0$ does not exist for
$SU(N\leqslant3)$. For $N=2,3$, $\Lambda(\mathbf{F})=1/24$ and%
\begin{equation}
S\operatorname*{Tr}(T_{\mathbf{F}}^{a}T_{\mathbf{F}}^{b}T_{\mathbf{F}}%
^{c}T_{\mathbf{F}}^{d})\overset{N=2,3}{=}\delta^{ab}\delta^{cd}+\delta
^{ac}\delta^{bd}+\delta^{ad}\delta^{bc}\ .
\end{equation}
This formula also provides a useful identity for the $SU(3)$ structure
constant:
\begin{subequations}
\begin{equation}
\frac{1}{4!}S\operatorname*{Tr}(T_{\mathbf{8}}^{a}T_{\mathbf{8}}%
^{b}T_{\mathbf{8}}^{c}T_{\mathbf{8}}^{b})=\frac{1}{4!}\sum_{perm(a,b,c,d)}%
f^{ax_{1}x_{2}}f^{bx_{2}x_{3}}f^{cx_{3}x_{4}}f^{dx_{4}x_{1}}=\frac{3}%
{4}(\delta^{ab}\delta^{cd}+\delta^{ac}\delta^{bd}+\delta^{ad}\delta^{bc})\ ,
\end{equation}
since $\Lambda(\mathbf{8})=3/4$.

The formula Eq.~(\ref{GenQuartic}) is valid for all unitary and orthogonal
algebras, except for $SO(8)$. Indeed, the $N$-dimensional Levi-Civita symbol is
an invariant for $SO(N)$, and when $N$ is even, it is possible to construct
out of it a symmetric symbol with $N/2$ indices. To see this, remember that
the adjoint $\mathbf{A}$ of $SO(N)$ is obtained as the antisymmetric tensor
product of the defining $N$-dimensional representation $\mathbf{F}$,
$\mathbf{A}=\mathbf{F}\otimes_{A}\mathbf{F}$. Thus, the $SO(N)$ generators can
be labelled by antisymmetric combinations of two indices $i,j=1,...,N$. If we
denote $a=(i,j)$, with $a=1,...,N(N-1)/2$, then
\end{subequations}
\begin{equation}
\Theta^{a_{1}...a_{N/2}}=\eta\varepsilon^{i_{1}...i_{N}}\ , \label{SymTensor}%
\end{equation}
with $\eta$ some constants, is a totally symmetric invariant tensor with $N/2$
indices. This explains one aspect of the isomorphism $SO(6)\sim SU(4)$. None
of the orthogonal algebras have a genuine $d^{abc}$ symbol, but the extra
invariant tensor $\Theta^{abc}$ of $SO(6)$ corresponds to the $d^{abc}$ symbol
of $SU(4)$. For $SO(8)$, $\Theta^{abcd}$ is an additional quartic symbol,
orthogonal to both tensor structures in Eq.~(\ref{GenQuartic}). Thus, the
totally symmetric trace over four $SO(8)$ generators projects not just on two
but three tensor structures.

\subsection*{Fourth-order trace reductions}

Any trace over four generators can be reduced and expressed entirely in terms
of the invariant tensors. For instance, for $SU(N)$ and $SO(N\neq8)$, we can
write
\begin{align}
\frac{1}{4!}S\operatorname*{Tr}(T_{\mathbf{R}}^{a}T_{\mathbf{R}}%
^{b}T_{\mathbf{R}}^{c}T_{\mathbf{R}}^{d})  &  =\frac{1}{6}\operatorname*{Tr}%
(T_{\mathbf{R}}^{a}T_{\mathbf{R}}^{b}T_{\mathbf{R}}^{c}T_{\mathbf{R}}%
^{d})+\frac{1}{6}\operatorname*{Tr}(T_{\mathbf{R}}^{a}T_{\mathbf{R}}%
^{b}T_{\mathbf{R}}^{d}T_{\mathbf{R}}^{c})+\frac{1}{6}\operatorname*{Tr}%
(T_{\mathbf{R}}^{a}T_{\mathbf{R}}^{c}T_{\mathbf{R}}^{b}T_{\mathbf{R}}%
^{d})\nonumber\\
&  +\frac{1}{6}\operatorname*{Tr}(T_{\mathbf{R}}^{a}T_{\mathbf{R}}%
^{c}T_{\mathbf{R}}^{d}T_{\mathbf{R}}^{b})+\frac{1}{6}\operatorname*{Tr}%
(T_{\mathbf{R}}^{a}T_{\mathbf{R}}^{d}T_{\mathbf{R}}^{b}T_{\mathbf{R}}%
^{c})+\frac{1}{6}\operatorname*{Tr}(T_{\mathbf{R}}^{a}T_{\mathbf{R}}%
^{d}T_{\mathbf{R}}^{c}T_{\mathbf{R}}^{b})\nonumber\\
&  =\operatorname*{Tr}(T_{\mathbf{R}}^{a}T_{\mathbf{R}}^{b}T_{\mathbf{R}}%
^{c}T_{\mathbf{R}}^{d})+\frac{2}{6}if^{dce}\operatorname*{Tr}(T_{\mathbf{R}%
}^{a}T_{\mathbf{R}}^{b}T_{\mathbf{R}}^{e})+\frac{3}{6}if^{cbe}%
\operatorname*{Tr}(T_{\mathbf{R}}^{a}T_{\mathbf{R}}^{e}T_{\mathbf{R}}%
^{d})\nonumber\\
&  +\frac{2}{6}if^{dbe}\operatorname*{Tr}(T_{\mathbf{R}}^{a}T_{\mathbf{R}}%
^{c}T_{\mathbf{R}}^{e})+\frac{1}{6}if^{dbe}\operatorname*{Tr}(T_{\mathbf{R}%
}^{a}T_{\mathbf{R}}^{e}T_{\mathbf{R}}^{c})+\frac{1}{6}if^{dce}%
\operatorname*{Tr}(T_{\mathbf{R}}^{a}T_{\mathbf{R}}^{e}T_{\mathbf{R}}%
^{b})\nonumber\\
&  =\operatorname*{Tr}(T_{\mathbf{R}}^{a}T_{\mathbf{R}}^{b}T_{\mathbf{R}}%
^{c}T_{\mathbf{R}}^{d})+i\frac{I_{3}(\mathbf{R})}{8}(f^{dce}d^{abe}%
+f^{cbe}d^{aed}+f^{dbe}d^{ace})\nonumber\\
&  +\frac{I_{2}(\mathbf{R})}{12}f^{abe}f^{cde}-\frac{I_{2}(\mathbf{R})}%
{4}f^{ade}f^{bce}+\frac{I_{2}(\mathbf{R})}{12}f^{ace}f^{bde}\ .
\end{align}
Or, introducing the quartic invariant:%
\begin{align}
\operatorname*{Tr}(T_{\mathbf{R}}^{a}T_{\mathbf{R}}^{b}T_{\mathbf{R}}%
^{c}T_{\mathbf{R}}^{d})  &  =I_{4}(\mathbf{R})d^{abcd}-i\frac{I_{3}%
(\mathbf{R})}{8}(f^{dce}d^{abe}+f^{cbe}d^{aed}+f^{dbe}d^{ace})\nonumber\\
&  -\frac{I_{2}(\mathbf{R})}{12}(f^{abe}f^{cde}-3f^{ade}f^{bce}+f^{ace}%
f^{bde})+\Lambda(\mathbf{R})(\delta^{ab}\delta^{cd}+\delta^{ac}\delta
^{bd}+\delta^{ad}\delta^{bc})\ .
\end{align}
As special cases, we can set $I_{3}(\mathbf{R})=0$ for $SO(N\neq6)$, $I_{4}%
(\mathbf{R})=0$ for $SU(3)$, and $I_{4}(\mathbf{R})=I_{3}(\mathbf{R})=0$ for
$SU(2)$. Note that the last two terms can be brought to a simpler though less
symmetric form using the Jacobi identities:%
\begin{align}
f^{cde}d^{abe}+f^{ade}d^{bce}+f^{bde}d^{ace}  &  =0\ ,\\
f^{abe}f^{cde}-f^{ace}f^{bde}+f^{ade}f^{bce}  &  =0\ .
\end{align}
Other identities sometimes useful in the computation of triangle graphs are :
\begin{align}
f^{ade}f^{bef}f^{cfd}  &  =+\frac{1}{2}I_{2}(\mathbf{A})f^{abc}\ ,\\
d^{ade}f^{bef}f^{cfd}  &  =-\frac{1}{2}I_{2}(\mathbf{A})d^{abc}\ .
\end{align}
The first identity derives from $\operatorname*{Tr}(T_{\mathbf{A}}%
^{a}T_{\mathbf{A}}^{b}T_{\mathbf{A}}^{c})=\operatorname*{Tr}(T_{\mathbf{A}%
}^{a}[T_{\mathbf{A}}^{b},T_{\mathbf{A}}^{c}])/2$ since $(T_{\mathbf{A}}%
^{a})^{T}=-T_{\mathbf{A}}^{a}$ for a real representation.

Specializing to $SU(N)$, there is another way to derive the fourth-order symmetric symbol.
First, remember that,%
\begin{equation}
T_{\mathbf{F}}^{a}T_{\mathbf{F}}^{b}=\frac{1}{N}I_{2}(\mathbf{F})\delta
^{ab}+\frac{I_{3}(\mathbf{F})}{4I_{2}(\mathbf{F})}d^{abc}T_{\mathbf{F}}%
^{c}+\frac{i}{2}I_{2}(\mathbf{F})f^{abc}T_{\mathbf{F}}^{c}\;.
\end{equation}
With this, we can derive%
\begin{equation}
Tr\left[  \{T_{\mathbf{F}}^{a}T_{\mathbf{F}}^{b}\}\{T_{\mathbf{F}}%
^{c}T_{\mathbf{F}}^{d}\}\right]  =\frac{4I_{2}(\mathbf{F})^{2}}{N}\delta
^{ab}\delta^{cd}+\frac{I_{3}(\mathbf{F})^{2}}{4I_{2}(\mathbf{F})}%
d^{abe}d^{cde}=\frac{1}{N}\delta^{ab}\delta^{cd}+\frac{1}{2}d^{abe}d^{cde}\ .
\end{equation}
On the other hand, this trace can be computed using the general reduction in
terms of invariant, giving%
\begin{equation}
Tr\left[  \{T_{\mathbf{F}}^{a}T_{\mathbf{F}}^{b}\}\{T_{\mathbf{F}}%
^{c}T_{\mathbf{F}}^{d}\}\right]  =4I_{4}(\mathbf{F})d^{abcd}+\frac{1}{3}%
I_{2}(\mathbf{F})(f^{ace}f^{bde}+f^{ade}f^{bce})+4\Lambda(\mathbf{F}%
)(\delta^{ab}\delta^{cd}+\delta^{ac}\delta^{bd}+\delta^{ad}\delta^{bc})\ .
\end{equation}
Combining the two,%
\begin{equation}
I_{4}(\mathbf{F})d^{abcd}=\frac{I_{3}(\mathbf{F})^{2}}{16I_{2}(\mathbf{F}%
)}d^{abe}d^{cde}-\frac{I_{2}(\mathbf{F})}{12}(f^{ace}f^{bde}+f^{ade}%
f^{bce})-\Lambda(\mathbf{F})(\delta^{ab}\delta^{cd}+\delta^{ac}\delta
^{bd}+\delta^{ad}\delta^{bc})+\frac{I_{2}(\mathbf{F})^{2}}{N}\delta^{ab}%
\delta^{cd}\ . \label{D4D3D3}%
\end{equation}
With the convention $I_{4}(\mathbf{F})=1$, this identity permits to compute
the quartic symbol $d^{abcd}$ directly out of the lower-rank invariants. We
can now check that for $N=3$, $I_{2}(\mathbf{F})=1/2$, $I_{3}(\mathbf{F})=1$,
$I_{4}(\mathbf{F})=0$ and $\Lambda(\mathbf{F})=1/24$,%
\begin{equation}
0=\frac{1}{8}d^{abe}d^{cde}-\frac{1}{24}(f^{ace}f^{bde}+f^{ade}f^{bce}%
)-\frac{1}{24}(\delta^{ac}\delta^{bd}-\delta^{ab}\delta^{cd}+\delta^{ad}%
\delta^{bc})\ ,
\end{equation}
which gives back the identity in Eq.~(\ref{IddSU3}). For $N=2$, $I_{2}%
(\mathbf{F})=1/2$, $I_{3}(\mathbf{F})=I_{4}(\mathbf{F})=0$, $\Lambda
(\mathbf{F})=1/24$, we recove the usual reduction formula for Levi-Civita
tensor:%
\begin{equation}
0=-\frac{1}{24}(\varepsilon^{ace}\varepsilon^{bde}+\varepsilon^{ade}%
\varepsilon^{bce})-\frac{1}{24}(\delta^{ac}\delta^{bd}-2\delta^{ab}\delta
^{cd}+\delta^{ad}\delta^{bc})\ .
\end{equation}

\begin{table}[t] \centering\small
\begin{tabular}[c]{lcccccccccc}\hline
\multicolumn{11}{l}{$SU(2)\rule[-0.06in]{0in}{0.2in}$}\\\hline
$\mathbf{R}$ & $(1)$ & $(\mathbf{2})$ & $(3)$ & $(4)$ & $(5)$ & $(6)$ & $(7)$
& $(8)$ & $(9)$ & $(10)\rule[-0.04in]{0in}{0.17in}$\\
$N$ & $\mathbf{2}$ & $\mathbf{3}$ & $\mathbf{4}$ & $\mathbf{5}$ & $\mathbf{6}$
& $\mathbf{7}$ & $\mathbf{8}$ & $\mathbf{9}$ & $\mathbf{10}$ & $\mathbf{11}%
\rule[-0.04in]{0in}{0.17in}$\\
$I_{2}$ & $1/2$ & $2$ & $5$ & $10$ & $35/2$ & $28$ & $42$ & $60$ & $165/2$ &
$110\rule[-0.04in]{0in}{0.17in}$\\
$\Lambda$ & $\dfrac{1}{24}$ & $\dfrac{2}{3}$ & $\dfrac{41}{12}$ & $\dfrac
{34}{3}$ & $\dfrac{707}{24}$ & $\dfrac{196}{3}$ & $\dfrac{259}{2}$ & $236$ &
$\dfrac{3223}{8}$ & $\dfrac{1958}{3}\rule[-0.12in]{0in}{0.32in}$\\\hline
\multicolumn{11}{l}{$SU(3)\rule[-0.06in]{0in}{0.2in}$}\\\hline
$\mathbf{R}$ & $(10)$ & $(20)$ & $(\mathbf{11})$ & $(30)$ & $(21)$ & $(40)$ &
$(05)$ & $(13)$ & $(22)$ & $(60)\rule[-0.04in]{0in}{0.17in}$\\
$N$ & $\mathbf{3}$ & $\mathbf{6}$ & $\mathbf{8}$ & $\mathbf{10}$ &
$\mathbf{15}$ & $\mathbf{15}^{\prime}$ & $\mathbf{21}$ & $\mathbf{24}$ &
$\mathbf{27}$ & $\mathbf{28}\rule[-0.04in]{0in}{0.17in}$\\
$I_{2}$ & $1/2$ & $5/2$ & $3$ & $15/2$ & $10$ & $35/2$ & $35$ & $25$ & $27$ &
$63\rule[-0.04in]{0in}{0.17in}$\\
$I_{3}$ & $1$ & $7$ & $0$ & $27$ & $14$ & $77$ & $-182$ & $-64$ & $0$ &
$378\rule[-0.04in]{0in}{0.17in}$\\
$\Lambda$ & $\dfrac{1}{24}$ & $\dfrac{17}{24}$ & $\dfrac{3}{4}$ & $\dfrac
{33}{8}$ & $\dfrac{29}{6}$ & $\dfrac{371}{24}$ & $\dfrac{539}{12}$ &
$\dfrac{235}{12}$ & $\dfrac{81}{4}$ & $\dfrac{441}{3}\rule[-0.12in]%
{0in}{0.32in}$\\\hline
\multicolumn{11}{l}{$SU(4)\rule[-0.06in]{0in}{0.2in}\rule[-0.06in]%
{0in}{0.2in}$}\\\hline
$\mathbf{R}$ & $(100)$ & $(010)$ & $(200)$ & $(\mathbf{101})$ & $(011)$ &
$(020)$ & $(003)$ & $(400)$ & $(201)$ & $(210)\rule[-0.04in]{0in}{0.17in}$\\
$N$ & $\mathbf{4}$ & $\mathbf{6}$ & $\mathbf{10}$ & $\mathbf{15}$ &
$\mathbf{20}$ & $\mathbf{20}^{\prime}$ & $\mathbf{20}^{\prime\prime}$ &
$\mathbf{35}$ & $\mathbf{36}$ & $\mathbf{45}\rule[-0.04in]{0in}{0.17in}$\\
$I_{2}$ & $1/2$ & $1$ & $3$ & $4$ & $13/2$ & $8$ & $21/2$ & $28$ & $33/2$ &
$24\rule[-0.04in]{0in}{0.17in}$\\
$I_{3}$ & $1$ & $0$ & $8$ & $0$ & $-7$ & $0$ & $-35$ & $112$ & $21$ &
$48\rule[-0.04in]{0in}{0.17in}$\\
$I_{4}$ & $1$ & $-4$ & $12$ & $8$ & $-11$ & $-56$ & $69$ & $272$ & $57$ &
$24\rule[-0.04in]{0in}{0.17in}$\\
$\Lambda$ & $\dfrac{29}{816}$ & $\dfrac{11}{102}$ & $\dfrac{23}{34}$ &
$\dfrac{40}{51}$ & $\dfrac{1313}{816}$ & $\dfrac{128}{51}$ & $\dfrac
{1211}{272}$ & $\dfrac{56}{3}$ & $\dfrac{1639}{272}$ & $\dfrac{176}%
{17}\rule[-0.12in]{0in}{0.32in}$\\\hline
\multicolumn{11}{l}{$SU(5)\rule[-0.06in]{0in}{0.2in}$}\\\hline
$\mathbf{R}$ & \multicolumn{1}{l}{$(1000)$} & \multicolumn{1}{l}{$(0100)$} &
\multicolumn{1}{l}{$(2000)$} & \multicolumn{1}{l}{$(\mathbf{1001})$} &
\multicolumn{1}{l}{$(0003)$} & \multicolumn{1}{l}{$(0011)$} &
\multicolumn{1}{l}{$(0101)$} & \multicolumn{1}{l}{$(0020)$} &
\multicolumn{1}{l}{$(2001)$} & \multicolumn{1}{l}{$(0110)\rule[-0.04in]%
{0in}{0.17in}$}\\
$N$ & $\mathbf{5}$ & $\mathbf{10}$ & $\mathbf{15}$ & $\mathbf{24}$ &
$\mathbf{35}$ & $\mathbf{40}$ & $\mathbf{45}$ & $\mathbf{50}$ & $\mathbf{70}$
& $\mathbf{75}\rule[-0.04in]{0in}{0.17in}$\\
$I_{2}$ & $1/2$ & $3/2$ & $7/2$ & $5$ & $14$ & $11$ & $12$ & $35/2$ & $49/2$ &
$25\rule[-0.04in]{0in}{0.17in}$\\
$I_{3}$ & $1$ & $1$ & $9$ & $0$ & $-44$ & $-16$ & $-6$ & $-15$ & $29$ &
$0\rule[-0.04in]{0in}{0.17in}$\\
$I_{4}$ & $1$ & $-3$ & $13$ & $10$ & $82$ & $-2$ & $-6$ & $-55$ & $79$ &
$-70\rule[-0.04in]{0in}{0.17in}$\\
$\Lambda$ & $\dfrac{47}{1560}$ & $\dfrac{83}{520}$ & $\dfrac{77}{120}$ &
$\dfrac{125}{156}$ & $\dfrac{1841}{390}$ & $\dfrac{1903}{780}$ & $\dfrac
{167}{65}$ & $\dfrac{1589}{312}$ & $\dfrac{11123}{1560}$ & $\dfrac{1075}%
{156}\rule[-0.12in]{0in}{0.32in}$\\\hline
\end{tabular}%
\caption{First few representations of $SU(N)$, $N=2,3,4,5$, labelled by their Dynkin index,
and their dimensions, quadratic, cubic, and quartic Casimir invariants, together with
$\Lambda(\mathbf{R})$ as given by Eq. (\ref{LambdaDef}).}%
\label{TableCasimirsSU}%
\end{table}%

\begin{table}[t] \centering\small
\begin{tabular}[c]{lcccccccccc}\hline
\multicolumn{11}{l}{$SO(5)\rule[-0.06in]{0in}{0.2in}$}\\\hline
$\mathbf{R}$ & $(10)$ & $(01)$ & $(\mathbf{02})$ & $(20)$ & $(11)$ & $(03)$ &
$(30)$ & $(12)$ & $(04)$ & $(21)\rule[-0.04in]{0in}{0.17in}$\\
$N$ & $\mathbf{5}$ & $\mathbf{4}$ & $\mathbf{10}$ & $\mathbf{14}$ &
$\mathbf{16}$ & $\mathbf{20}$ & $\mathbf{30}$ & $\mathbf{35}$ & $\mathbf{35}%
^{\prime}$ & $\mathbf{40}\rule[-0.04in]{0in}{0.17in}$\\
$I_{2}$ & $1$ & $1/2$ & $3$ & $7$ & $6$ & $21/2$ & $27$ & $21$ & $28$ &
$29\rule[-0.04in]{0in}{0.17in}$\\
$I_{4}$ & $2$ & $-1/2$ & $-6$ & $26$ & $6$ & $-69/2$ & $162$ & $-6$ & $-132$ &
$91\rule[-0.04in]{0in}{0.17in}$\\
$\Lambda$ & $\dfrac{1}{8}$ & $\dfrac{1}{32}$ & $\dfrac{5}{8}$ & $\dfrac{21}%
{8}$ & $\dfrac{13}{8}$ & $\dfrac{133}{32}$ & $\dfrac{153}{8}$ & $\dfrac{77}%
{8}$ & $\dfrac{35}{2}$ & $\dfrac{261}{16}\rule[-0.12in]{0in}{0.32in}$\\\hline
\multicolumn{11}{l}{$SO(7)\rule[-0.06in]{0in}{0.2in}$}\\\hline
$\mathbf{R}$ & $(100)$ & $(001)$ & $(\mathbf{010})$ & $(200)$ & $(002)$ &
$(101)$ & $(300)$ & $(110)$ & $(011)$ & $(003)\rule[-0.04in]{0in}{0.17in}$\\
$N$ & $\mathbf{7}$ & $\mathbf{8}$ & $\mathbf{21}$ & $\mathbf{27}$ &
$\mathbf{35}$ & $\mathbf{48}$ & $\mathbf{77}$ & $\mathbf{105}$ &
$\mathbf{112}$ & $\mathbf{112}\rule[-0.04in]{0in}{0.17in}$\\
$I_{2}$ & $1$ & $1$ & $5$ & $9$ & $10$ & $14$ & $44$ & $45$ & $46$ &
$54\rule[-0.04in]{0in}{0.17in}$\\
$I_{4}$ & $2$ & $-1$ & $-2$ & $30$ & $-16$ & $10$ & $220$ & $42$ & $-46$ &
$-126\rule[-0.04in]{0in}{0.17in}$\\
$\Lambda$ & $\dfrac{13}{138}$ & $\dfrac{43}{552}$ & $\dfrac{125}{138}$ &
$\dfrac{111}{46}$ & $\dfrac{155}{69}$ & $\dfrac{889}{276}$ & $\dfrac{1474}%
{69}$ & $\dfrac{735}{46}$ & $\dfrac{187}{12}$ & $\dfrac{2007}{92}%
\rule[-0.12in]{0in}{0.32in}$\\\hline
\multicolumn{11}{l}{$SO(9)\rule[-0.06in]{0in}{0.2in}\rule[-0.06in]%
{0in}{0.2in}$}\\\hline
$\mathbf{R}$ & $(1000)$ & $(0001)$ & $(\mathbf{0100})$ & $(2000)$ & $(0010)$ &
$(0002)$ & $(1001)$ & $(3000)$ & $(1100)$ & $(0101)\rule[-0.04in]%
{0in}{0.17in}$\\
$N$ & $\mathbf{9}$ & $\mathbf{16}$ & $\mathbf{36}$ & $\mathbf{44}$ &
$\mathbf{84}$ & $\mathbf{126}$ & $\mathbf{128}$ & $\mathbf{156}$ &
$\mathbf{231}$ & $\mathbf{432}\rule[-0.04in]{0in}{0.17in}$\\
$I_{2}$ & $1$ & $2$ & $7$ & $11$ & $21$ & $35$ & $32$ & $65$ & $77$ &
$150\rule[-0.04in]{0in}{0.17in}$\\
$I_{4}$ & $2$ & $-2$ & $2$ & $34$ & $-18$ & $-50$ & $16$ & $286$ & $106$ &
$-54\rule[-0.04in]{0in}{0.17in}$\\
$\Lambda$ & $\dfrac{17}{228}$ & $\dfrac{10}{57}$ & $\dfrac{245}{228}$ &
$\dfrac{517}{228}$ & $\dfrac{329}{76}$ & $\dfrac{1855}{228}$ & $\dfrac
{376}{57}$ & $\dfrac{5395}{228}$ & $\dfrac{5005}{228}$ & $\dfrac{850}%
{19}\rule[-0.12in]{0in}{0.32in}$\\\hline
\multicolumn{11}{l}{$SO(10)\rule[-0.06in]{0in}{0.2in}$}\\\hline
$\mathbf{R}$ & $(10000)$ & $(00001)$ & $(\mathbf{01000})$ & $(20000)$ &
$(00100)$ & $(00002)$ & $(10010)$ & $(00011)$ & $(30000)$ &
$(11000)\rule[-0.04in]{0in}{0.17in}$\\
$N$ & $\mathbf{10}$ & $\mathbf{16}$ & $\mathbf{45}$ & $\mathbf{54}$ &
$\mathbf{120}$ & $\mathbf{126}$ & $\mathbf{144}$ & $\mathbf{210}$ &
$\mathbf{210}^{\prime}$ & $\mathbf{320}\rule[-0.04in]{0in}{0.17in}$\\
$I_{2}$ & $1$ & $2$ & $8$ & $12$ & $28$ & $35$ & $34$ & $56$ & $77$ &
$96\rule[-0.04in]{0in}{0.17in}$\\
$I_{4}$ & $2$ & $-2$ & $4$ & $36$ & $-16$ & $-50$ & $14$ & $-68$ & $322$ &
$144\rule[-0.04in]{0in}{0.17in}$\\
$\Lambda$ & $\dfrac{19}{282}$ & $\dfrac{103}{564}$ & $\dfrac{160}{141}$ &
$\dfrac{104}{47}$ & $\dfrac{770}{141}$ & $\dfrac{2345}{282}$ & $\dfrac
{3791}{564}$ & $\dfrac{1792}{141}$ & $\dfrac{7007}{282}$ & $\dfrac{1168}%
{47}\rule[-0.12in]{0in}{0.32in}$\\\hline
\end{tabular}%
\caption{First few representations of $SO(N)$, $N=5,7,9,10$, labelled by their Dynkin index, and their dimensions, quadratic, and quartic Casimir invariants, together with $\Lambda(\mathbf{R})$ as given by Eq. (\ref{LambdaDef}). The cubic invariant vanishes for all these algebras. The normalizations of the generators and of the quartic symbols is fixed in terms of that adopted for $SU(N)$ algebras, using Eq.~(\ref{InBR}). The $SO(4)$ and $SO(6)$ algebras are not included since they are isomorphic to $SU(2)\otimes SU(2)$ and $SU(4)$, respectively. Note that the normalizations does not necessarily match, with for example $I_2(SU(4)) = I_2(SO(6))$ but $I_4(SU(4)) = - 2 I_4(SO(6))$.}%
\label{TableCasimirsSO}%
\end{table}%

\begin{table}[t] \centering\small
\begin{tabular}[c]{|c|ccc|ccc|ccc|c|}\hline
\multicolumn{11}{|l|}{$SO(8)$\rule[-0.06in]{0in}{0.2in}\rule[-0.06in]%
{0in}{0.2in}}\\\hline
& \multicolumn{3}{|c|}{$\mathbf{8}\ (I_{2},\Lambda=1,1/12)$} &
\multicolumn{3}{|c|}{$\mathbf{112}\ (I_{2},\Lambda=54,45/2)$} &
\multicolumn{3}{|c|}{$\mathbf{224}\ (I_{2},\Lambda=100,115/3)$} &
$\mathbf{28}(\ I_{2},\Lambda=6,1)$\\
& $(1000)$ & $(0001)$ & $(0010)$ & $(2000)$ & $(0002)$ & $(0020)$ & $(1002)$ &
$(1020)$ & $(2001)$ & $(\mathbf{0100)}\rule[-0.04in]{0in}{0.17in}$\\
$I_{4}$ & $2$ & $-1$ & $-1$ & $252$ & $-126$ & $-126$ & $-40$ & $-40$ & $212$
& $0\rule[-0.04in]{0in}{0.17in}$\\
$I_{4}^{\prime}$ & $0$ & $-1$ & $1$ & $0$ & $-126$ & $126$ & $-128$ & $128$ &
$-44$ & $0\rule[-0.04in]{0in}{0.17in}$\\\cline{2-7}\cline{5-7}\cline{11-11}
& \multicolumn{3}{|c|}{$\mathbf{35}\ (I_{2},\Lambda=10,7/3)$} &
\multicolumn{3}{|c|}{$\mathbf{160}\ (I_{2},\Lambda=60,19)$} &  &  &  &
$\mathbf{300}\ (I_{2},\Lambda=150,65)$\\
& $(2000)$ & $(0002)$ & $(0020)$ & $(1100)$ & $(0101)$ & $(0110)$ & $(0012)$ &
$(0021)$ & $(2010)$ & $(0200)\rule[-0.04in]{0in}{0.17in}$\\
$I_{4}$ & $32$ & $-16$ & $-16$ & $72$ & $-36$ & $-36$ & $-172$ & $-172$ &
$212$ & $0\rule[-0.04in]{0in}{0.17in}$\\
$I_{4}^{\prime}$ & $0$ & $-16$ & $16$ & $0$ & $-36$ & $36$ & $-84$ & $84$ &
$44$ & $0\rule[-0.04in]{0in}{0.17in}$\\\cline{2-11}\cline{5-11}
& \multicolumn{3}{|c|}{$\mathbf{56}\ (I_{2},\Lambda=15,13/4)$} &
\multicolumn{3}{|c|}{$\mathbf{294}\ (I_{2},\Lambda=210,133)$} &
\multicolumn{3}{|c|}{$\mathbf{567}\ (I_{2},\Lambda=324,162)$} & $\mathbf{350}%
\ (I_{2},\Lambda=150,55)$\\
& $(0011)$ & $(1001)$ & $(1010)$ & $(4000)$ & $(0004)$ & $(0040)$ & $(2100)$ &
$(0102)$ & $(0120)$ & $(1011)\rule[-0.04in]{0in}{0.17in}$\\
$I_{4}$ & $-18$ & $9$ & $9$ & $1344$ & $-672$ & $-672$ & $864$ & $-432$ &
$-432$ & $0\rule[-0.04in]{0in}{0.17in}$\\
$I_{4}^{\prime}$ & $0$ & $9$ & $-9$ & $0$ & $-672$ & $672$ & $0$ & $-432$ &
$432$ & $0\rule[-0.04in]{0in}{0.17in}$\\\hline
\end{tabular}%
\caption{First few representations of $SO(8)$, labelled by their Dynkin index. Because of the invariance of the eight-dimensional Levi-Civita tensor, this algebra has a second quartic invariant tensor. Its normalization is fixed to make manifest the relationship between the values of both quartic Casimir invariants, and corresponds to $\eta=-1/8$ in Eq.~(\ref{SymTensor}). A second feature of $SO(8)$ is its triality symmetry: dimensions and quadratic Casimir invariants are the same under permutations of the first, third, and fourth simple root. Both quartic Casimir invariants vanish when summed over representations linked by the permutation symmetry~\cite{Okubo81}. This means in particular that they vanish identically for the $\mathbf{28}$, $\mathbf{300}$, and $\mathbf{350}$.}%
\label{TableCasimirsSO8}%
\end{table}%

\subsection*{Casimir invariants for simple groups}

Thanks to the orthogonality condition adopted to fix $\Lambda(\mathbf{R}%
)$~\cite{Okubo81}, the usual formula can be employed to get the explicit
values of the invariant $I_{4}(\mathbf{R})$ for various representations,%
\begin{align}
I_{n}(\mathbf{R})  &  =(-1)^{n}I_{n}(\mathbf{R}^{\dagger})\ ,\\
I_{n}(\mathbf{R}_{1}\oplus\mathbf{R}_{2})  &  =I_{n}(\mathbf{R}_{1}%
)+I_{n}(\mathbf{R}_{2})\ ,\\
I_{n}(\mathbf{R}_{1}\otimes\mathbf{R}_{2})  &  =I_{n}(\mathbf{R}%
_{1})N(\mathbf{R}_{2})+I_{n}(\mathbf{R}_{2})N(\mathbf{R}_{1})=\sum
I_{n}(\mathbf{R}_{i}^{\prime})\;,
\end{align}
with $n=2,3,4$ and where $\mathbf{R}_{1}\otimes\mathbf{R}_{2}=\sum
_{i}\mathbf{R}_{i}^{\prime}$. Altogether, these relations are more than
sufficient to derive the Casimir invariants for any of the standard Lie
algebra. We give in Tables~\ref{TableCasimirsSU},~\ref{TableCasimirsSO}
and~\ref{TableCasimirsSO8} their values for the first few representations of
some unitary and orthogonal algebras of rank $r\leq5$, along with
$\Lambda(\mathbf{R})$. We also checked these numbers by computing
$I_{2,3,4}(\mathbf{R})$ directly using explicit matrix representations for the
first few representations of each algebra. These numbers are compatible with
the explicit formula in terms of Dynkin indices given in Ref.~\cite{Okubo81},
up to the normalization conventions.

The normalization of the generators adopted for $SO(N)$ algebras in
Table~\ref{TableCasimirsSO} and~\ref{TableCasimirsSO8} is not standard but
physically inspired. Specifically, the invariants of an algebra $M$ can be
expressed in terms of that of its subalgebra $N$. For instance, if a
representation $\mathbf{R}_{M}$ branches into the sum of representations
$\mathbf{R}_{N}$, we have the simple sum rule:%
\begin{equation}
I_{n}(\mathbf{R}_{M})=\eta\sum_{\mathbf{R}_{N}\subset\mathbf{R}_{M}}%
I_{n}(\mathbf{R}_{N})\ , \label{InBR}%
\end{equation}
where $\eta$ is a constant reflecting the normalization convention adopted for
the generators of $M$ and $N$. In Table~\ref{TableCasimirsSO}, we chose to fix
$\eta=1$. For example, the generators in the defining representation of
$SO(10)$ are normalized so that%
\begin{equation}
I_{2}(\mathbf{10})^{SO(10)}=I_{2}(\mathbf{\bar{5}})^{SU(5)}+I_{2}%
(\mathbf{5})^{SU(5)}=1\ ,
\end{equation}
since $\mathbf{10}\rightarrow\mathbf{\bar{5}}+\mathbf{5}$. Similarly, the
normalization of the quartic symbol of $SO(10)$ is then fixed by imposing
$I_{4}(\mathbf{10})^{SO(10)}=2I_{4}(\mathbf{5})^{SU(5)}=2$. This makes sense
physically if one thinks of a field in a given $SO(10)$ representation
circulating in some loop. Our normalization conventions make the matching of
this amplitude to that computed in terms of the fields of the subalgebra most
transparent. Note that the generators and quartic symbols of all $SO(N)$
algebras are fixed once that of $SO(10)$ is, since $SO(N)\subset SO(N+1)$.
Further, we also checked that these conventions are compatible with
$SO(3)\otimes SO(7)\subset SO(10)$ and $SO(4)\otimes SO(6)\subset SO(10)$,
with $SO(4)\sim SU(2)\otimes SU(2)$.

Other relations between the invariants of an algebra and that of its
subalgebras are given in the text, see in particular Eq.~(\ref{BranchingI4})
which gives $I_{n}(\mathbf{R}_{M})$ in terms of $I_{n-1}(\mathbf{R}_{N})$,
Eq.~(\ref{IdI4I2}) which fixes $I_{4}(\mathbf{R}_{M})$ in terms of
$I_{2}(\mathbf{R}_{N})$, or Eq.~(\ref{ReducU1}) which gives $I_{n}%
(\mathbf{R}_{M})$ in terms of the $U(1)$ charges of the $\mathbf{R}_{M}$
states. To close this section, let us give a few illustrations for these relations.

Consider first the reduction of $SU(2)$ down to the $U(1)$ subgroup of $SU(2)$
generated by $T^{3}$. Since there is no quartic invariant for $SU(2)$,
Eq.~(\ref{ReducU1}) is easy to check. The fundamental $SU(2)$ representation
corresponds to two complex states of charge $|T^{3}|=1/2$, so we can identify
$2(1/2)^{4}=3\Lambda^{SU(2)}(\mathbf{2})$ since $\Lambda^{SU(2)}%
(\mathbf{2})=1/24$. Similarly, the complex adjoint representation of $SU(2)$
contains two states of unit charge, hence $2=3\Lambda^{SU(2)}(\mathbf{3})$,
and the isospin $3/2$ decomposes into four states such that $2((1/2)^{4}%
+(3/2)^{4})=3\Lambda^{SU(2)}(\mathbf{4})$, which give back the correct values
$\Lambda^{SU(2)}(\mathbf{3})=2/3$ and $\Lambda^{SU(2)}(\mathbf{4})=41/12$. The
same exercise can be repeated for $SU(3)$, for which the absence of the
quartic invariant ensures that $\operatorname*{Tr}((T_{\mathbf{R}}^{3}%
)^{4})=\operatorname*{Tr}((T_{\mathbf{R}}^{8})^{4})$ if $T^{3}$ and $T^{8}$
are the conventional Cartan generators (equal to half the corresponding
Gell-Mann matrices in the fundamental representation). To apply the same
method for $SU(5)$, we need to first fix two free parameters. Specifically, we
start from%
\begin{equation}
3\Lambda(\mathbf{R})+d^{\alpha\alpha\alpha\alpha}I_{4}(\mathbf{R})=\delta
\sum_{q_{\alpha}\in\mathbf{R}}q_{\alpha}^{4}\ ,
\end{equation}
The value of $d^{\alpha\alpha\alpha\alpha}$ and the $U(1)$ generator
normalization $\delta$ (which was coincidentally equal to one in the previous
$SU(2)$ example) need to be fixed. If we identify $T^{\alpha}$ as the
hypercharge generator in the subalgebra $SU(3)\otimes SU(2)\otimes U(1)\subset
SU(5)$, we can use the branching rules~\cite{Slansky}
\begin{subequations}
\label{SU5BR}%
\begin{align}
\mathbf{5}  & =(\mathbf{3},\mathbf{1})_{2}+(\mathbf{1},\mathbf{2})_{-3}\ ,\\
\mathbf{10}  & =(\mathbf{\bar{3}},\mathbf{1})_{4}+(\mathbf{3},\mathbf{2}%
)_{-1}+(\mathbf{1},\mathbf{1})_{-6}\ ,
\end{align}
and these two constants are fixed as
\end{subequations}
\begin{equation}
\left\{
\begin{array}
[c]{c}%
I_{4}(\mathbf{5})d^{\alpha\alpha\alpha\alpha}+3\Lambda(\mathbf{5}%
)=\delta(2\times3^{4}+3\times2^{4})\\
I_{4}(\mathbf{10})d^{\alpha\alpha\alpha\alpha}+3\Lambda(\mathbf{10}%
)=\delta(6^{4}+3\times4^{4}+3\times2\times1^{4})
\end{array}
\right.  \rightarrow\left\{
\begin{array}
[c]{l}%
\delta=1/60^{2}\ ,\\
d^{\alpha\alpha\alpha\alpha}=-5/156\ .
\end{array}
\right.
\end{equation}
One can then check that the $I_{4}$ for the other $SU(5)$ representations are
correctly reproduced.

The same branching rules can be used in connection with Eq.~(\ref{BranchingI4}), 
which we write as%
\begin{equation}
I_{4}(\mathbf{R}_{M})=\eta\sum_{\mathbf{R}_{N}\subset\mathbf{R}_{M}}q_{\alpha
}(\mathbf{R}_{N})I_{3}(\mathbf{R}_{N})\ .
\end{equation}
The subalgebra needs not be maximal so we consider $U(1)\otimes SU(3)\subset
SU(5)$. Using the values quoted in Table~\ref{TableCasimirsSU}, the first rule
of Eq.~(\ref{SU5BR}) translates into $I_{4}^{SU(5)}(\mathbf{5})=2\eta
I_{3}^{SU(3)}(\mathbf{3})$ and fixes $\eta=1/2$. Then, we can check that this
equation is valid for all the other $SU(5)$ representations listed in
Table~\ref{TableCasimirsSU}.

\end{document}